%% file: main.tex
\documentclass[twocolumn]{aastex631}

\usepackage{hyperref}
\usepackage{graphicx}
\usepackage{subcaption}

\begin{document}

\title{Galaxies and Their Environment at $z \gtrsim 10$ --- I: Primordial Chemical Enrichment, Accretion, Cooling, and Virialization of Gas in Dark Matter Halos}

\author{William M. Hicks}
\affiliation{Center for Astrophysics and Space Sciences\\
University of California, San Diego, La Jolla, CA, 92093}

\author{Michael L. Norman}
\affiliation{Center for Astrophysics and Space Sciences\\
University of California, San Diego, La Jolla, CA, 92093}
\affiliation{San Diego Supercomputer Center\\
University of California, San Diego, La Jolla, CA, 92093}

\author{Azton I. Wells}
\affiliation{Argonne National Laboratory\\
University of Chicago, Chicago, IL}

\author{James O. Bordner}
\affiliation{San Diego Supercomputer Center\\
University of California, San Diego, La Jolla, CA, 92093}



\begin{abstract}
Recent observations made using the \textit{James Webb Space Telescope} have identified a number of high-redshift galaxies that are unexpectedly luminous. In light of this, it is clear that a more detailed understanding of the high redshift, pre-reionization universe is required for us to obtain the complete story of galaxy formation. This study is the first in a series that seeks to tell the story of galaxy formation at $z \gtrsim 10$ using a suite of large-scale adaptive mesh refinement cosmological simulations. Our machine-learning-accelerated surrogate model for Population III star formation and feedback, \texttt{StarNet}, gives us an unprecedented ability to obtain physically accurate, inhomogeneous chemical initial conditions for a statistically significant number of galaxies. We find that of the 12,423 halos in the mass range of $10^6\,\,M_\odot < M_\mathrm{vir} < 10^9\,\, M_\odot$ that form in our fiducial simulation, $16\%$ are chemically enriched by Population III supernovae by $z\sim12$. Primordial hypernovae and pair-instability supernovae contribute in approximately equal amounts to the enrichment of halos as a whole, while Type II supernova enrichment is less important. The mean Population III metallicity of gas within a halo is $2\times10^{-3}\,\,Z_\odot$. We then profile and compare various cooling processes at the centers of halos, and find a complete absence of atomic cooling halos. All of our halos with central cooling gas are dominated by H$_2$ cooling, metal cooling, or a mixture of the two, even in the presence of a strong H$_2$-photodissociating Lyman-Werner background. We also find that accretion through the virial radius is not driven by cooling, as gas flowing in through filaments is typically not yet able to cool efficiently by the time it reaches a given halo. We then investigate the virialization state of the halos, and confirm that both the dark matter and the gas are virialized for the halos in our sample. We find that gas virialization in halos with $M_\mathrm{vir}\gtrsim10^7\,\,M_\odot$ is supported by bulk turbulent flows, and that thermal energy accounts for only a small fraction of the total kinetic energy. Because of this, the mean gas temperature is well below the virial temperature for these halos. We then compute the mass of gas that is available for Population II star formation, and infer star formation rates for each potential star-forming halo. We find good agreement in our inferred Population II statistics with the those in the \textit{Renaissance Simulations}.
\end{abstract}



\input{Introduction}

\input{Methodology}

\input{Results}

\input{Discussion}

\input{Conclusion}

\begin{acknowledgments}

\end{acknowledgments}


\bigskip
\noindent
This research was partially supported by National Science Foundation CDS\&E grant AST-2108076 to M.L.N. The simulations
were carried out on the Frontera supercomputer operated
by the Texas Advanced Computing Center with LRAC allocation AST20007.  Simulations were performed with {\tt Enzo-E} (\href{https://github.com/enzo-project/enzo-e}{https://github.com/enzo-project/enzo-e}) coupled to {\tt Grackle} \citep{Grackle}. Analysis and plot generation was done using {\tt YT} \citep{YT}. {\tt Enzo-E}, {\tt Grackle}, and {\tt YT} are all collaborative open source codes representing efforts from many independent scientists around the world.



\appendix

\section{Validating Enzo-E}
\input{Validating}

\label{app:enzo_vs_enzo-e}

\clearpage

\bibliography{bib}{}
\bibliographystyle{aasjournal}



\end{document}

%% file: Introduction.tex
\section{Introduction}
\label{sec:introduction}

%themes
%impossibly massive galaxies at high z
%star formation history in galaxy halos
%role of Pop III
%pristine halo survival
%uncertainty: Pop III IMF
%uncertainty: Pop III multiplicity
%uncertainty: +/- feedback effects
%progress with SAMs
%progress with direct simulation
%StarNet
%purpose and design of these simulations
%plan for the paper

The claim of the existence of ``impossibly massive" galaxies observed by JWST at redshift $z > 10$ \citep{Labbe2023} has spurred a re-evaluation of our standard model of cosmology and our understanding of the physics of galaxy formation and stimulated renewed interest in the earliest evolutionary stages of galaxies growth \citep{FirstBillionYears}. The high inferred stellar masses imply high star formation rates averaged over the entire formation history of the galaxy and its progenitors. It also implies high rates of mass accretion to feed the star formation process. Even if the stellar mass estimates are high due to unfounded assumptions about the stellar mass function or some other systematic, some high-z galaxies are nonetheless unusually luminous, which implies high star formation rates \citep{FirstBillionYears}. This raises two fundamental questions: (1) when does star formation become efficient in high redshift galaxy halos? and (2) how can high star formation rates be sustained in the presence of feedback?  

Regarding the first question, the conventional wisdom is that star formation efficiency jumps in halos with a virial temperature above $10^4$ K because then the gas can cool efficiently by Ly $\alpha$ emission. At z = 10, $T_\mathrm{vir}=10^4$ K corresponds to a halo of mass of $M_\mathrm{vir}=5 \times 10^7 M_{\odot}$ \citep{Greif2008}. According to the conventional wisdom, halos of lower mass (so-called minihalos) cool inefficiently by H$_2$ and HD if chemically pristine, and additionally by fine structure lines of C and O if chemically enriched by the first generation of stars \citep{BrommLoeb2003}. In either case, star formation would be inefficient in minihalos due to the low concentration of coolants.

%assembly of the first galaxies
Halos with $T_\mathrm{vir} > 10^4$ K are often referred to as {\em atomic cooling halos} (ACH) since they can cool by collisionally excited H and He lines. From a practical point of view, ACHs can be taken as the first galaxies and the building blocks of more massive galaxies \citep{Greif2015}. The reasoning is that minihalos would  be starless after massive Population III stars (hereafter Pop III) have exploded or collapsed to black holes, whereas ACHs have sufficiently deep potential wells to retain photoionized gas that can cool and form stars. If we want to understand the buildup of massive high redshift galaxies, it is therefore imperative that we understand the formation and properties of ACHs. 

The first detailed cosmological simulations of ACHs were performed by \citet{WiseAbel2007} and \citet{Greif2008}. They simulated the formation of an individual halo near the threshold mass of $M_\mathrm{vir}=5 \times 10^7 M_{\odot}(\frac{1+z}{10})^{-3/2}$ using high resolution zoom-in simulations. In the absence of star formation and feedback but including primordial gas chemistry and cooling, they found that most of the gas accretes along filaments at $T < T_\mathrm{vir}$ and then virializes into fluid turbulence in the halo interior. Gas accreting directly from voids passes through an accretion shock near the virial radius and is heated to $T \approx 10^4$ K, where it becomes partially ionized, activating H line cooling. Hydrogen line cooling is found to be confined to just inside the accretion shock, while H$_2$ cooling dominates in the cooler, denser halo core. \citet{Greif2008} included deuterium chemistry in their primordial gas chemistry network, while \citet{WiseAbel2007} did not. The former authors found that HD is formed in appreciable concentrations in the warm, partially ionized gas near the halo's edge, and is efficiently mixed into the halo's interior by the turbulence. This is important because whereas H$_2$ cooling alone can only cool the gas to $T \approx 200$ K, HD can cool the gas to the CMB temperature floor, potentially reducing the Pop III stellar mass scale.   

%chemical enrichment by Pop III supernovae and transition to Pop II
The stellar content of ACHs depends upon whether the gas is pristine or chemically enriched by Pop III supernovae and subsequent generations of metal-enriched star formation. This topic has received considerable attention by modelers and simulators in the past years. These investigations can be broadly divided into two related subjects: (1) the creation and survival of pristine halos, and hence Pop III star formation through cosmic time; and (2) the internal and external chemical enrichment of individual halos by stellar sources. Investigations of type 1 are typically carried out with semi-analytic models (SAMs) of increasing sophistication, while investigations of type 2 tend to employ hydrodynamic and radiation hydrodynamic cosmological simulations. 

%survival of pristine halos

A pioneering study by \cite{Trenti_2009a} showed, using Press-Schechter modeling combined with a probabilistic model for self-enrichment, that Pop III stars could continue to form in pristine minihalos and ACHs to z=10. This is due to the fact that chemical enrichment is a local phenomenon, whereas hierarchical halo growth occurs everywhere in a largely pristine universe. A key uncertainty of their model is whether a pristine halo of ACH mass forms a single or multiple Pop III stars. They investigated both cases, and found that if multiple Pop III stars form with a fixed star formation efficiency (defined as the fraction of baryons within the halo that are converted to stars), the Pop III star formation rate density (SFRD) dominates the Pop II SFRD to {\em lower redshifts} than the single star case because the former rate is boosted. Of course, increasing the production of Pop III stars increases the likelihood that one will explode as a supernova, enriching the halo and terminating any further primordial star formation. 

%Visbal
More recent SAMs (e.g., \citet{Visbal_2020}) use dark matter halo merger trees taken from N-body simulations and add analytic prescriptions to model the effects of H$_2$ photodissociation from Lyman-Werner (LW) radiation, suppression of star formation due to inhomogeneous reionization, and metal enrichment by supernova-driven winds.  \citet{Visbal_2020} computed both the Pop III and metal-enriched star formation histories from $z \approx 30$ to 6 using a novel grid-based method that places feedback spheres of different size for radiative and supernova feedback zones using the location of halos taken from the merger trees. They found that initially long-range LW feedback, local metal enrichment, and photoionization of halos control the Pop III SFRD, but that for $z \leq 15$, external enrichment of pristine halos by nearby star forming halos and inhomogeneous reionization begins to dominate. They find that the combination of reionization feedback and external enrichment reduces the Pop III SFRD at $z=6$ by an order of magnitude compared to LW feedback alone. These results highlight the importance of including at least these three processes in models of high redshift galaxy evolution. 

%BOG, Renaissance, Phoenix
The first attempts to numerically simulate the formation of ACHs including Pop III and metal-enriched star formation and feedback (hereafter Pop II) were limited by small volumes and number statistics due to the severe range of scales present. In the {\em Birth of a Galaxy} simulations, \citet{wise2012b,wise2012a,wise2014} simulated a 1 cMpc$^3$ volume the with the AMR code \texttt{Enzo} including 9-species primordial gas chemistry, subgrid recipes for Pop III and II star formation and feedback, metal injection and cooling, EUV radiative transfer, and a LW background. With a dark matter mass resolution of 1840 $M_{\odot}$ and maximum spatial resolution of 1 comoving pc, the formation of several protogalaxies in halos of mass $M_\mathrm{vir} \approx 10^8 M_{\odot}$ were well resolved and evolved to $z=7$. Assuming a top-heavy Pop III stellar IMF and pair-instability supernova (PISN) yields, they found that the transition from Pop III to Pop II star formation is sudden in halos of mass $M_\mathrm{vir} > 10^7 M_{\odot}$, which marks the transition from H$_2$ to metal line cooling. They found that a single PISN raises the halo metallicity to approximately $10^{-3} Z_{\odot}$, confirming and reinforcing earlier predictions \citep{BrommLoeb2003,Greif2007,WiseAbel2008b}. 

The co-evolution of Pop III and Pop II star formation in larger volumes and more massive halos was investigated by Xu and collaborators with the {\em Renaissance Simulations} suite using \texttt{Enzo} running on the NCSA Blue Waters petascale supercomputer \citep{Xu13,Xu14,chen2014,OShea15,Xu16a,Xu16b,Xu16c}. The {\em Renaissance Simulations} employed the same physical prescriptions as the {\em Birth of a Galaxy} simulations in a much larger volume of (40 cMpc)$^3$, with 3 zoom-in regions enclosing low, high, and average density environments referred to as Void, Rarepeak, and Normal. With a dark matter resolution of $2.34 \times 10^4 M_{\odot}$ and maximum spatial resolution 19 comoving pc, the evolution of halos over a mass range of $10^{6.5} < M_\mathrm{vir}/M_{\odot} < 10^9$ and their star formation histories, both primordial and enriched, could be simulated in detail. 

\citet{Xu13} examined the Pop III {\em multiplicity} as a function of halo mass in the Rarepeak simulation, and found that the majority of Pop III stars formed in halos of mass $10^7 < M_\mathrm{vir}/M_{\odot} < 10^8$ with a typical multiplicity of $\sim 10$. The most massive starless pristine halo was found to have a mass $M_\mathrm{vir} = 7 \times 10^7 M_{\odot}$, and every halo more massive than that was enriched with Pop III supernova ejecta. 

The larger halo mass range and sample size of the {\em Renaissance Simulations} permitted a statistical analysis of the scaling properties of the first galaxies \citep{chen2014}. 
They found star formation efficiency significantly increases at $M_\mathrm{vir} \sim 10^8 M_{\odot}$, which they attributed to the ACH threshold. However, this mass is significantly above the ACH threshold at that redshift, suggesting that the efficiency boost is driven by internal enrichment by Pop II feedback, and not by the onset of H line cooling. 

In this paper we study the accretion, virialization, and cooling of intergalactic gas onto halos spanning the atomic cooling threshold in the absence of metal-enriched star formation and feedback. We employ a new, more scalable version of the \texttt{Enzo} code called \texttt{Enzo-E}, which provides more uniform mass resolution in the baryonic fluid in the vicinity of protogalactic halos, allowing us to better characterize the gas dynamics there. By virtue of \texttt{Enzo-E}’s parallel scalability, we also can survey larger cosmological volumes improving our statistical coverage. We take into account the pre-enrichment of protogalactic halos by Pop III supernovae using our recently developed machine learning-based surrogate model \texttt{StarNet} \citep{Wells_2022b}. We also take into account improved models for the evolving LW radiation background \citep{incatasciato_2023} and H$_2$ self-shielding \citep{Krumholz_2011}.  We find that gas accretes onto halos predominantly by cold mode accretion and virializes via turbulence well below the halo’s virial temperature for its given mass, in agreement with the results of \citet{WiseAbel2007} and \citet{Greif2008}. Consequently, hydrogen atomic line cooling is found to be unimportant to the thermal evolution of the gas in halos as massive as $5 \times 10^8\,\,M_{\odot}$. Rather, a combination of metal line cooling due to Pop III pre-enrichment and H$_2$ cooling are dominant across the halo mass range surveyed, calling into question the conventional notion we have of the importance of atomic cooling halos to galaxy formation and evolution. 

The plan of the paper is as follows. In Sec. \ref{sec:methodology} we describe our suite of \texttt{Enzo-E} simulations and our numerical methodology, and summarize the salient features of \texttt{StarNet} as well as the {\it Phoenix Simulations} used to train the surrogate model. In Sec. \ref{sec:results}, we present our main results. Sec \ref{sec:popIII_statistics} analyzes the statistics of the Pop III stellar associations produced by \texttt{StarNet} and compares our star formation rate density with the semi-analytic models of \citet{Visbal_2020}. Sec. \ref{sec:chemical_enrichment} quantifies the chemical enrichment by Pop III supernovae of over 2,000 halos in our sample of 12,423 halos. In Sec. \ref{sec:accretion_cooling}, we analyze the accretion and cooling of gas in our sample of halos; in Sec. \ref{sec:virialization}, we discuss the topic of incomplete virialization of gas in halos and how that effects how the gas cools. In Sec. \ref{sec:warm_filaments}, we analyze the large scale environment from which the halos are accreting, and characterize it as a cosmic web of cool-warm filaments with a typical temperature $10^3$ K. This result, coupled with incomplete virialization of gas, explains the absence of significant H and He line cooling in ACHs. In Sec. \ref{sec:individual_halo_analysis}, we analyze the properties of 5 of the most massive halos in our sample. In Sec. \ref{sec:popII_inferences}, we estimate the Pop II stellar masses and star formation rates in our sample using a post-processing approach. Finally, in Sec. \ref{sec:sensitivity_study}, we discuss the sensitivity of our results to variations in feedback and chemistry prescriptions. In Sec. \ref{sec:discussion}, we discuss our key results and make comparisons to results in the published literature where relevant. Sec. \ref{sec:discussion} is organized according to key processes which have an effect on the formation and evolution of pre-galactic halos, and includes a discussion of the limitations and caveats of our models. A summary of our key results and conclusions are presented in Sec. \ref{sec:conclusion}. 

%threshold to form Pop III

%metal cooling halos (transition to metal enriched SF)

%scaling properties show no discontinuity at ACH threshold

%discussion of Wells & Norman: motivation, results, Pop III supernova-driven winds

%eventual goal is to probe more massive end of galaxy mass function "Our eventual goal is to simulate the formation of the most massive galaxies at a given redshift using the combined benefits of a more scalable cosmoligical code and deep learning-based surrogate models for subgrid physics."

%% file: Methodology.tex
\section{Methodology}
\label{sec:methodology}

%\begin{figure}
%    \centering
%    \includegraphics[width=0.46\textwidth]{Figures/Methodology/k31_vs_z.png}
%    \caption{\textcolor{red}{Remove? Can refer reader to figure in Incatasciato that compares all LWB curves}}
%    \label{fig:lwb_rates}
%\end{figure}

Our simulation suite is run using the \texttt{Enzo-E} code\footnote{\href{https://github.com/enzo-project/enzo-e}{https://github.com/enzo-project/enzo-e}}. \texttt{Enzo-E} is a port of \texttt{Enzo} \citep{Enzo2014} onto an entirely new parallel adaptive mesh refinement framework called \texttt{Cello} \citep{bordner_2018}. \texttt{Enzo-E/Cello} is parallelized with \texttt{Charm++} \citep{Kale2011}. \texttt{Enzo-E} and \texttt{Cello} are developed in tandem, but are each their own separate application. \texttt{Cello} organizes the computational mesh as a fully distributed array-of-octrees, where the computational domain is decomposed into ``blocks" of field data. Each block is a cubic Cartesian mesh of size $n_b^3$, where $n_b \geq 4$ is an input parameter. For our simulations, $n_b = 16$. Each block has an associated list of Lagrangian particles (dark matter, stars, etc.) which varies in length. When refinement is triggered in a block, 8 child blocks of the same dimension are spawned within the parent block's volume, and the field data is interpolated such that the resolution increases by a factor of 2 in the child blocks. Particles are always associated with the finest level block enclosing them.

\texttt{Enzo-E} encapsulates many of the physics solvers originally developed for \texttt{Enzo}. These include multispecies gas dynamics using a Piecewise Parabolic Method (PPM) adapted for cosmological flows \citep{Bryan1995}, collisionless N-body dynamics using the Particle-Mesh method, and various star formation and feedback recipes (see \citet{Enzo2014} for more details). The Poisson equation for gravity is solved using an entirely new method from that used in \texttt{Enzo}. The large-scale gravitational field is solved on the root grid using a V-cycle geometric multigrid method assuming periodic boundary conditions. The local gravitational field in each refined octree is then solved using the bi-conjugate gradient method taking as its boundary conditions the large-scale potential interpolated onto the external facets of the octree mesh. Nine-species (HI, HII, HeI, HeII, HeIII, H$_2$, H$_2+$, H$^-$, $e^-$) non-equilibrium chemistry and cooling is handled using the \texttt{Grackle} library \citep{Grackle}. We include an H$_2$ photodissociating Lyman-Werner background (LWB) using fits derived in \citep{incatasciato_2023}. Specifically, we calculate the global H$_2$ photodissociation rate as $k_\mathrm{H_2}=1.38\times10^{-12} J_{21}\,\,\mathrm{s^{-1}}$ \citep{Abel_1997}, where $J_{21}$ is given by equation 9 in \cite{incatasciato_2023}. The H$^-$ photodetachment rate and H$_2^+$ photodissociation rates are then calculated using equation 10, with the fitting parameters listed in Table 5 of \cite{incatasciato_2023}. We also track metals sourced from primordial supernovae using an implementation of the machine-learning accelerated surrogate model for Pop III star formation and feedback, \texttt{StarNet} \citep{Wells_2022b, Wells_2022a, Wells_2021}. Metal line cooling is computed in each enriched cell using precomputed tables assuming ionization equilibrium as described in \citep{Grackle}. To demonstrate the similarities and differences between \texttt{Enzo} and \texttt{Enzo-E}, comparisons between two sets of identical cosmological simulations run with each code are presented and analyzed in Appendix \ref{app:enzo_vs_enzo-e}.

In total, we perform 6 simulations in this study, the parameters for which are listed in Table \ref{table: params}. Our fiducial simulation, \texttt{N512\_fiducial}, has $512^3$ cells and dark matter particles on the root grid, up to 5 levels of AMR, and a box of length 5.12 comoving Mpc on a side, providing a maximum spatial resolution of 312.5 comoving pc (26 proper pc at $z=11$). Refinement is triggered in a block when the mass in a cell reaches a value of $M>\delta_\mathrm{thresh}(\Delta x_\mathrm{root})^3$. Here, $\delta_\mathrm{thresh}=8$ is the chosen overdensity threshold, where $\delta=1$ corresponds to the cosmic mean density at a given redshift, and $\Delta x_\mathrm{root}$ is the cell width on the root grid. Cosmological initial conditions are generated using \texttt{MUSIC} \citep{MUSIC} at $z=199$ using the Planck 2014 \citep{Planck2014} cosmological parameters: $\Omega_\mathrm{m}=0.3111$, $\Omega_\mathrm{b}=0.048975$, $\Omega_k=0$, $\Omega_\lambda=0.6889$, $H_0=0.6766$, $\sigma_8=0.811$, $n=0.965$. With these parameters, the dark matter particle mass is $3.34\times10^4\,\,M_\odot$, which comparable to that in the {\em Renaissance Simulations}. The simulation is run to a final redshift of $z=11.92$. 

The remaining simulations listed in Table \ref{table: params} are run with the same resolution settings, but with varied physics and box sizes. These are meant to test the sensitivity of our results to variations in our physics models, and to provide more statistics. \texttt{N256\_fiducial} is a scaled-down version of \texttt{N512\_fiducial} that includes all the same physics, but with the simulated volume and root grid both reduced by a factor of 8. \texttt{N256\_noLWB} is a re-run of \texttt{N256\_fiducial}, but without a LWB. \texttt{N256\_adiabatic} is a version that does not include chemistry and cooling, and does not source metals from Pop III supernovae. \texttt{N256\_6species} includes chemistry and cooling only for atomic hydrogen and helium species: HI, HII, HeI, HeII, HeIII, $e^-$. \texttt{N256\_9species} includes the full nine-species chemistry network with metal cooling and a LWB, but metals are sourced using a uniform metallicity floor of $Z_\mathrm{floor}=10^{-5.5}\,\,Z_\odot$. 

This study focuses on the physical processes leading up to Pop II star and galaxy formation, and we thus do not include an explicit model for Pop II star formation and feedback in any of our simulations. This is done so that we may isolate the effects of primordial pre-enrichment on the accretion, virialization, and cooling gas in protogalaxies. It is also in the spirit of the pioneering simulations of \citet{WiseAbel2007} and \citet{Greif2008}, who also did not include metal-enriched star formation and feedback. In a follow-up paper, we will re-run our simulation suite with the inclusion of Pop II star formation and feedback, and will discuss how the findings in this study are affected by the additional physics. We will also investigate the properties of the galaxies that form following enrichment by Pop III supernovae.

\begin{table*}
    \centering
    \begin{tabular}{ |p{2.5cm}||p{1cm}|p{2cm}|p{3cm}|p{3cm}|p{1cm}|p{2cm}| }
        \hline
        \multicolumn{7}{|c|}{Simulations} \\
        \hline
        Name & Root & $L_\mathrm{box}$ [cMpc] & Species tracked & $Z_\mathrm{source}$ &$z_\mathrm{final}$ & $\mathrm{MMH}\,\,[M_\odot]$\\
        \hline
        \texttt{N512\_fiducial}  & $512^3$ & 5.12 & H, He, H$_2$, $e^-$, ions & \texttt{StarNet} & 11.92 & $5.67\times10^8$\\
        \texttt{N256\_fiducial}  & $256^3$ & 2.56 & H, He, H$_2$, $e^-$, ions & \texttt{StarNet} & 11.34 & $4.90\times10^8$\\
        \texttt{N256\_noLWB}     & $256^3$ & 2.56 & H, He, H$_2$, $e^-$, ions & \texttt{StarNet} & 11.35 & $4.60\times10^8$\\
        \texttt{N256\_adiabatic} & $256^3$ & 2.56 & None                      & None             & 11.58 & $5.62\times10^8$\\
        \texttt{N256\_6species}  & $256^3$ & 2.56 & H, He, $e^-$, ions        & None             & 12.03 & $4.63\times10^8$\\
        \texttt{N256\_9species}  & $256^3$ & 2.56 & H, He, H$_2$, $e^-$, ions & $Z_\mathrm{floor}=10^{-5.5}\,\,Z_\odot$ & 12.10 & $4.55\times10^8$\\
        \hline
    \end{tabular}
    \caption{List of relevant parameters for each of the simulations run for this study. From left to right, columns show simulation name, root grid dimension, box length in comoving Mpc, chemical species participating in chemistry and cooling with \texttt{Grackle}, the source of metals (either predicted with \texttt{StarNet}, set using a uniform metallicity floor, or no metals), final redshift, and the virial mass of the most massive halo at the final redshift. All simulations have a maximum AMR level of 5, and reach the same maximum spatial resolution of 312.5 comoving pc (26 proper pc at z=11). We use a dark matter particle mass of $3.34\times10^4\,\,M_\odot$. The column labelled ``MMH" shows the virial mass of the most massive halo in each simulation. The $256^3$ series of simulations start from identical cosmological initial conditions.}
    \label{table: params}
\end{table*}

\subsection{Primordial Chemical Enrichment with \texttt{StarNet}}
Our simulations are scoped to generate a statistically significant number of halos with $T_\mathrm{vir} \gtrsim 10^4$ K while capturing inhomogeneous chemical enrichment from primordial supernovae. To accomplish this, we call the surrogate model, \texttt{StarNet} \citep{Wells_2022b}, once every 5 Myr. \texttt{StarNet} utilizes deep convolutional neural networks to identify Pop III star-forming regions within subvolumes of size 10 proper kpc throughout the computational domain. \texttt{StarNet} flags cells in the subdomain that are potential sites of Pop III star formation using the hydrodynamic fields as input. If any cell within a given predicted star-forming region has a metallicity above $Z_\mathrm{crit}=10^{-5.5}\,\,Z_\odot$, the prediction is discarded. In this way, Pop III stellar associations only form in pristine gas. For more details, see \citep{Wells_2022b}.  

Once the star-forming regions are identified, population statistics for each region are randomized following statistics from the \textit{Phoenix Simulations} (see Sec. \ref{sec:phx}), and a simple linear regression model is used to predict the size of the composite supernova remnant once all stars within the population reach their endpoint. Metal yields are calculated based off of the masses of the stars within the population, and the metals are then uniformly deposited onto the mesh within a sphere of the predicted radius. The metal bubbles predicted by \texttt{StarNet} typically have radii of 1 to 3 kpc. The temperature of the gas within the sphere is set to \texttt{max($T_i$, $10^4$ K)}, where $i$ denotes the cell index. In our implementation of \texttt{StarNet}, we also photodissociate all H$_2$ within the sphere, ionize all hydrogen, and singly ionize all helium (i.e. all chemical species are converted to HII, HeII, and $e^-$, while conserving proton and electron counts). Details of each population (number of stars, stellar masses, metal yields, etc.) are saved in massless \texttt{popIII\_remnant} particles that are free to move around the mesh after their corresponding supernova remnants are deposited. These particles are placed at the center of mass of the supernova remnant at the time of deposition. Because the \texttt{popIII\_remnant} particles and the gas in their vicinity fall under the influence of the same gravitational potential, the \texttt{popIII\_remnant} particles serve as tracers that generally follow the supernova remnants as the enriched gas is advected. With this, it is possible to analyze the stellar population statistics in post, as well as track the separate progenitor stellar populations as supernova remnants inevitably merge. For a detailed account of \texttt{StarNet}, its training, and its capabilities, we refer the reader to \citet{Wells_2021, Wells_2022a, Wells_2022b}. 

\subsection{The Phoenix Simulations}
\label{sec:phx}

\texttt{StarNet} was developed in \texttt{Enzo} using training data from the \textit{Phoenix Simulations} \citep{Wells_2022a}. All predictions from \texttt{StarNet} are thus consistent with those observed in the \textit{Phoenix Simulations} (PHX hereafter). We briefly describe these simulations here.  The PHX simulations are a set of three radiation hydrodynamic cosmological simulations run with the \texttt{Enzo} code --- one with a root grid dimension of $512^3$ and a box length of 5.21 comoving Mpc, and two with dimension $256^3$ and box length 2.61 comoving Mpc. The PHX simulations have the same root grid resolution as our simulation suite, but are run with 9 levels of AMR. It should be noted that while the maximum resolution of the PHX simulations is much higher than the simulations considered in this study, the differences in AMR approaches between \texttt{Enzo} and \texttt{Enzo-E} make it such that the \texttt{Enzo-E} simulations obtain better resolution outside of halo centers. 

The PHX simulations include direct models for both Pop III and Pop II star formation and feedback, as well as radiative transfer using the adaptive ray tracing algorithm, \texttt{Moray} \citep{wise_moray}. It is found in \citep{Wells_2022a} that Pop III stars form in associations with up to $\sim150$ individual stars. A detailed analysis is then performed to connect Pop II star clusters to their progenitor Pop III supernova remnants, and the influence of different configurations of Pop III supernova type on the properties of the Pop II stars that form in their wake is explored. Finally, a piecewise linear regression model is fit to relate the size of a conglomerate Pop III supernova remnant to the number of stars, their masses, and their relative formation times in the association. This model is the very same model that is used in \texttt{StarNet} to determine the radius of the evolved remnant, assuming it evolves spherically, once the location and properties of the association are predicted. 

%% file: Results.tex
\section{Results}
\label{sec:results}

\input{Results/PopIIIStatistics}

\input{Results/ChemicalEnrichment}

\input{Results/AccretionCooling}

\input{Results/IndividualHaloAnalysis}

\input{Results/PopII_inferences}

\input{Results/SensitivityStudy}

%% file: Results/PopIIIStatistics.tex
\subsection{Population III Statistics}
\label{sec:popIII_statistics}

%%%% PROJECTION WITH ZOOM-INS
\begin{figure*}
    \centering
    \includegraphics[width=0.96\textwidth]{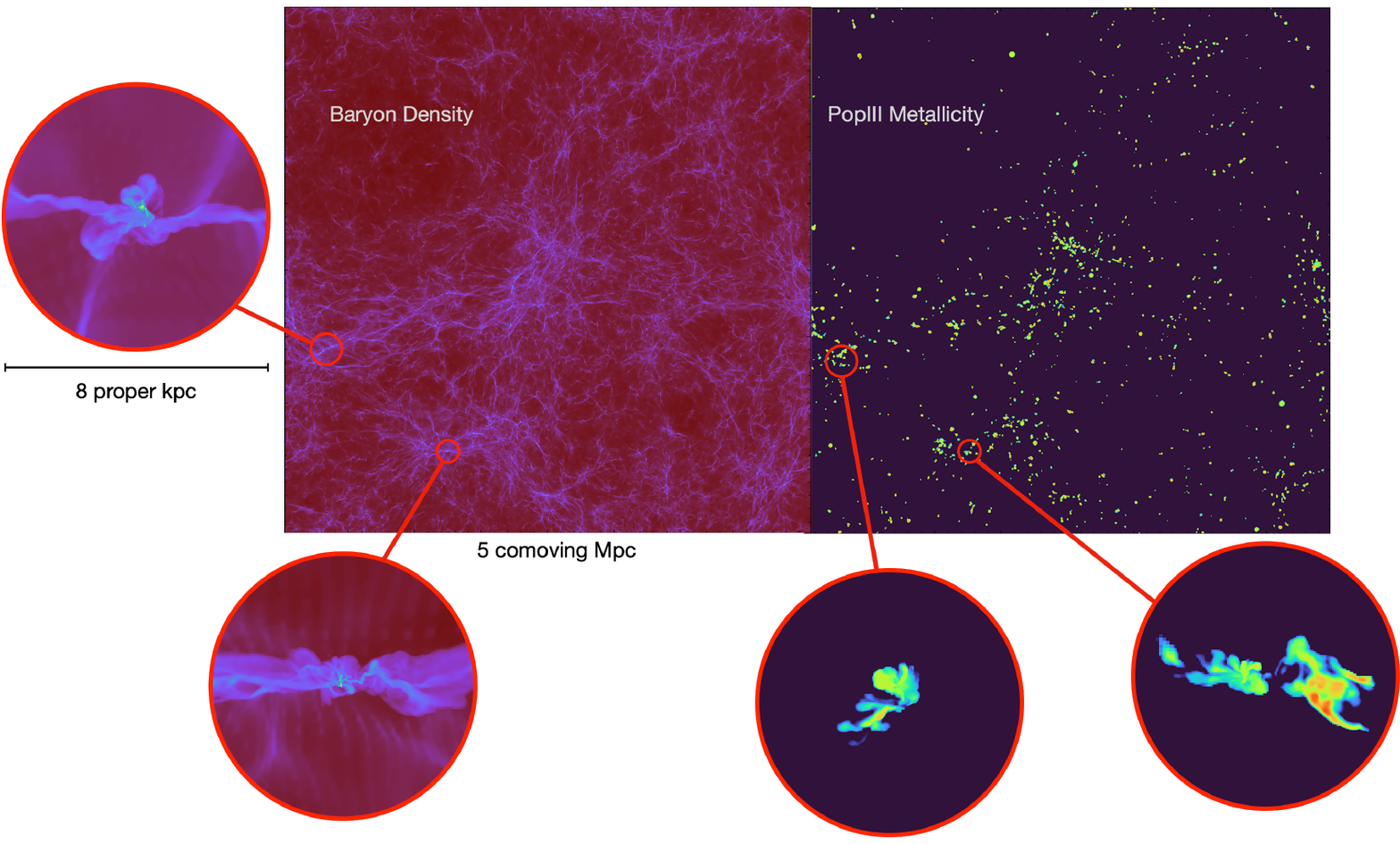}
    \caption{Full-box projections of baryon density and Pop III gas metallicity for the \texttt{N512\_fiducial} simulation, which has a simulated volume of $\sim 125\,\,\mathrm{comoving\,\,Mpc^3}$. The small red circles show the locations of the two most massive halos in the box. Connected to the small red circles via red lines are slices through the center of the two most massive halos, for both baryon density and metallicity. The halos have virial radii of $\sim2$ kpc. Metal enrichment in this simulation is purely from Pop III feedback predicted by \texttt{StarNet}.}
    \label{fig:projections_with_zooms}
\end{figure*}
%%%%

\begin{figure}
    \centering
    \includegraphics[width=0.46\textwidth]{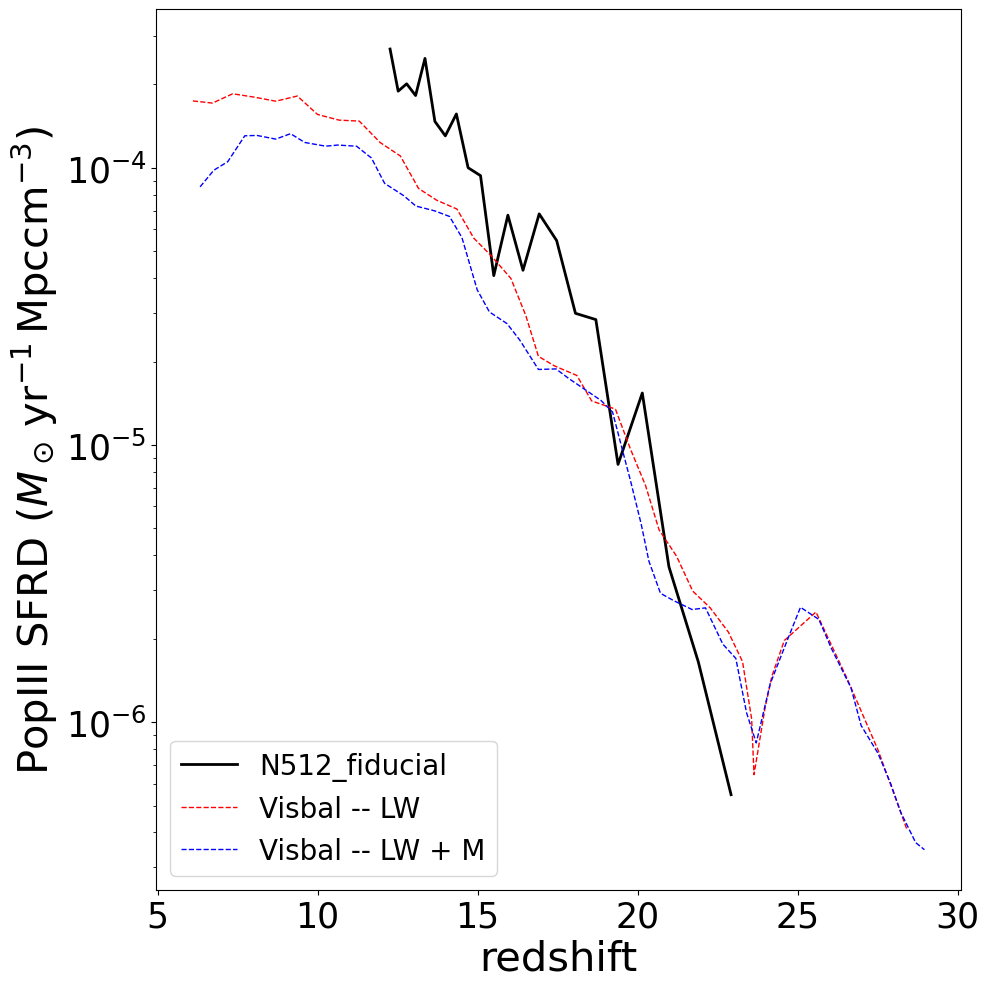}
    \caption{Global Pop III SFR vs. redshift for the \texttt{N512\_fiducial} simulation. The red and blue dashed lines show the rates obtained in \cite{Visbal_2020} for the LW and LW+M cases, respectively. LW refers to the semianalytic model that includes the LWB, and LW+M refers to the model that includes both the LWB and external enrichment. Our rates agree very well with those in \cite{Visbal_2020} at early times, though our rates are higher at $z\sim12$ by a factor of 2-3.}
    \label{fig:popIII_SFR}
\end{figure}

\begin{figure*}[t!]
    \centering
    \includegraphics[width=0.96\textwidth]{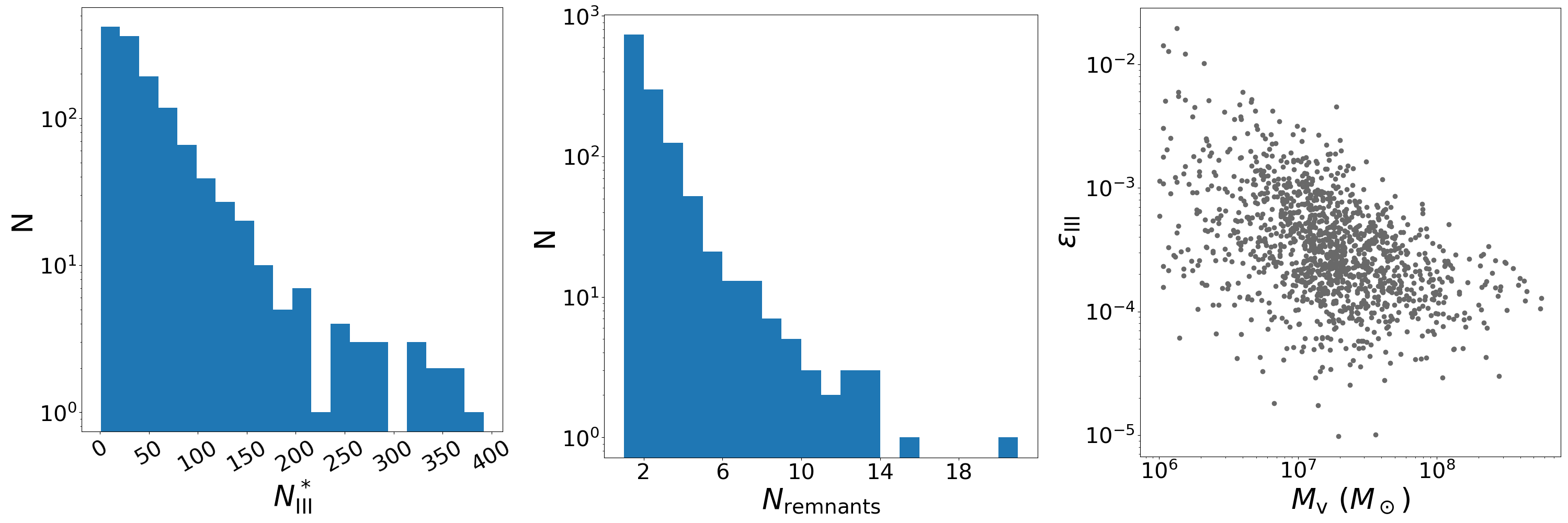}
    \caption{\textit{Left:} Distribution of the number of Pop III stars associated with supernova remnants within the virial radius of each halo. Most halos are associated to $\lesssim 50$ Pop III stars, while some have $\gtrsim 300$. Note that the count for the number of Pop III stars here includes those in the mass ranges of $40\,\,M_\odot < M_\mathrm{III}^* < 140\,\,M_\odot$ or $260\,\,M_\odot < M_\mathrm{III}^*$ that do not explode as supernovae, but instead collapse into an inert black hole. \textit{Middle:} Distribution of the number of \texttt{popIII\_remnant} particles within the virial radius of each halo. The number of \texttt{popIII\_remnant} particles is a measure of the minimum number of separate Pop III associations that contribute to the enrichment of the halo. The actual number of enriching associations could be higher than the number of \texttt{popIII\_remnant} particles within the virial radius if the halo is externally enriched. \textit{Right:} Pop III star-forming efficiency vs. virial mass for each halo, where the star-forming efficiency is defined as the ratio between Pop III stellar mass and baryon mass.}
    \label{fig:popIII_halo_stats}
\end{figure*}

We begin by discussing the Pop III star formation history in our fiducial simulation over the redshift range $23 \geq z \geq 12$. Fig. \ref{fig:projections_with_zooms} provides a visual impression of their formation sites and feedback effects. Specifically, Figure \ref{fig:projections_with_zooms} shows full-box projections of baryon density and metallicity for the \texttt{N512\_fiducial} simulation at $z=11.92$, with zoomed-in slices through the centers of the two most massive halos shown around the perimeter. The collections of high-metallicity gas are entirely a result of primordial chemical enrichment with \texttt{StarNet}. Although metals are initially deposited in uniform spheres, the local morphologies of the deposited remnants lose their sphericity over time due to a combination of hydrodynamic mixing and tidal forces. Examples of the non-spherical morphologies of deposited remnants can be seen in the halo slice plots in Figure \ref{fig:projections_with_zooms}. 

By the final output of the \texttt{N512\_fiducial} simulation, there have been 2,009 separate Pop III star-forming events in the volume predicted by \texttt{StarNet} over the course of the simulation. The global Pop III star formation rate density (SFRD) is plotted versus redshift in Figure \ref{fig:popIII_SFR}, assuming stellar populations predicted by \texttt{StarNet} form uniformly over a period of 10 Myr. By $z=12$, $\mathrm{SFRD} \approx 3\times10^{-4}\,\,M_\odot\,\mathrm{Myr^{-1}\,Mpccm^{-3}}$. To verify that the SFRD is reasonable, we overplot two Pop III SFRD curves from Figure 2 of \cite{Visbal_2020} for comparison. Our SFRD's generally agree with those in \cite{Visbal_2020}, though there is a difference of a factor of 2-3 at $z=12$. This is likely due to our inclusion of H$_2$ self-shielding, which is not included in \cite{Visbal_2020}. In the absence of H$_2$ self-shielding, the LWB will suppress further Pop III star formation. This causes in a turnover of the Pop III SFRD, which can be seen in the blue and red dashed lines of Figure \ref{fig:popIII_SFR}. An effect of H$_2$ self-shielding is to cause a delay in the turnover of the Pop III SFRD, as it allows for efficient H$_2$ cooling to take place in star-forming cores that would not take place otherwise. Another explanation for the flattening of the SFRD at lower redshifts in the Visbal models is their inclusion of radiative and supernova feedback from Pop II star formation which we do not take into account. According to \cite{Visbal_2020}, growing spheres of photoionized and chemically enriched gas from normal star formation in protogalaxies exclude an increasing volume of the IGM from forming Pop III stars.   

By inspecting the number of massless \texttt{popIII\_remnant} particles within the virial radius of a given halo, we can obtain information about the number of separate supernova remnants that have merged within the halo. Because we explicitly throw out positive \texttt{StarNet} predictions in high metallicity gas, it is unlikely that a given halo would experience multiple internal star-forming events. Figure \ref{fig:popIII_halo_stats} shows histograms of the number of Pop III associations contributing to the enrichment of each halo, as well as the total number of stars belonging to the associations, and a plot of the Pop III star-forming efficiency for each halo. Halos are identified using the \texttt{ROCKSTAR} halo finder \citep{Behroozi2013}. While most of the halos contain $<5$ \texttt{popIII\_remnant} particles as expected, many halos contain $>10$ \texttt{popIII\_remnant} particles, with the most massive halo in the simulation containing 20 by the final output. This means that the enriched gas in this halo is the result of a merger between at least 20 separate supernova remnants, with the majority of remnants being deposited initially outside the most massive halo's virial radius and being accreted.

The Pop III star-forming efficiency shown in Figure \ref{fig:popIII_halo_stats} is defined as
\begin{equation}
    \varepsilon_\mathrm{III}=\frac{M_\mathrm{III}^*}{M_\mathrm{b}}.
\end{equation}
In this case, $M_\mathrm{III}^*$ refers to the total mass of stars associated with all remnants within the virial radius. Since these stars are assumed to have all reached their endpoint, $\varepsilon_\mathrm{III}$ is a direct measure of the fraction of gas within the halo that was once contained within Pop III stars. There is a wide scatter due to the stochastic nature of the population statistics, but there is generally a downwards trend as baryon accretion rates increase with halo mass. The geometric mean of $\varepsilon_\mathrm{III}$ is approximately $3\times10^{-4}$.

%% file: Results/ChemicalEnrichment.tex
\subsection{Chemical Enrichment}
\label{sec:chemical_enrichment}

\begin{figure}
    \centering
    \includegraphics[width=0.47\textwidth]{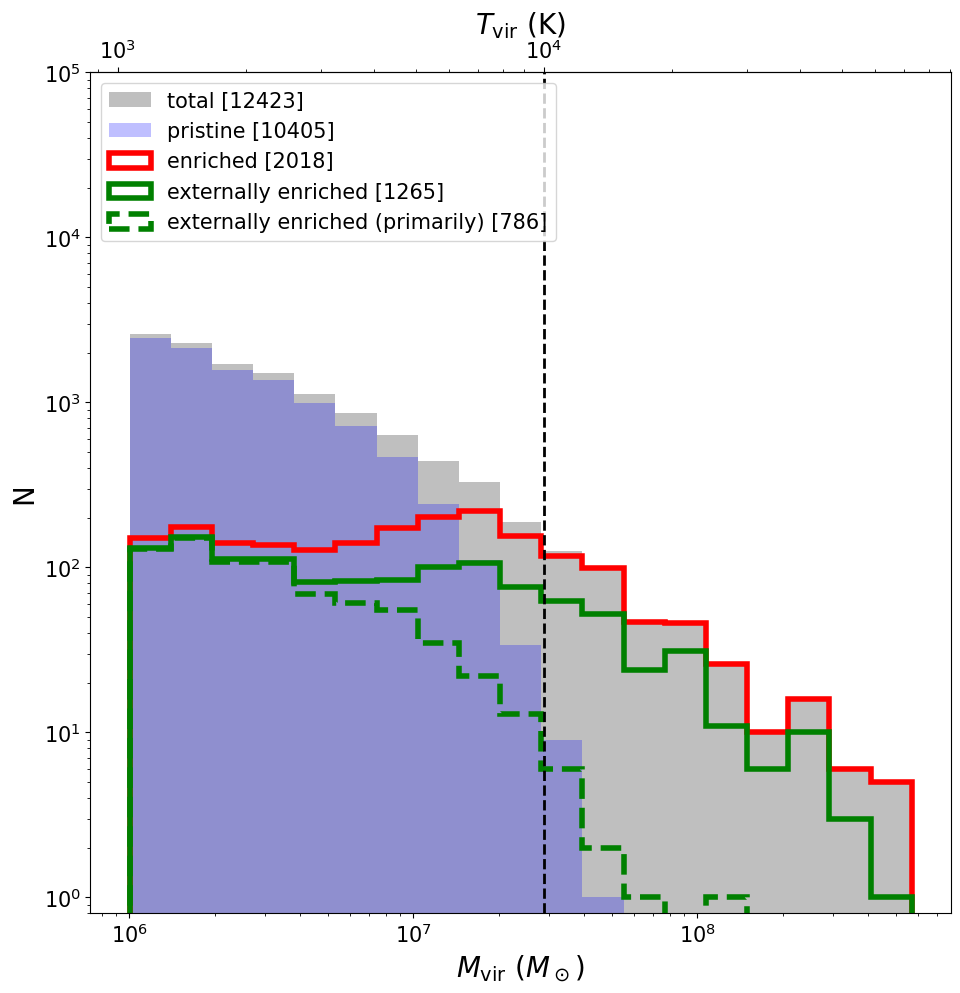}
    \caption{Halo mass functions for halos in the \texttt{N512\_fiducial} simulation at $z\sim12$. The grey filled histogram shows the mass function for all halos. The blue filled histogram shows the mass function for pristine halos (i.e. halos that have not been chemically enriched by Pop III supernovae). The red curve shows the mass function for halos that have been chemically enriched by Pop III supernovae. The green solid curve shows the mass function for the subset of enriched halos that have been externally enriched. The green dashed curve shows the mass function for the subset of externally enriched halos that have been enriched \textit{primarily} by external sources. The bracketed numbers in the figure legend show the total number of halos for each label. The vertical dashed line shows the virial mass corresponding to a virial temperature of $10^4$ K at the final redshift. All but 11 halos with virial temperatures above $10^4$ K are chemically enriched by the final redshift.}
    \label{fig:hmf_enrichment}
\end{figure}

\begin{figure}
    \centering
    \includegraphics[width=0.47\textwidth]{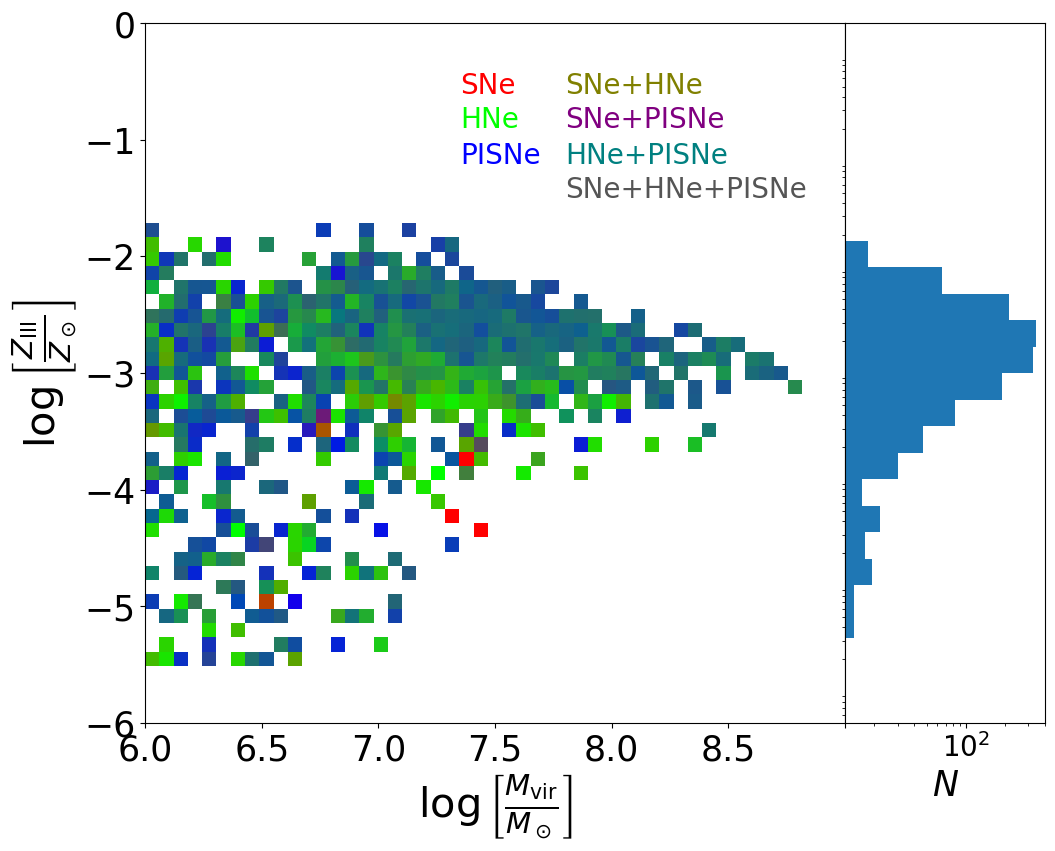}
    \caption{\textit{Left:} 2D histogram of Pop III metallicity vs. virial mass for chemically enriched halos, where a given bin is colored by the relative contribution by type II SNe, HNe, and PISNe to the enrichment of halos within that bin. To compare the relative contributions in a bin, the total mass of metals sourced from supernovae of each type within the virial radius are summed separately over all halos within the bin. The masses are normalized by the total, and mapped to an RGB color index, given by, $\left(\frac{M_\mathrm{SNe}}{M_\mathrm{tot}}, \frac{M_\mathrm{HNe}}{M_\mathrm{tot}}, \frac{M_\mathrm{PISNe}}{M_\mathrm{tot}}\right)$, where $M_\mathrm{tot}=M_\mathrm{SNe}+M_\mathrm{HNe}+M_\mathrm{PISNe}$. Colors corresponding to various combinations of supernova contributions in equal amounts are shown in the text in the upper right-hand corner. The image is made up mostly of shades of blue and green, which suggests that enrichment is overall dominated by both HNe and PISNe. Subhalos are filtered out of the sample for this figure to avoid the double counting of metals.  As halo mass increases, metallicities tend towards $10^{-3}\,\,Z_\odot$. \textit{Right:} Distribution of the average halo metallicity within the virial radius. The distribution peaks at $2\times10^{-3}\,\,Z_\odot$.}
    \label{fig:metallicity_vs_Mvir_RGB}
\end{figure}

By the final output of \texttt{N512\_fiducial}, there are 12,423 halos in the sample---2,018 ($16\%$) of which are chemically enriched. Of the chemically enriched halos, 1,265 ($63 \%$) are enriched externally, where an externally enriched halo is defined to be a halo where
\begin{equation}
    M_Z > M_{Z,\mathrm{p}}.
\end{equation}
$M_Z$ is represents the total metal mass within the virial radius, and $M_{Z,\mathrm{p}}$ represents the total mass of metals produced by internal supernovae. The fraction of enriched halos that are externally enriched is in agreement with \cite{Hicks_2021}, which achieves a similar result using an explicit model for Pop III star formation and feedback that tracks the formation and destruction of individual stars. Going further, 786 (62\%) of the externally enriched halos are \textit{primarily} externally enriched, meaning that $>50\%$ of the metals observed inside the virial radius originate from external sources. 

Figure \ref{fig:hmf_enrichment} shows halo mass functions for each subset of halos previously discussed. While 10,405 ($84\%$) of halos are pristine by the final output, only 10 pristine halos have $T_\mathrm{vir} > 10^4$ K. Apart from these 10, all halos with $T_\mathrm{vir} > 10^4$ K are chemically enriched. This is significant because the gas within halos with $T_\mathrm{vir} > 10^4$ K is expected to cool predominantly through radiative line transitions in atomic hydrogen and helium \citep{Omukai_2001}. However, Figure \ref{fig:hmf_enrichment} shows that metal-line cooling is almost always present in halos with $T_\mathrm{vir} > 10^4$ K.

\texttt{Enzo-E} uniquely tracks separate metallicity fields for each type of supernova considered by \texttt{StarNet}. As such, it is possible to disentangle the relative contributions to the enrichment of halos from each type of supernova. Figure \ref{fig:metallicity_vs_Mvir_RGB} shows a phase diagram of halo-averaged metallicity vs. virial mass, where the set of fractional contributions from each type of supernova to the total metal mass within each bin are mapped to RGB color values (red = core-collapse SNe, green = HNe, blue = PISNe). The diagram is almost entirely shades of blue and green, meaning that enrichment is dominated by HNe and PISNe. There are many halos with equal contributions of HNe and PISNe metals. These results are highly dependent on the chosen characteristic mass of Pop III stars because it determines the relative fraction of supernovae of each type. The distribution of combined halo-averaged metallicities peaks at $10^{-3}\,\,Z_\odot$ (see right panel of Fig. \ref{fig:metallicity_vs_Mvir_RGB}), with halos at the high-mass end having a higher tendency towards this value.

%% file: Results/AccretionCooling.tex
\subsection{Accretion and Cooling}
\label{sec:accretion_cooling}

\begin{figure}
    \centering
    \includegraphics[width=0.46\textwidth]{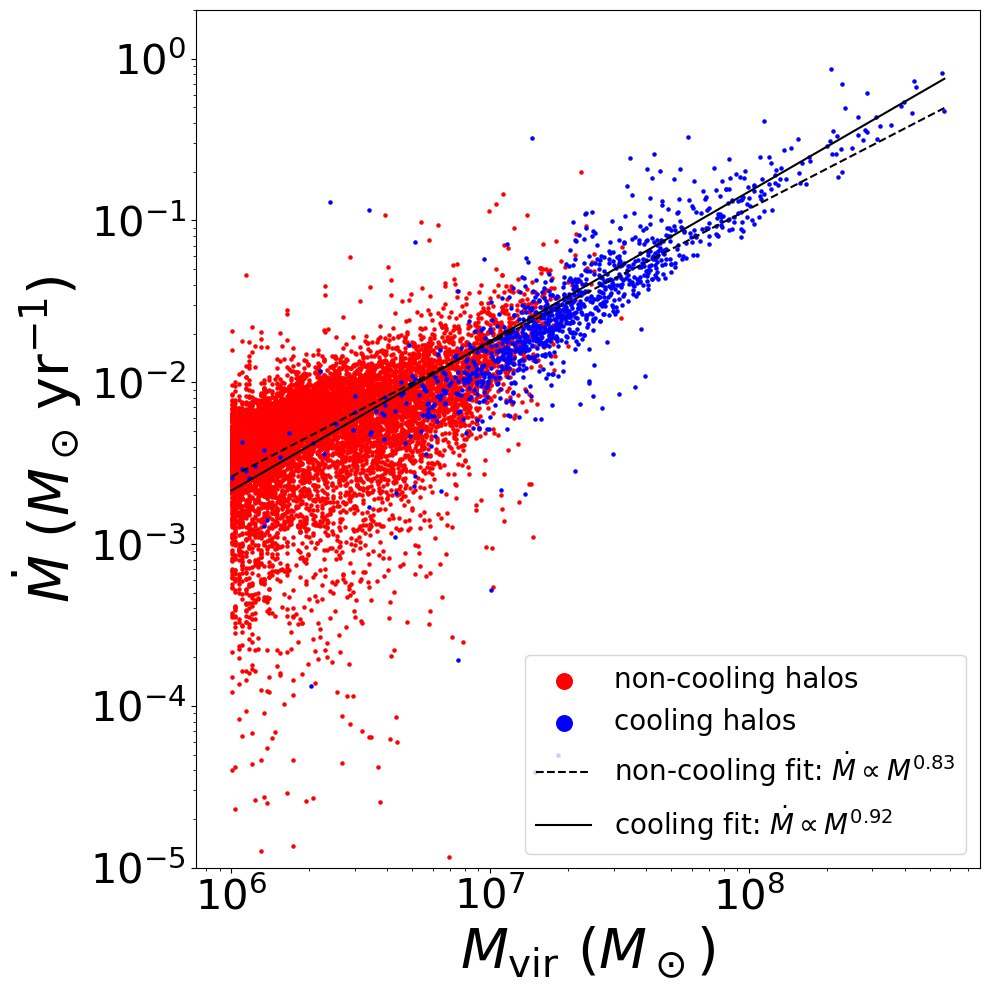}
    \caption{Accretion rate through the virial radius vs. virial mass for cooling halos (blue) and non-cooling halos (red). Each sample is fit separately to a power law. The slopes are 0.92 for cooling halos and 0.83 for the non-cooling halos. The similarity in the two slopes suggests that cooling does not drive gas accretion through the virial radius. For consistency, we have also made similar power law fits for H$_2$ cooling halos, metal cooling halos, and all halos. These additional fits are not plotted, but the slopes are 0.99 for H$_2$ cooling halos, 0.82 for metal cooling halos, and 0.92 for all halos. Subhalos are filtered out of the sample for this analysis.}
    \label{fig:accretion_rates}
\end{figure}

\begin{figure}
    \centering
    \includegraphics[width=0.46\textwidth]{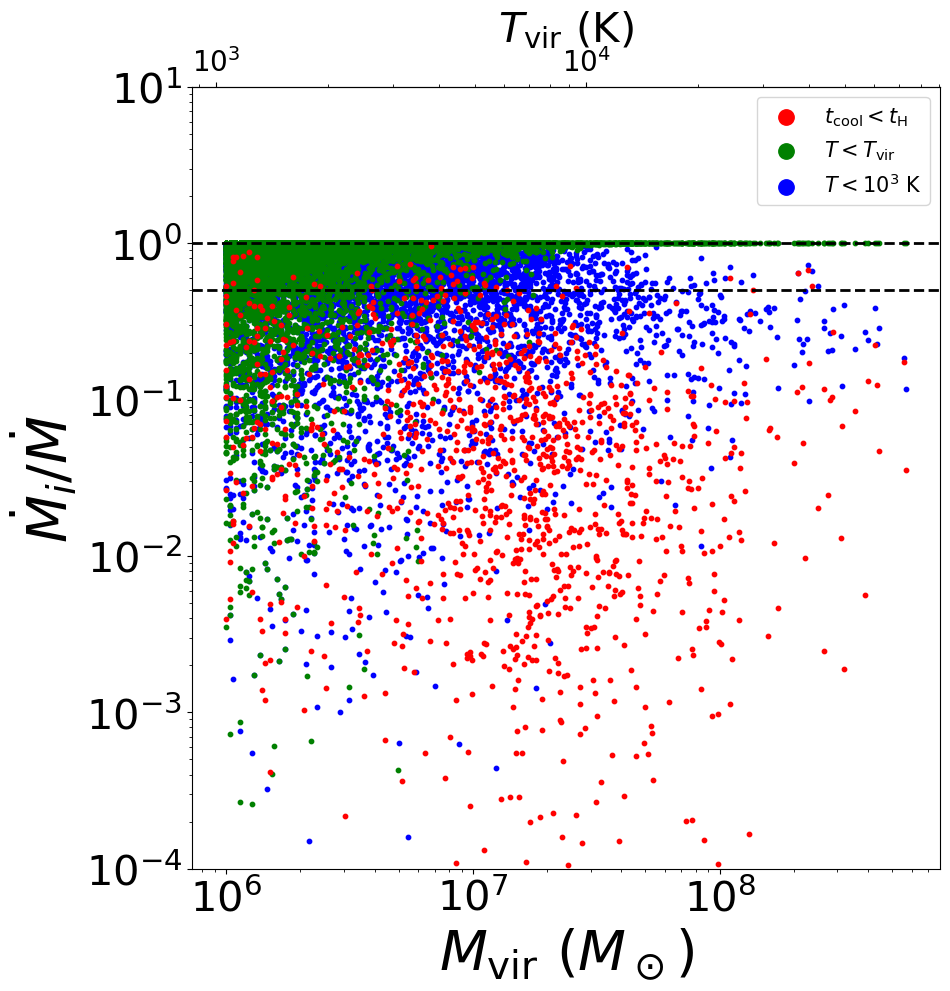}
    \caption{Fractional inflow rates for gas at the virial radius that is cooling (red), below $T_\mathrm{vir}$ (green), and/or below $10^3$ K (blue). Horizontal dashed lines are at fractional rates of 0.5 and 1.0. Blue (green) dots between the horizontal lines show halos where most of the gas flowing into the virial radius has $T < 10^3$ K ($T < T_\mathrm{vir}$). Red dots between the horizontal lines show halos where the majority of the gas flowing into the virial radius is able to cool in less than a Hubble time.}
    \label{fig:inflow_rate_fractions}
\end{figure}

\begin{figure}
    \centering
    \includegraphics[width=0.47\textwidth]{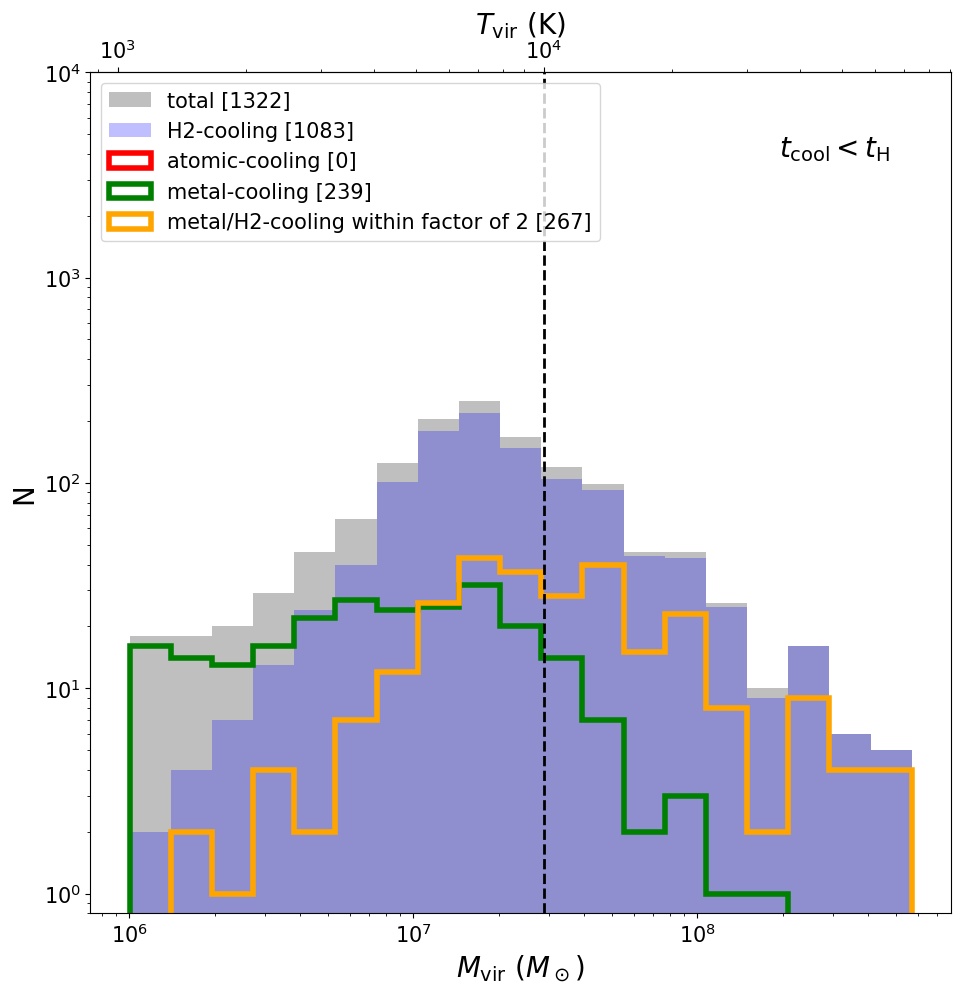}
    \caption{Halo mass functions for cooling halos. The grey histogram shows the mass function for all cooling halos. The blue histogram shows the mass function for H$_2$ cooling halos. The green curve shows the mass function for metal cooling halos. The orange curve shows cooling halos that have relative mean H$_2$ and metal cooling rates within $R_\mathrm{1000}$ that are within a factor of 2 of eachother. The vertical dashed line corresponds to a virial temperature of $10^4$ K. The total number of each type of halo is placed in brackets in the figure legend. Atomic cooling halos would show up in this plot as a red curve; however, we find zero atomic cooling halos in the sample.}
    \label{fig:hmf_cooling}
\end{figure}

\begin{figure*}[hbtp]
    \centering
    \includegraphics[width=0.96\textwidth]{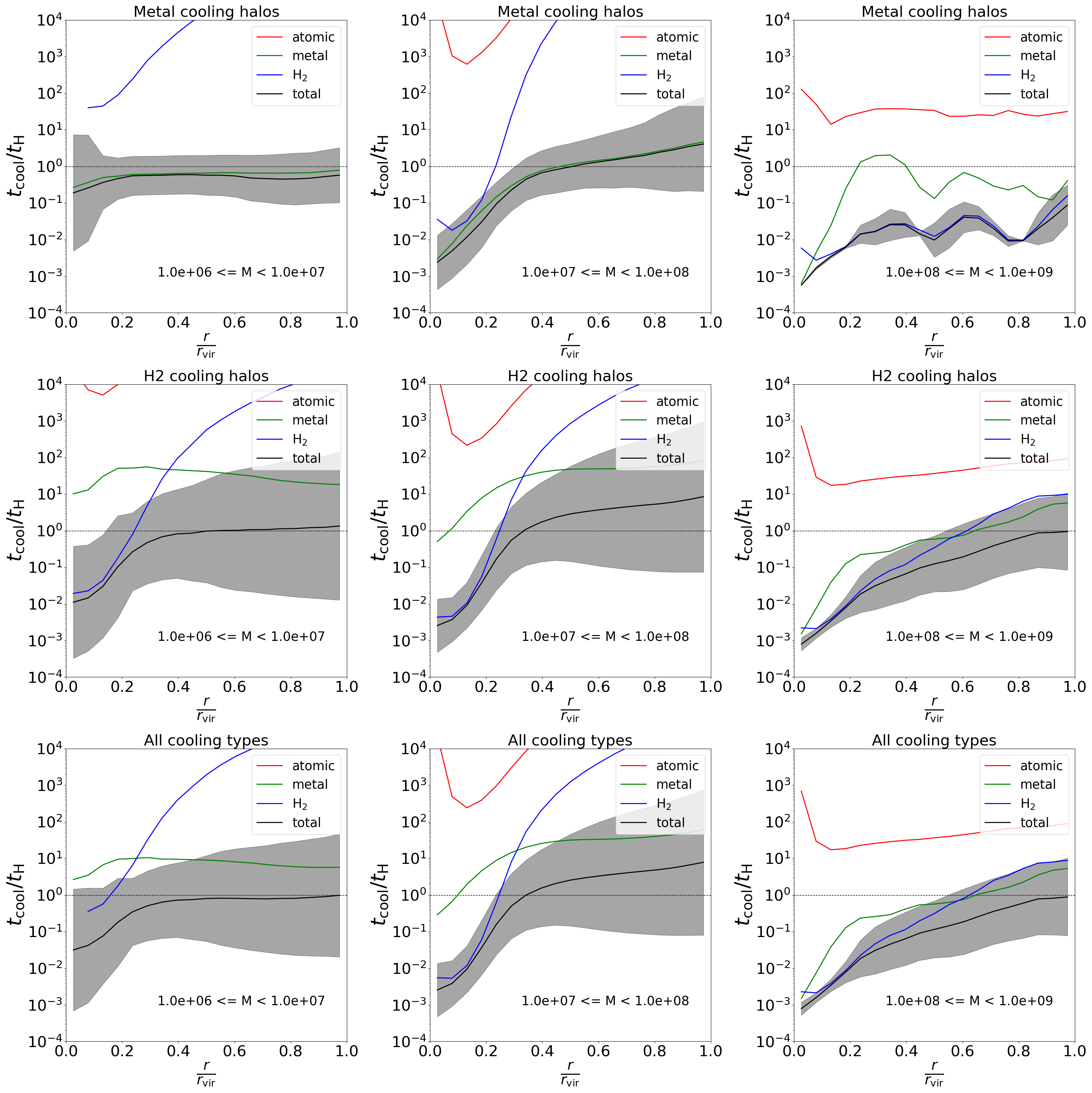}
    \caption{Radial profiles of average cooling time, normalized by the Hubble time at $z\sim12$, for halos in bins of virial mass and dominant cooling type within $R_{1000}$. The top and middle rows show profiles for halos that cool primarily though metals and H$_2$, respectively, while the bottom row shows profiles for all cooling halos, regardless of which coolant is dominant. The columns separate halos into mass bins, with mass increasing from left to right. The text in the plot denotes the mass of halos within a given bin. Atomic cooling rates are shown in red, metal cooling rates are shown in green, H$_2$ cooling rates are shown in blue, and the total cooling rates are shown in black. The regions filled in with grey extend out to 1 standard deviation in log space from of the total cooling rate curve. The gas in the most massive halos is, on average, cooling at all radii.}
    \label{fig:cooling_times_vs_radius_average}
\end{figure*}

\begin{figure*}[hbtp]
    \centering
    \includegraphics[width=0.97\textwidth]{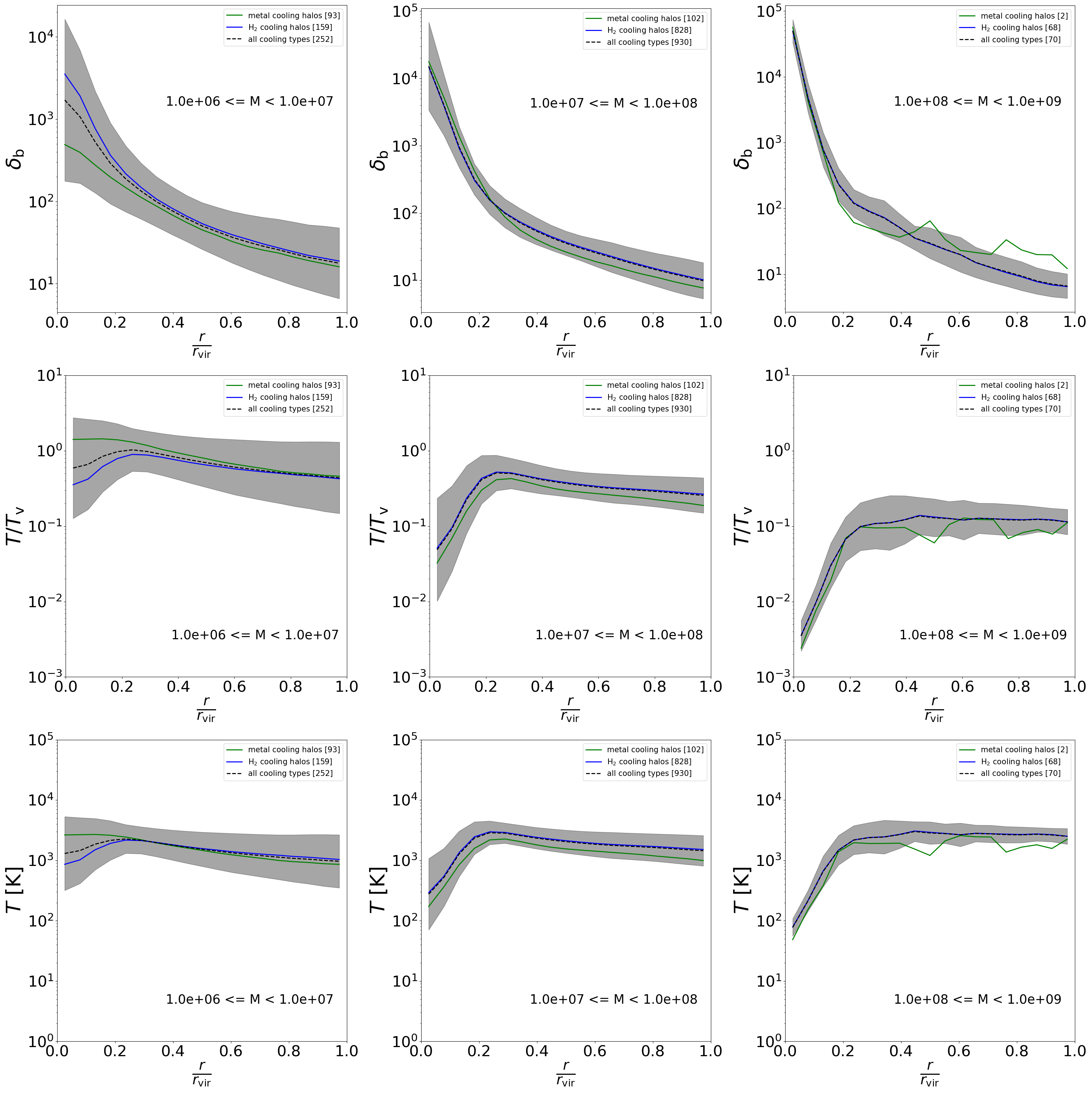}
    \caption{Average radial profiles of various gas properties for halos, binned by cooling type and virial mass. Each row shows a different gas property, and each column corresponds to halos in a given mass range. Metal cooling halos are shown in green, H$_2$ cooling halos are shown in blue, and the combined sample of cooling halos is shown in black. From top to bottom, profiles are plotted for baryon overdensity, normalized gas temperature, and un-normalized gas temperature, metallicity. The grey filled region extends out to 1 standard deviation in log space for the ``all cooling halos" sample.}
    \label{fig:gas_properties_vs_radius_average_0}
\end{figure*}

\begin{figure*}[hbtp]
    \centering
    \includegraphics[width=0.97\textwidth]{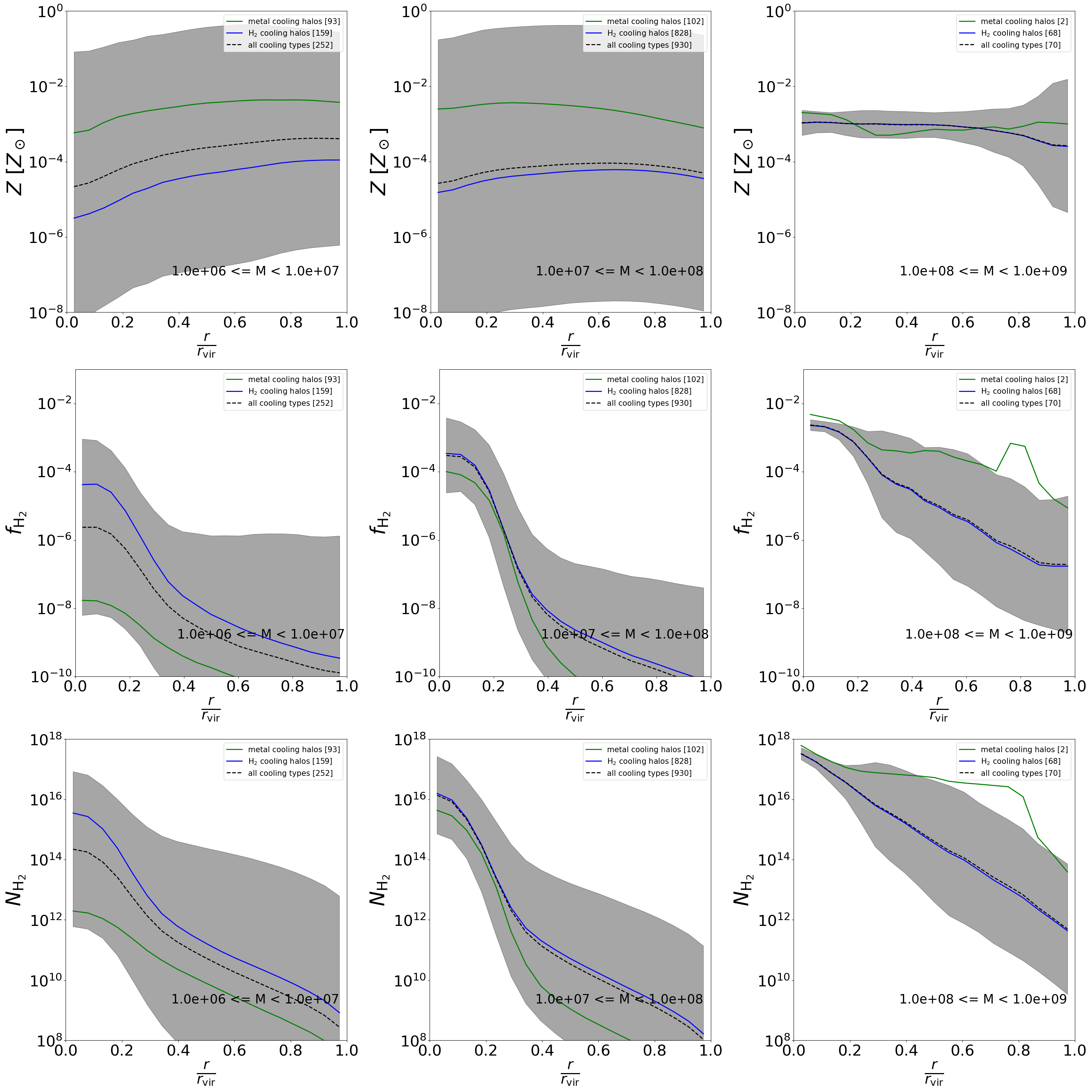}
    \caption{Same as Figure \ref{fig:gas_properties_vs_radius_average_0}, but showing profiles for metallicity, H$_2$ fraction, and H$_2$ column density.}
    \label{fig:gas_properties_vs_radius_average_1}
\end{figure*}

\begin{figure*}[t]
    \centering
    \includegraphics[width=0.96\textwidth]{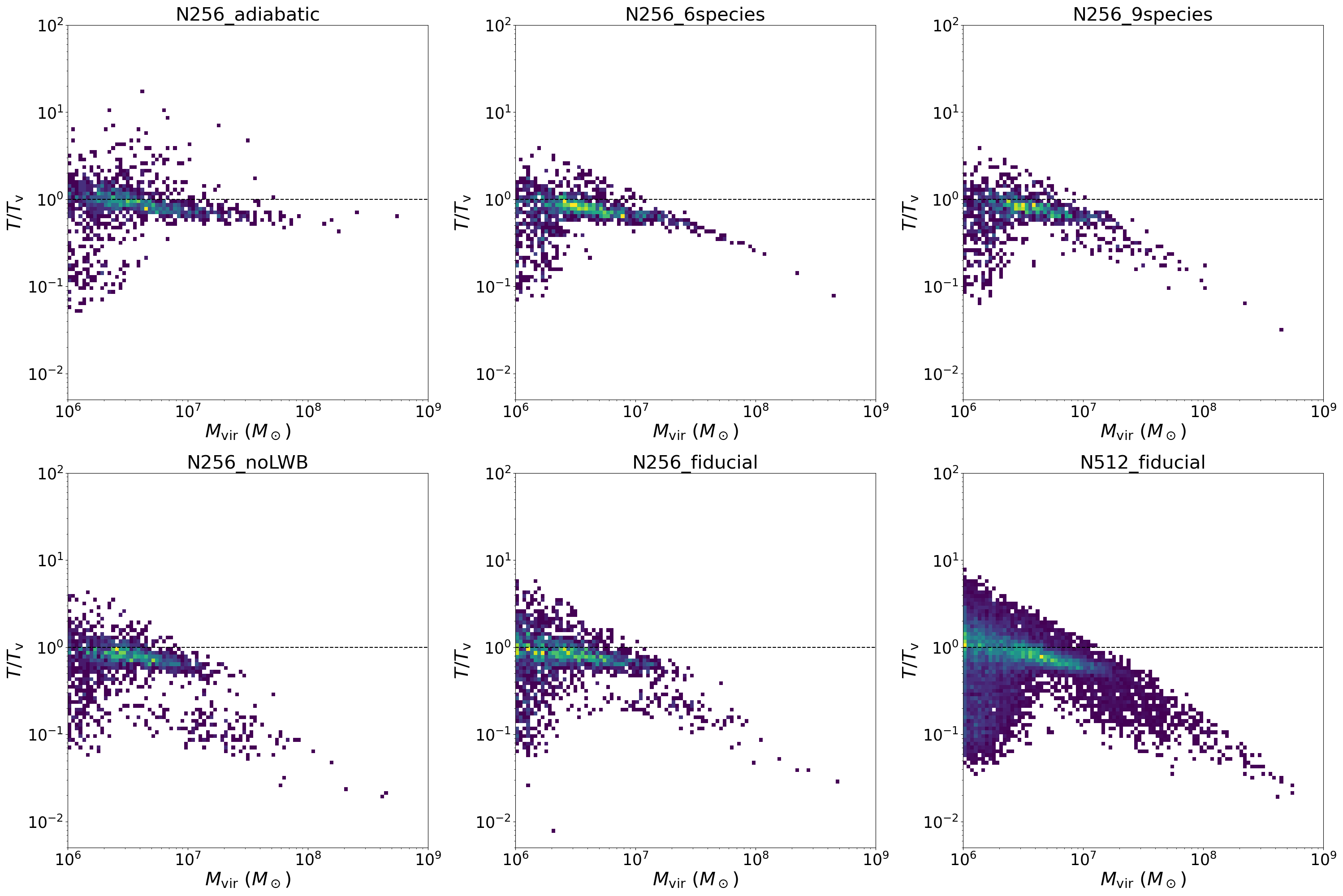}
    \caption{ Normalized gas temperature within the virial radius for all halos in the simulations listed in Table \ref{table: params}. An ideal virialized halo would have $T/T_\mathrm{vir}\approx1$.}
    \label{fig:T_Tvir_Mvir_comparison}
\end{figure*}

\begin{figure*}[t]
    \centering
    \includegraphics[width=0.96\textwidth]{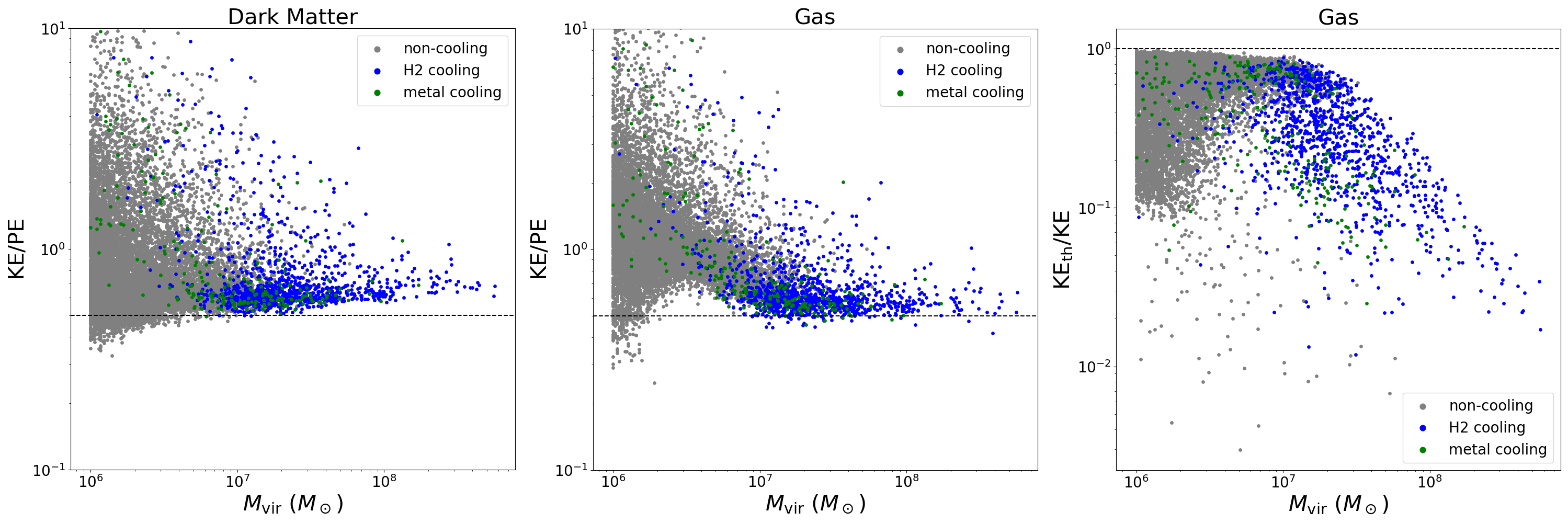}
    \caption{\textit{Left:} Ratio of total kinetic to total potential energy for dark matter particles within the virial radius. The horizontal dashed line corresponds to KE/PE=0.5, which is the value corresponding to an ideally virialized halo. \textit{Middle:} Ratio of total kinetic, including thermal and bulk motions, to total potential energy for gas within the virial radius. \textit{Right:} Ratio of thermal energy to total kinetic energy. For each panel, grey points represent non-cooling halos, blue points represent H$_2$ cooling halos, and green points represent metal cooling halos.}
    \label{fig:virial_check}
\end{figure*}

\begin{figure}[t]
    \centering
    \includegraphics[width=0.47\textwidth]{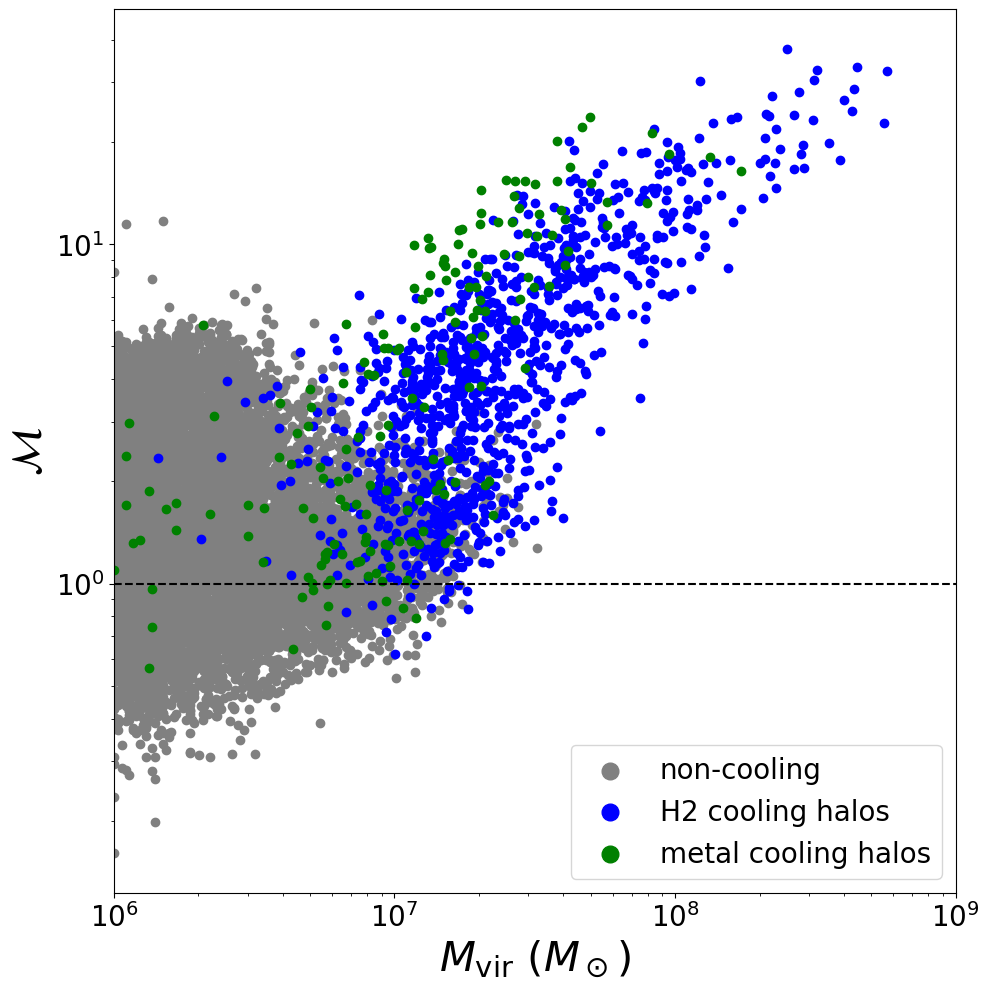}
    \caption{Mass-weighted average Mach number within the virial radius vs. virial mass for halos in the \texttt{N512\_fiducial} simulation. Flows become increasingly supersonic for $M_\mathrm{vir}\gtrsim10^7\,\,M_\odot$ as halo mass increases.}
    \label{fig:mach_scatter}
\end{figure}

\begin{figure*}
    \centering
    \includegraphics[width=0.96\textwidth]{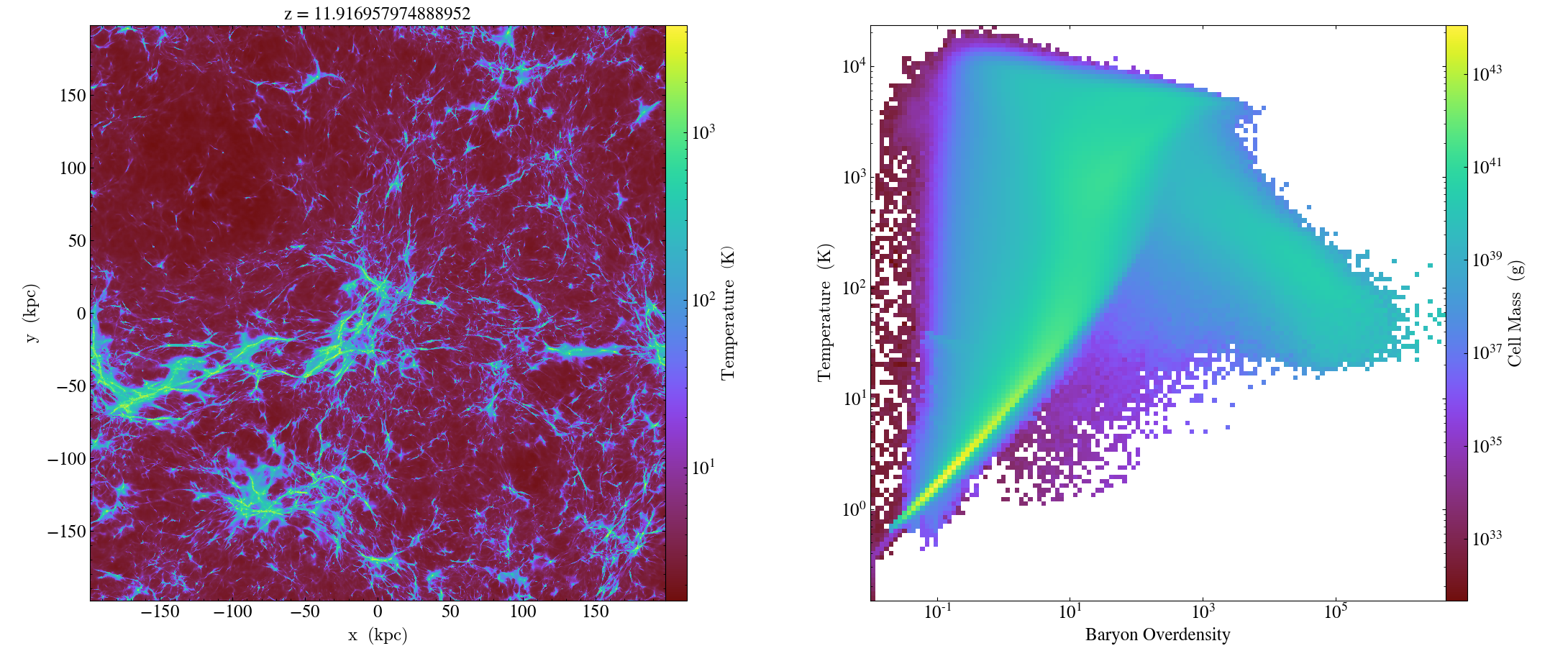}
    \caption{\textit{Left:} Full-box temperature projection for the \texttt{N512\_fiducial} simulation. Typical temperatures of gaseous filaments can be observed to be between $10^2$ K and $10^4$ K. \textit{Right:} Temperature vs. baryon overdensity phase diagram for all gas in the \texttt{N512\_fiducial} simulation. Overdensities between $10^1$ and $10^2$ correspond to gas in filaments.}
    \label{fig:temperature_projection}
\end{figure*}

\begin{figure*}
    \centering
    \includegraphics[width=0.96\textwidth]{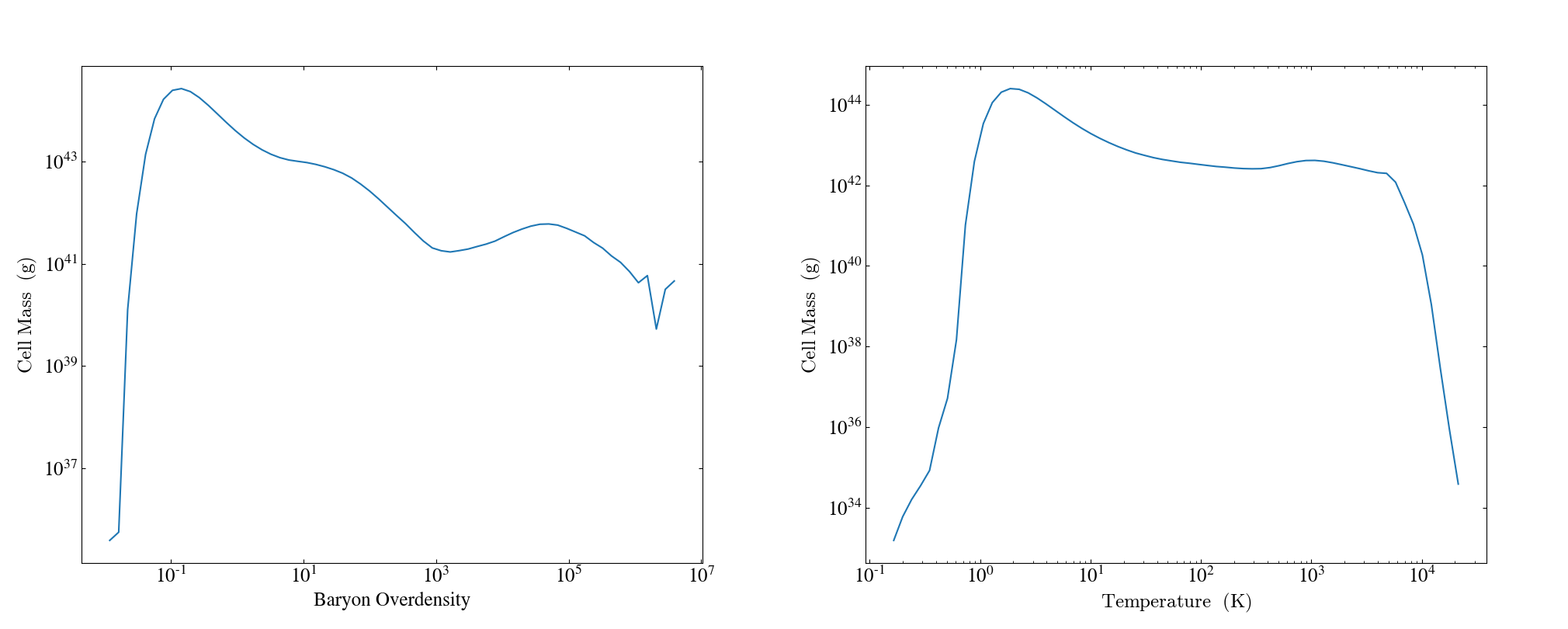}
    \caption{1D profiles for baryon overdensity (left) and temperature (right) over the full volume of the \texttt{N512\_fiducial} simulation.}
    \label{fig:1d_profiles}
\end{figure*}

For the following analysis, halos in the sample are classified as ``cooling" if the average cooling time within $R_{1000}$ is less than a Hubble time. Cooling rates and times are calculated in post using \texttt{PyGrackle}\footnote{\href{https://grackle.readthedocs.io/en/latest/Python.html}{https://grackle.readthedocs.io/en/latest/Python.html}}, the Python frontend of the \texttt{GRACKLE} chemistry and cooling library. We consider the following three modes of radiative cooling: (1) line transitions in atomic hydrogen and helium, (2) rovibrational line transitions in molecular hydrogen, and (3) metal-line transitions. The average volumetric cooling rate of each type is calculated within $R_{1000}$ as
\begin{equation}
    \overline{\Lambda} = \frac{\sum_i{\Lambda_i \Delta x_i^3}}{\sum_i{\Delta x_i^3}},
\end{equation}
where $i$ denotes the cell index and $\Delta x_i$ is the width of cell $i$.
A corresponding cooling time is then calculated as 
\begin{equation}
    t_\mathrm{cool} = \frac{\overline{\varepsilon}_\mathrm{int}}{\overline{\Lambda}},    
\end{equation} 
where $\overline{\varepsilon}$ is the average volumetric internal energy density within $R_{1000}$. The cooling halos are then further classified as either (1) atomic cooling halos, (2) H$_2$ cooling halos, or (3) metal cooling halos, depending on which of the three averaged cooling rates is largest within $R_{1000}$.

For each halo, we compute the accretion rate as the baryon mass flux through the virial radius:
\begin{equation}
    \dot{M}_\mathrm{acc} = -4\pi R_\mathrm{vir}^2 \frac{\sum_{i(r=R_\mathrm{vir})}{\mathbf{p}_i \cdot \hat{r}}}{\sum_{i(r=R_\mathrm{vir})}{\Delta x_i^3}}.
    \label{eq:accretion_rate}
\end{equation}
Here, $\hat{r}$ is the radial unit vector, centered at the halo's center of mass, and $\mathbf{p}_i$ is the gas momentum for a given cell, $i$, which is at the virial radius. The velocity used in computing the gas momentum is taken to be the value relative to the halo's center-of-mass velocity. 

Figure \ref{fig:accretion_rates} plots the accretion rates versus virial mass for halos in the \texttt{N512\_fiducial} simulation. There is a clear positive correlation between the two quantities. The data for cooling and non-cooling halos are separately fit to power laws of the form $dM/dt \propto M^\beta$. The two curves have similar slopes, indicating the physics of accretion is independent of cooling. The cooling halo curve has a slope of $\beta=0.92$, and the non-cooling curve has a slope of $\beta=0.83$. The majority of halos with $M_\mathrm{vir}>10^7\,\,M_\odot$ are cooling within $R_{1000}$.

Figure \ref{fig:inflow_rate_fractions} shows fractional inflow rates for gas at the virial radius that is cooling ($t_\mathrm{cool} < t_\mathrm{Hubble}$, red points), has $T < T_\mathrm{vir}$ (green points), and has $T < 10^3$ K (blue points). Fractional inflow rates are computed for each halo in the \texttt{N512\_fiducial} simulation. Total inflow rates are calculated in the same way as accretion rates with Equation \ref{eq:accretion_rate}, except that only gas flowing inwards (i.e. cells with $\mathbf{p}_i\cdot\hat{r} < 0$) contributes to the summation in the numerator. The inflow rates for gas satisfying each of the three conditions listed above are then calculated and normalized by the total. An immediate takeaway from this figure is that for most halos, the inflowing gas is non-cooling. For halos with $M_\mathrm{vir} \gtrsim 3\times10^7\,\,M_\odot$ that we observe, most of the gas flowing in has $10^3\,\,\mathrm{K} < T < T_\mathrm{vir}$. For most halos below this mass, the gas flows in with $T < 10^3$ K, though there is more scatter as halo mass decreases.

Figure \ref{fig:hmf_cooling} shows halo mass functions for cooling halos of various types. Cooling halos account for $12\%$ of halos in the sample. The most striking result from Figure \ref{fig:hmf_cooling} is the complete absence of atomic cooling halos. Halos with $T_\mathrm{vir}>10^4$ K are often colloquially referred to as ``atomic cooling halos", as it is typically assumed that gas at the center of these will cool primarily through atomic line transitions of H and He; however we do not observe this behavior in any of our halos, even those that are not yet metal enriched by supernovae. Instead, we observe that H$_2$ cooling is overall dominant for halos with $M_\mathrm{vir} > 10^4$ K. H$_2$ cooling halos account for $82\%$ of all cooling halos. The remaining $18\%$ of halos are dominated by metal-line cooling. $20\%$ of cooling halos have average metal and H$_2$ cooling rates that are within a factor of 2 of each other, meaning that there are significant contributions from both metal and H$_2$ cooling.

While the most massive halos tend to be dominated by H$_2$ cooling within $R_{1000}$, metal cooling tends to exceed H$_2$ cooling further in towards the center. This can be seen in the third column of Figure \ref{fig:cooling_times_vs_radius_average}, which shows average profiles of cooling time, normalized by the Hubble time, for halos of each cooling type across three mass ranges. For metal cooling halos with $M_\mathrm{vir} < 10^8\,\,M_\odot$, metal cooling is dominant at all radii. For H$_2$ cooling halos in this mass range, metal cooling times, while lower than H$_2$ cooling times at large radii, generally do not drop below the Hubble time. The H$_2$ cooling times for these halos drop below the Hubble time for $r/R_\mathrm{vir} < 0.2$.

Cooling halos with virial mass between $10^6\,\,M_\odot$ and $10^7\,\,M_\odot$ tend to have flat cooling time profiles, with $t_\mathrm{cool}<t_\mathrm{H}$ at all radii. As halo mass increases, the average profiles steepen for $r/R_\mathrm{vir} < 0.3$. For metal cooling halos with $M_\mathrm{vir} > 10^7\,\,M_\odot$, gas within within $R_{1000}$ is still able to cool in less than a Hubble time through H$_2$ cooling alone.

Figures \ref{fig:gas_properties_vs_radius_average_0} and \ref{fig:gas_properties_vs_radius_average_1} show average radial profiles for a variety of physical properties of the gas within cooling halos of different mass ranges. The baryon overdensity profiles generally steepen towards the center as halo mass increases, as can be expected, with metal and H$_2$ cooling halos reaching similar overdensities on average. In regards to the temperature of the gas, the average temperature at the virial radius does not change significantly as virial mass increases. When normalized by $T_\mathrm{vir}$, then, the temperature profiles generally shift downwards as virial mass increases. One would na\"ively expect that the average gas temperature within a virialized halo will be equal to the virial temperature, corresponding to $T/T_\mathrm{vir}=1$ within the virial radius. This is approximately true for the lowest mass bin; however, the two higher mass bins show average values of $T/T_\mathrm{vir} < 1$ for all radii. At face value, this suggests that halos with $10^7\,\,M_\odot < M_\mathrm{vir} < 10^8\,\,M_\odot$ have not yet reached virial equilibrium at $z\sim12$, and are actively accreting gas that is not shock heated to the virial temperature. 

The presence of a strong LWB serves to photodissociate H$_2$. Intuitively, one would expect H$_2$ cooling to be suppressed by the LWB. This is generally true in low-density environments; however, we do not observe suppressed H$_2$ cooling in dense cores. This is primarily due to H$_2$ self-shielding. The bottom two rows of Figure \ref{fig:gas_properties_vs_radius_average_1} show average profiles for H$_2$ fraction ($f_\mathrm{H_2}$) and H$_2$ column density ($N_\mathrm{H_2}$). At large radii, $f_\mathrm{H_2}$ is generally low, with values $\leq 10^{-8}$ for the central mass bin. At $r/R_\mathrm{vir} \approx 0.3$, the $f_\mathrm{H_2}$ and $N_\mathrm{H_2}$ curves both sharply increase. H$_2$ self-shielding becomes non-negligible at $N_\mathrm{H_2} \approx 5\times10^{14}\,\,\mathrm{cm^{-2}}$ for the self-shielding model used in our simulation \citep{WolcottGreen2019}.

\subsection{Cooling vs. Incomplete Virialization}
\label{sec:virialization}

The temperature profiles in Figure \ref{fig:gas_properties_vs_radius_average_0} suggest that the halos in the \texttt{N512\_fiducial} simulation are generally not in true virial equilibrium, using the virial temperature as a metric. \cite{Lochhaas_2021} shows that the virial temperature is not an accurate description of virialized gas, as bulk flows at the center of the halo provides a non-negligible contribution of non-thermal kinetic energy to the system, which is not taken into account in the formal definition of the virial temperature. Using a set of cosmological zoom-in simulations of star-forming galaxies, they find that the true mean temperature within a ``virialized" halo is typically lower than $T_\mathrm{vir}$ by a factor of $\sim2$. It should be noted that the simulations in \cite{Lochhaas_2021} include thermal energy injection due to various forms of Pop II stellar feedback, thus providing efficient means of heating the gas to offset radiative cooling, while our simulations do not. The most massive halos in our simulations experience runaway cooling at their center, which drives the mean halo gas temperature down to values $\sim 50\times$ smaller than $T_\mathrm{vir}$. This can be seen in Figure \ref{fig:T_Tvir_Mvir_comparison}, which shows the mean gas temperature within the virial radius, scaled by $T_\mathrm{vir}$, for every halo across all of our simulations. In each case, there is a large scatter for halos with $M_\mathrm{vir} \lesssim 5\times10^6\,M_\odot$. The scatter then decreases for $M_\mathrm{vir}\gtrsim10^7\,M_\odot$. Beyond this point, the values either level out as $M_\mathrm{vir}$ increases, as is the case in the \texttt{N256\_adiabatic} simulation, or steadily drop as gas cooling becomes more efficient at the halo centers. In the adiabatic case, the values level out at $T/T_\mathrm{vir}\sim0.6$, which is in agreement with the findings in \cite{Lochhaas_2021}. For the \texttt{N256\_6species} simulation, the temperatures drop smoothly with very little spread as $M_\mathrm{vir}$ increases. The \texttt{N256\_9species}, \texttt{N256\_fiducial}, and \texttt{N512\_fiducial} values have more scatter at high masses because there is more complexity in the physics contributing to cooling, and reach values of $T/T_\mathrm{vir}$ as low as $10^{-2}$.

As a direct measure of virialization, we show the ratio of the total kinetic energy to total potential energy separately for dark matter and gas for each halo in Figure \ref{fig:virial_check}. For the gas, the kinetic energy is taken to be the sum of both the thermal and non-thermal components, where the non-thermal component is computed as
\begin{equation}
    \mathrm{KE}=\sum_{i=0}^{N_\mathrm{cells}}{\frac{1}{2}M_{i} v_i^2},
    \label{eq:KE}
\end{equation}
where $i$ loops over cells within the virial radius. The dark matter KE is computed in the same way as the gas, except instead of looping over cells in Equation \ref{eq:KE}, $i$ loops over dark matter particles. Potential energies are computed pairwise. The dark matter potential energy takes into account DM-DM gravitational interactions, as well as DM-gas, and vice versa for the gas potential energy. Specifically,
\begin{equation}
    \mathrm{PE_{gas}} = -\frac{1}{2}\left[\sum_{j=0}^{N_\mathrm{cells}} \sum_{i=0}^{N_\mathrm{cells}}{\delta_{ij}} + \sum_{j=0}^{N_\mathrm{dm}} \sum_{i=0}^{N_\mathrm{cells}}\right] {G\frac{M_iM_j}{r_{ij}}}
    \label{eq:PE_gas}
\end{equation}
for the gas, and
\begin{equation}
    \mathrm{PE_{dm}} = -\frac{1}{2}\left[\sum_{j=0}^{N_\mathrm{dm}} \sum_{i=0}^{N_\mathrm{dm}}{\delta_{ij}} + \sum_{j=0}^{N_\mathrm{cells}} \sum_{i=0}^{N_\mathrm{dm}}\right] {G\frac{M_iM_j}{r_{ij}}}
    \label{eq:PE_dm}
\end{equation}
for the dark matter. Here, $r_{ij}=\mathrm{\texttt{max}\left( \left|\mathbf{r}_i-\mathbf{r}_j\right|,\,\,\varepsilon \right)}$, where $\varepsilon=1$ kpccm (0.08 kpc at $z=12$, which is roughly equal to 3 cell widths at the highest AMR refinement level). $\delta_{ij}$ evaluates to zero if $i=j$. The summations are multiplied by a factor of 1/2 to avoid double-counting.

For an ideal virialized halo, the magnitude of the time-averaged ratio of kinetic to potential energy should equal 0.5, as denoted by the dashed lines in the left and middle panel of Figure \ref{fig:virial_check}. We see in Figure \ref{fig:virial_check} that there is a lot of scatter in KE/PE at the low virial mass, but values concentrate at the high mass end around $\mathrm{KE/PE}\approx0.6$ for both the dark matter and the gas, suggesting that both the dark matter and gas are in a dynamical equilibrium consistent with virialization. We show the ratio between the thermal energy and total kinetic energy of the gas in the right panel of Figure \ref{fig:virial_check}. As halo mass increases, we observe that the gas KE becomes increasingly dominated by the non-thermal component. This is related to the runaway cooling effect that we discuss above. The shape of the gas profile mirrors that of the temperature profile in the \texttt{N512\_fiducial} panel of Figure \ref{fig:T_Tvir_Mvir_comparison}. From this, we can infer that while the gas is virialized for $M_\mathrm{vir}\gtrsim10^7\,\,M_\odot$, virialization is supported by bulk turbulent flows, and not by thermal motions. For this reason, we describe these halos as being in a state of ``incomplete virialization". If the gas KE was instead dominated by the thermal component, we would expect the halo-averaged gas temperatures in Figure \ref{fig:T_Tvir_Mvir_comparison} to be roughly equal to the virial temperature.

Due to the low gas temperatures in halos with $M_\mathrm{vir}\gtrsim10^7\,\,M_\odot$, internal flows become increasingly supersonic on average as halo mass increases. This is shown in Figure \ref{fig:mach_scatter}, which plots mean Mach number within the virial radius versus virial mass for all halos in the \texttt{N512\_fiducial} simulation. At the high mass end, mean Mach numbers have values of $\mathcal{M} > 10$. The shocks that result from colliding supersonic flows within the virial radius drive the turbulence that supports the virialization state of the gas.

\subsection{The Cosmic Web at Redshift 12 Consists of Warm Filaments}
\label{sec:warm_filaments}

We now take a short detour to discuss the properties of the medium that connects our dark matter halos together. We have seen that the majority of gas flowing into these halos flows in through filaments with temperatures between $10^2$ K and $10^4$ K at $z\sim12$, with $10^3$ K being typical (Fig. \ref{fig:gas_properties_vs_radius_average_0}). It is not unsurprising, then, that this is the temperature range that generally describes the entire network of filaments at this redshift. Figure \ref{fig:temperature_projection} shows a projection of temperature for the \texttt{N512\_fiducial} simulation, as well as a global phase diagram of temperature versus overdensity. To accompany this, Figure \ref{fig:1d_profiles} shows 1D profiles of baryon overdensity and temperature. Baryon overdensities typical of filaments are $\delta_\mathrm{b}\simeq3-10$. The mean temperature of the IGM at a given redshift is given by
\begin{equation}
    \label{eq:T_IGM}
    T(z) = \frac{T_0}{1+200}(1+z)^2,
\end{equation}
where $T_0=2.73$ K is the IGM temperature at $z=0$ \citep{anninos_1996}. At $z=12$, Equation \ref{eq:T_IGM} evaluates to 2.3 K, which is consistent with the peak temperature value in the right panel of Figure \ref{fig:1d_profiles}.

At $z>10$, the IGM has not yet been heated and ionized by the buildup of a UV background. Because of the low temperatures, and thus low sound speeds, gas flows in the IGM are highly supersonic, with Mach numbers of $\mathcal{M} > 100$. Since $\mathcal{M} \gg 1$, a flow encountering a filament from the outside will be strongly shocked. We have previously seen in Figure \ref{fig:inflow_rate_fractions} the gas flowing through filaments into the most massive halos generally has cooling times greater than a Hubble time when it reaches the virial radius. As such, it is reasonable to assume that the shock will be adiabatic. For a strong shock in neutral gas, the post-shock temperature can be written as
\begin{equation}
    T_\mathrm{post} \approx \frac{2(\gamma-1)}{(\gamma+1)^2}\mathcal{M}_\mathrm{pre}^2 T_\mathrm{pre}
    \label{eq:strong_shock_temp}
\end{equation}
\citep{Draine_textbook}, where $\mathcal{M}_\mathrm{pre}$ and $T_\mathrm{pre}$ are the pre-shock Mach number and temperature, respectively. For $\mathcal{M}_\mathrm{pre}=100$, $T_\mathrm{pre}=2.3$ K, and $\gamma=5/3$, Equation \ref{eq:strong_shock_temp} evaluates to $5\times10^3$ K. This temperature is roughly the maximum temperature observed at the baryon overdensities consistent with filaments in Figure \ref{fig:temperature_projection}.

%% file: Results/IndividualHaloAnalysis.tex
\subsection{The Five Most Massive Halos}
\label{sec:individual_halo_analysis}

\begin{figure*}
    \centering
    \includegraphics[width=0.96\textwidth]{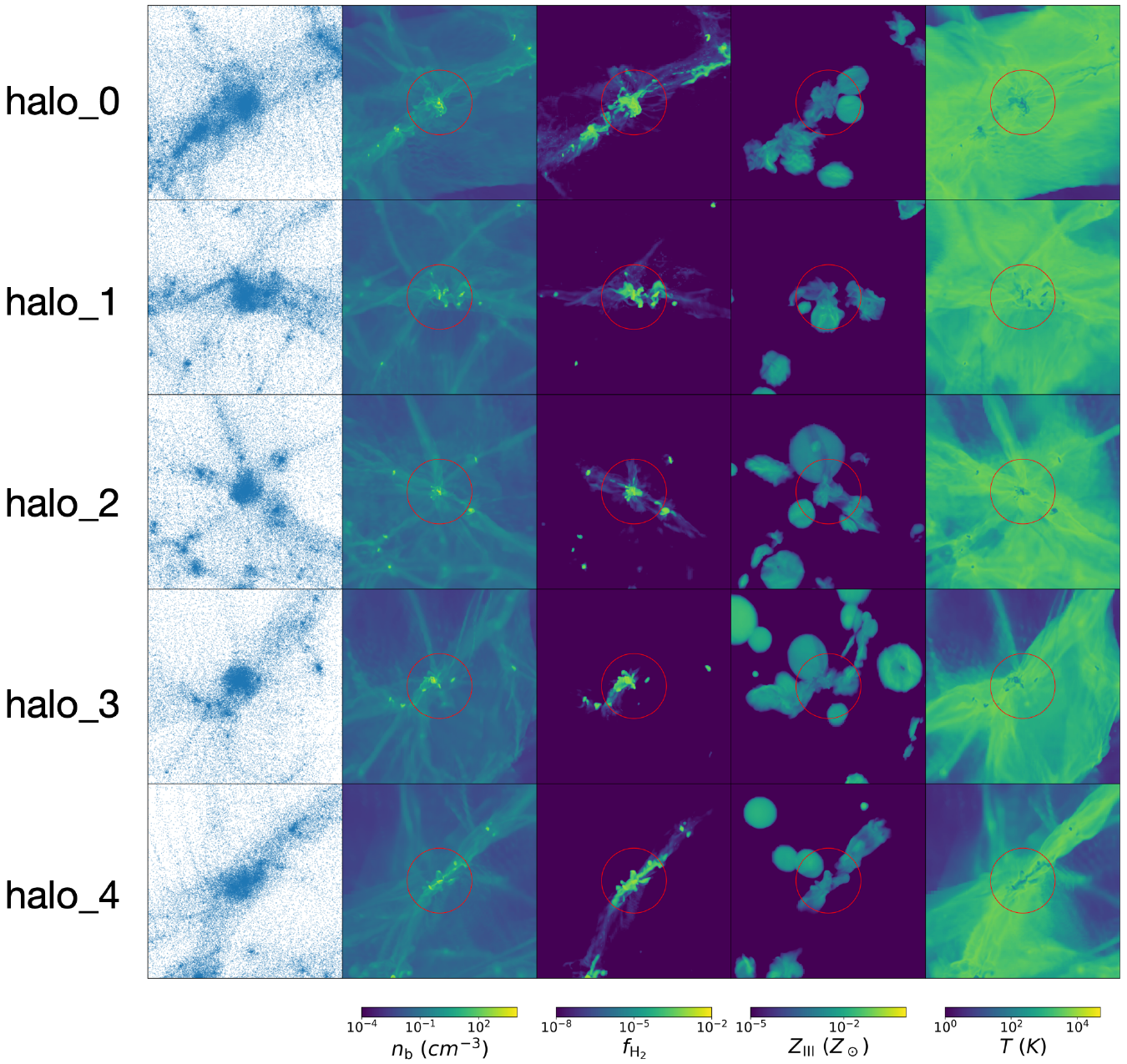}
    \caption{Projections of various quantities around the five most massive halos in the \texttt{N512\_fiducial} simulation. The red circles extend out to the virial radius, where each halo has a virial radius of approximately 2 proper kpc. From left to right, the images show dark matter particles, baryon density, H$_2$ fraction, metallicity, and temperature.}
    \label{fig:halo_projections}
\end{figure*}

\begin{figure*}
    \centering
    \includegraphics[width=0.96\textwidth]{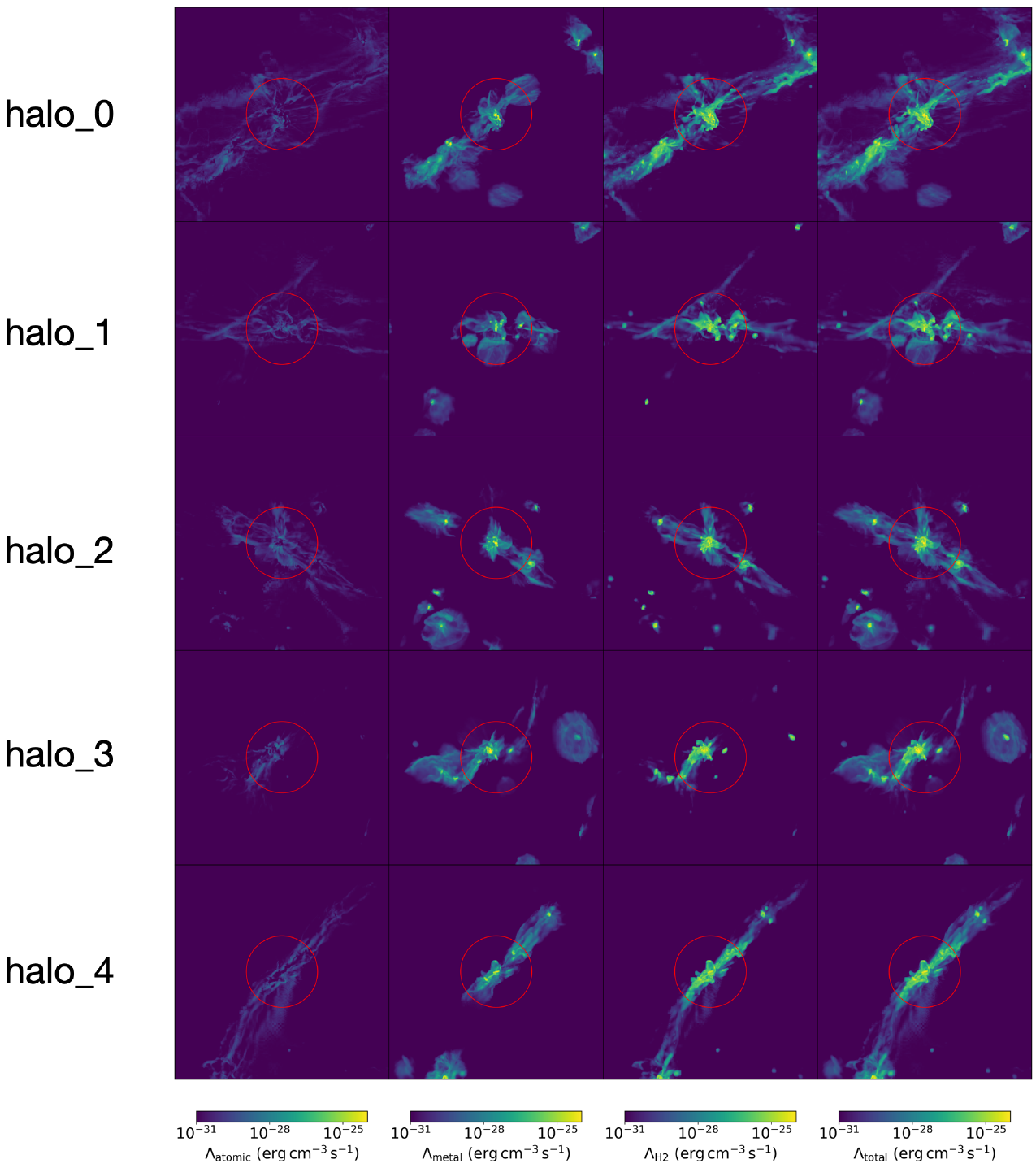}
    \caption{Projections of volumetric cooling rates for the five most massive halos in the \texttt{N512\_fiducial} simulation. From left to right, the images show atomic cooling rate, H$_2$ cooling rate, metal cooling rate, and total cooling rate. The similarities between H$_2$ cooling rates and total cooling rates demonstrate that H$_2$ cooling is overall dominant for these halos. The red circles extend out to the virial radius, where each halo has a virial radius of approximately 2 proper kpc.}
    \label{fig:halo_projections_cooling}
\end{figure*}

\begin{figure*}
    \centering
    \includegraphics[width=0.96\textwidth]{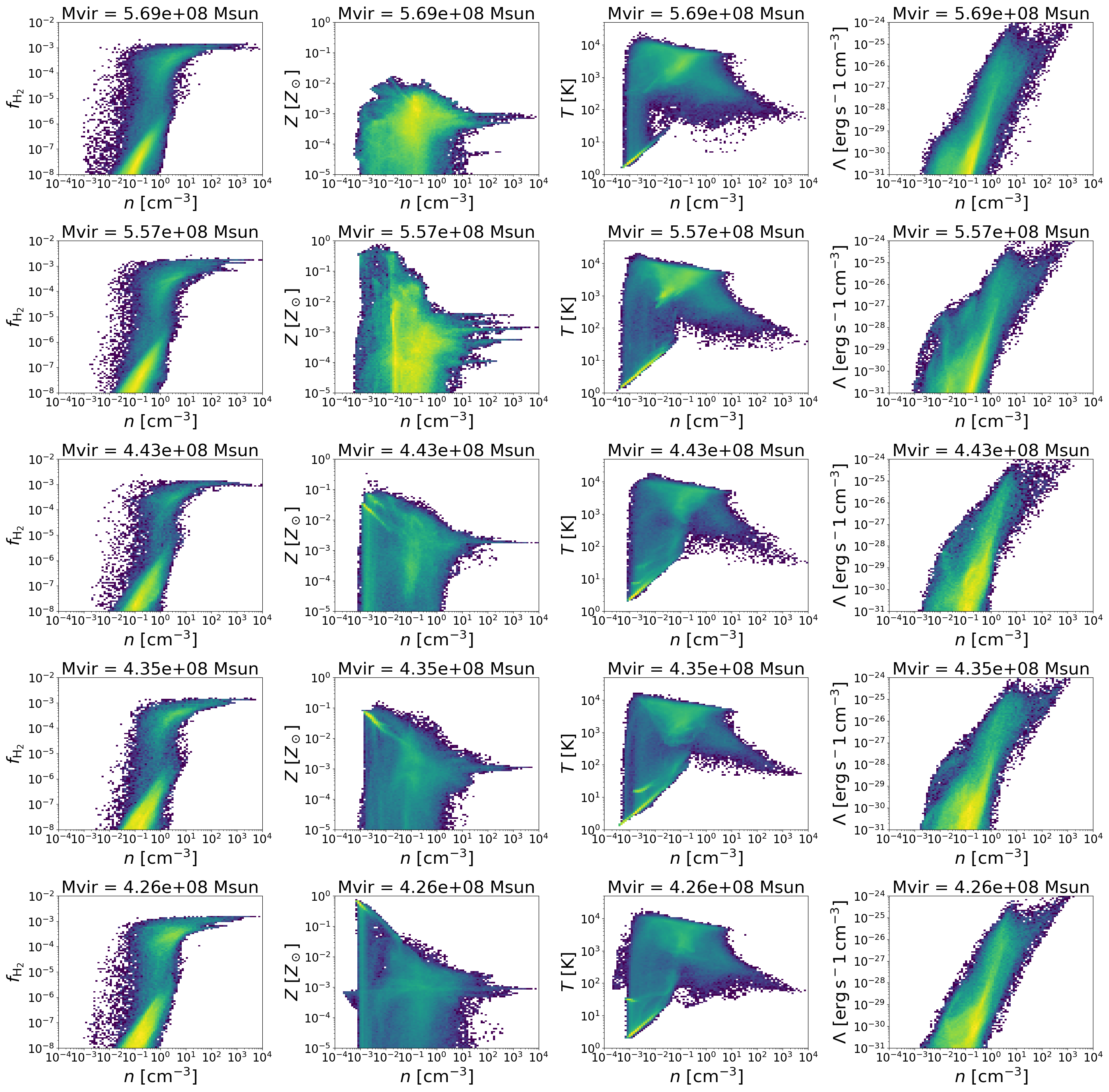}
    \caption{Phase diagrams of various properties of the gas within the five most massive halos vs. baryon number density. From left to right, diagrams are shown for H$_2$ fraction, metallicity, volumetric cooling rate, and temperature. Each row corresponds to a different halo. From top to bottom, the diagrams correspond to halos 0, 1, 2, 3, and 4. The distributions of the chosen properties look very similar for each of the halos.}
    \label{fig:halo_phases}
\end{figure*}

\begin{table*}
    \centering
    \begin{tabular}{ |p{1cm}||p{1.5cm}|p{1.5cm}|p{1.5cm}|p{1.5cm}|p{2cm}|p{1cm}|p{1.75cm}|p{1.5cm}| }
        \hline
        \multicolumn{9}{|c|}{Halo Properties} \\
        \hline
        Name & $M_\mathrm{vir}\,\,[M_\odot]$ & $M_\mathrm{b}\,\,[M_\odot]$ & $M^\mathrm{*}_\mathrm{II}\,\,[M_\odot]$ & $R_\mathrm{vir}$ [kpc] & $\left< Z \right>_{1000}\,\,[Z_\odot]$ & $\Lambda_{Z}/\Lambda_{\mathrm{H}_2}$ & $N_\mathrm{remnants}$ & $N_\mathrm{cells}$ \\
        \hline
        \texttt{Halo\_0} & $5.67\times10^8$ & $1.21\times10^8$ & $1.44\times10^6$ & 2.08 & $7.19\times10^{-4}$ & 0.97 & 20 & 993,088\\
        \texttt{Halo\_1} & $5.57\times10^8$ & $1.09\times10^8$ & $1.80\times10^6$ & 2.06 & $1.25\times10^{-3}$ & 0.93 & 13 & 821,874\\
        \texttt{Halo\_2} & $4.43\times10^8$ & $7.76\times10^7$ & $1.02\times10^6$ & 1.91 & $1.74\times10^{-3}$ & 0.96 & 13 & 451,572\\
        \texttt{Halo\_3} & $4.35\times10^8$ & $8.22\times10^7$ & $8.24\times10^5$ & 1.90 & $9.31\times10^{-4}$ & 0.37 & 10 & 675,178\\
        \texttt{Halo\_4} & $4.26\times10^8$ & $8.17\times10^7$ & $8.90\times10^5$ & 1.89 & $9.90\times10^{-4}$ & 0.69 & 12 & 820,252\\
        \hline
    \end{tabular}
    \caption{Properties of the five most massive halos in the \texttt{N512\_fiducial} simulation. From left to right, columns show halo name, dark matter virial mass, total baryon mass within $R_\mathrm{vir}$, predicted Pop II stellar mass (see Sec. \ref{sec:popII_inferences}), virial radius in proper kpc, mean metallicity within $R_{1000}$, the ratio of mean metal cooling rate to mean H$_2$ cooling rate within $R_{1000}$, the number of \texttt{popIII\_remnant} particles within $R_\mathrm{vir}$, and the total number of computational cells within $R_\mathrm{vir}$.}
    \label{table:halo_properties}
\end{table*}

This study is ultimately motivated by the topic of high redshift galaxy formation, which would take place preferentially in the most massive halos. In this section, we analyze the five most massive halos in the \texttt{N512\_fiducial} simulation, and discuss their potential for galaxy formation. Table \ref{table:halo_properties} lists some basic properties of these 5 halos, which have virial masses between $4\times10^8\,\,M_\odot$ and $6\times10^8\,\,M_\odot$. All 5 halos contain metal-enriched gas from $> 10$ Pop III remnants, yet H$_2$ cooling rates exceed metal cooling rates within $R_\mathrm{1000}$, though halos 0, 1, 2, and 4 have metal cooling rates that are within a factor of 2 of the H$_2$ cooling rates. For halos 0, 1, and 2, the fraction of mean central metal cooling rates to H$_2$ cooling rates 0.97, 0.93, and 0.96, respectively, meaning that metal cooling rates are on par with H$_2$ cooling rates at the centers of these halos.

We show zoomed-in projections of dark matter, baryon number density, H$_2$ fraction, metallicity, and temperature for the 5 most massive halos in Figure \ref{fig:halo_projections}. In the dark matter projections, much substructure can be seen. The halos look to be in various stages of merging with nearby halos. As a result, the baryon distributions have highly irregular morphologies. Halos 0 and 1 in particular are undergoing major mergers, as is evident by the strong tidal disruption in both the dark matter and the baryon density projections. Halo 2 is perhaps the most stable system of the batch, with a quasi-spherical distribution of particles at the center, and multiple satellite halos. H$_2$ flows in through filaments, and gas with $n_\mathrm{b} \gtrsim 5\times10^{-1} \,\,\mathrm{cm^{-3}}$ is self-shielded to H$_2$ photodissociation by the LWB, resulting in large H$_2$ fractions within the halos.

Projections of atomic, metal, H$_2$, and total cooling rates for the five most massive halos are shown in Figure \ref{fig:halo_projections_cooling}. The H$_2$ cooling rates are visibly very similar to the total cooling rates, which again demonstrates the importance of H$_2$ cooling. The regions where metal cooling is important appear to be more concentrated towards the halo centers. Atomic cooling rates appear to be non-zero along the filament boundaries, where some amount of shock heating has taken place. However, atomic cooling does not make any significant contribution to cooling when compared with the total cooling rates.

Figure \ref{fig:halo_phases} shows phase diagrams of H$_2$ fraction, metallicity, volumetric cooling rate, and temperature versus baryon number density for the gas within each of the five most massive halos. The images look very similar for each of the chosen halos. The effects of H$_2$ self-shielding can be observed in the H$_2$ fraction diagrams (leftmost column), where the distribution for $n \gtrsim 10^{-1}\,\,\mathrm{cm^{-3}}$ becomes bimodal, with values tending towards $f_\mathrm{H_2}=10^{-3}$ at large densities. Metallicities (second column from the left) tend towards $Z=10^{-3}\,\,Z_\odot$ at large densities. This tendency of metallicities towards $10^{-3} Z_\odot$ also shows up in the halo-averaged metallicities of Figure \ref{fig:metallicity_vs_Mvir_RGB}, where the mean metallicities of gas within high mass halos are also concentrated around $10^{-3}\,\,Z_\odot$.

%% file: Results/PopII_inferences.tex
\subsection{Estimates Of Pop II Star Formation}
\label{sec:popII_inferences}

\begin{figure*}
    \centering
    \includegraphics[width=0.96\textwidth]{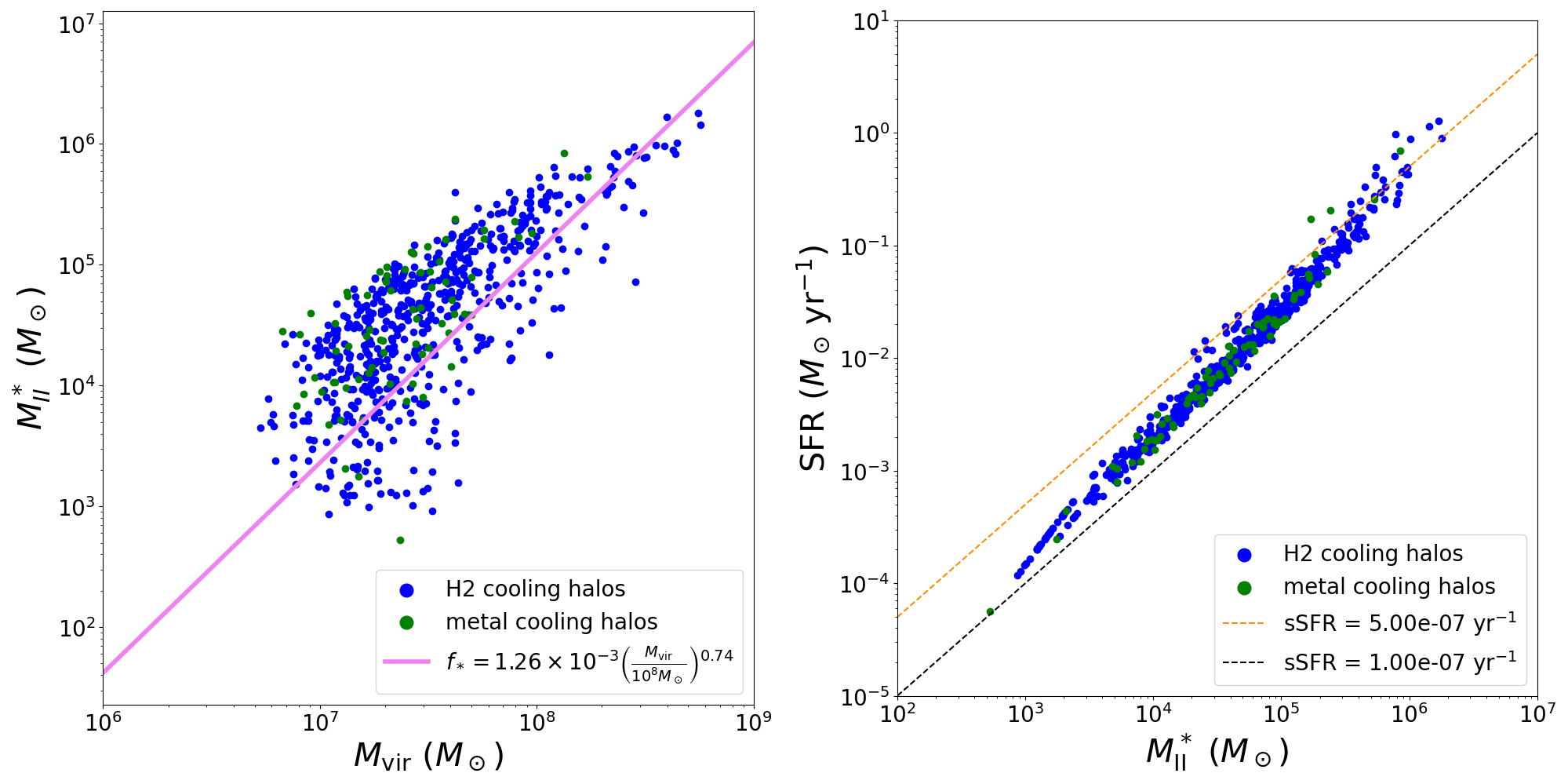}
    \caption{\textit{Left:} Predicted Pop II stellar mass vs. virial mass. Blue points correspond to H$_2$ cooling halos, and green points correspond to metal cooling halos. The pink line shows a fit derived in \cite{oshea2015} using the \textit{Renaissance Simulations}. \textit{Right:} Predicted SFR vs. predicted Pop II stellar mass for H$_2$ cooling and metal cooling halos. The dashed lines show the expected values given a constant specific SFR.}
    \label{fig:predicted_stellar_masses}
\end{figure*}

We estimate the mass of Pop II stars, $M^*_\mathrm{II}$, that would potentially form in the most massive halos if a model for explicit Pop II star formation was included in the \texttt{N512\_fiducial} simulation. To do this, we apply the following set of set of star-forming criteria in post to identify potential star-forming cells:
\begin{enumerate}
    \item \textit{gas flow is converging:} $\mathbf{\nabla}\cdot\mathbf{v} < 0$
    \item \textit{gas is self-gravitating:} $\frac{\varepsilon}{\varphi} < 1$, where $\varepsilon$ is the specific total energy of the gas, and $\varphi$ is the gravitational potential. The ratio of these two quantities is akin to the virial parameter described in \cite{Hopkins_2018}.
    \item \textit{cell mass is above a Jeans mass:} $M_\mathrm{cell} > M_\mathrm{Jeans}$
    \item \textit{gas is enriched:} $Z_\mathrm{cell} > Z_\mathrm{crit}$, where we take $Z_\mathrm{crit} = 10^{-5.5}\,\,Z_\odot$
\end{enumerate}
 The set of star formation criteria is inspired by the set used in \cite{Hopkins_2022}, and are those we will use in a followup study that will include an explicit recipe for Pop II star formation and feedback in a re-run of \texttt{N512\_fiducial}. We have verified that the inclusion of a criterion enforcing a minimum gas density for star formation is not necessary, as gas density is implicitly taken into account in the Jeans mass and self-gravitating criteria (items 2 and 3, respectively). The flagged star-forming cells typically have baryon overdensities between $10^4$ and $10^5$, which is consistent with the central overdensities seen in Figure \ref{fig:gas_properties_vs_radius_average_0}. We assume a conversion fraction of gas into stars of 0.05, and multiply the gas mass within a cell that satisfies all criteria by this value to obtain an estimate of the Pop II stellar mass. We then compute the Pop II SFR as
 \begin{equation}
     \mathrm{SFR}=\frac{\eta_{sf}M_\mathrm{b}}{t_\mathrm{ff}},
 \end{equation}
 where $\eta_\mathrm{s}=0.05$ is the conversion fraction of gas into stars, $M_\mathrm{b}$ and $t_\mathrm{ff}$ are the baryon mass and freefall time of the star forming cell, respectively. If multiple cells are flagged as star-forming in a halo, we report the total stellar mass and SFR summed over all flagged cells.
 
 The Pop II stellar mass estimates for the 5 most massive halos are listed in Table \ref{table:halo_properties}. It should be stressed that the true stellar masses for these halos would be affected by feedback. Additionally, star formation will happen continuously, as opposed to at one instant, which could likely result in more the 5\% of the gas being converted into stars. Nevertheless, this analysis provides us with a value for the \textit{amount} of gas that has collected within the halos by $z\sim12$ that is currently available for star formation. The stellar masses for the top 5 most massive halos are of order $10^6\,\,M_\odot$. Extending this analysis to all halos yields Figure \ref{fig:predicted_stellar_masses}, which shows a scatter plot of the estimated Pop II stellar mass versus virial mass. Stellar masses range from as low as $5\times10^2\,\,M_\odot$ to as high as $2\times10^6\,\,M_\odot$. The stellar masses are in good agreement with those obtained in \cite{oshea2015} using the \textit{Renaissance Simulations}. Figure \ref{fig:predicted_stellar_masses} also shows Pop II SFR versus stellar mass. Specific SFRs for galaxies formed in the \textit{Renaissance Simulations} are shown in Figure 4 of \cite{McCaffrey_2023}. This is to be expected since we have ignored feedback effects.  Our sSFRs are slightly higher than those observed in the \textit{Renaissance Simulations}, of which $1\times10^7\,\,\mathrm{yr^{-1}}$ is the largest sSFR observed. The specific SFRs obtained by our analysis are between $1\times10^{-7}\,\,\mathrm{yr^{-1}}$ and $6\times10^{-7}\,\,\mathrm{yr^{-1}}$.

%% file: Results/SensitivityStudy.tex
\subsection{Sensitivity to Variations in Feedback and Chemistry Prescriptions}
\label{sec:sensitivity_study}

%\begin{figure*}
%    \centering
%    \includegraphics[width=0.96\textwidth]{Figures/SensitivityStudy/gas_properties_vs_radius_comparison_cooling.png}
%    \caption{Caption}
%    \label{fig:comparison_average_gas_properties_cooling}
%\end{figure*}

%\begin{figure*}
%    \centering
%    \includegraphics[width=0.96\textwidth]{Figures/SensitivityStudy/MMH_density_temperature.png}
%    \caption{Projections and slices of density and temperature for the most massive halo in each of the $256^3$ simulations. From left to right, images are shown for \texttt{N256\_adiabatic}, \texttt{N256\_6species}, \texttt{N256\_9species}, \texttt{N256\_noLWB}, and \texttt{N256\_fiducial}. From top to bottom, images are density projections, temperature projections, density slices, and temperature slices. Similarities and differences between the gas morphologies can be observed here, with the most drastic differences coming from the \texttt{N256\_adiabatic}.}
%    \label{fig:MMH_density_temperature_comparison}
%\end{figure*}

\begin{figure*}
    \centering
    \includegraphics[width=0.96\textwidth]{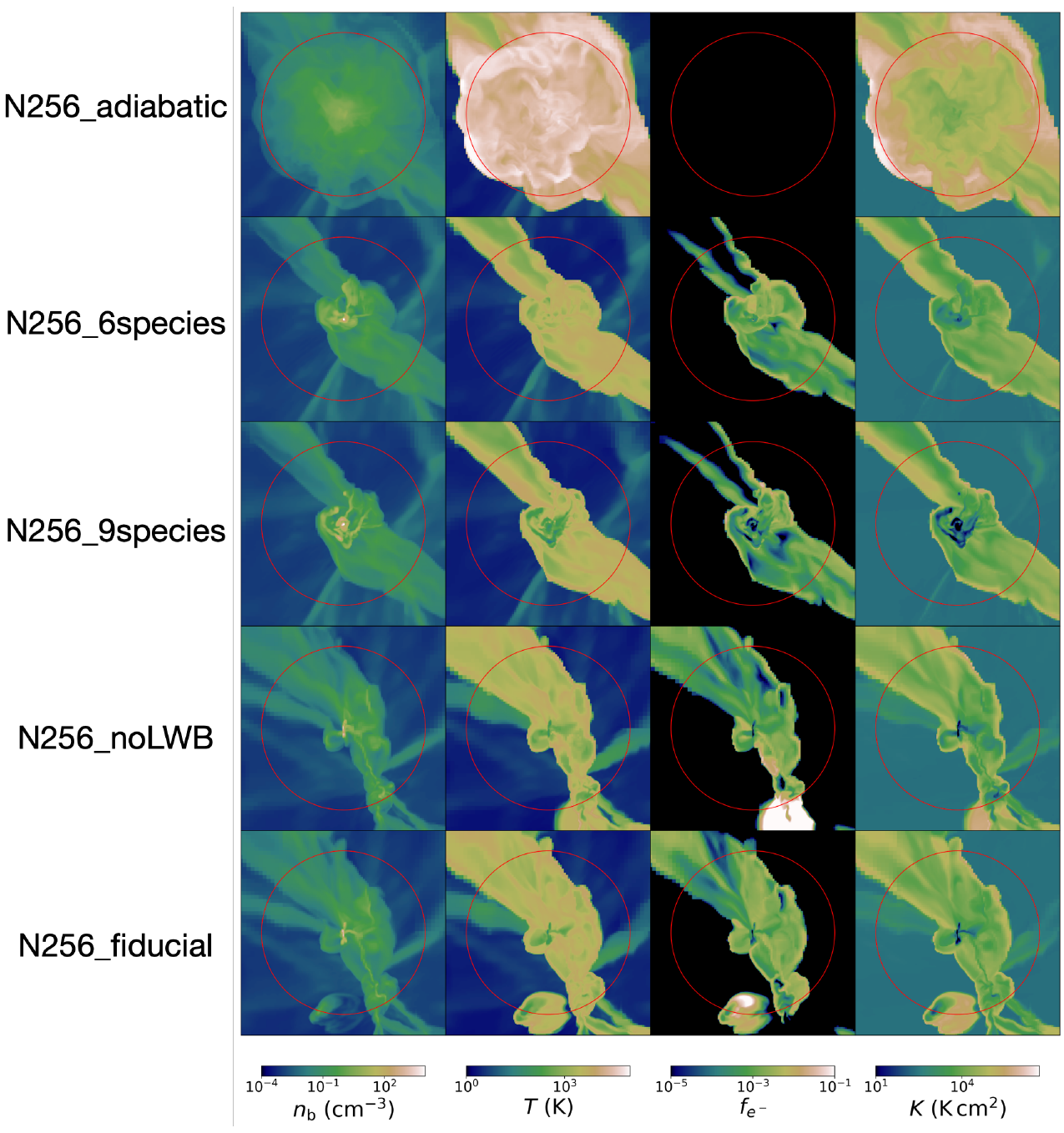}
    \caption{Slices of various quantities through the center the most massive halo in each of the $256^3$ simulations. From top to bottom, images are shown for \texttt{N256\_adiabatic}, \texttt{N256\_6species}, \texttt{N256\_9species}, \texttt{N256\_noLWB}, and \texttt{N256\_fiducial}. From left to right, slices are of the baryon number density, temperature, electron fraction, and entropy ($K=T/n_\mathrm{b}^{2/3}$) fields. The red circles denote the virial radius of each halo. Similarities and differences between the gas morphologies can be observed here, with the most drastic differences coming from the \texttt{N256\_adiabatic}.}
    \label{fig:MMH_density_temperature_comparison}
\end{figure*}

\begin{figure*}
    \centering
    \includegraphics[width=0.96\textwidth]{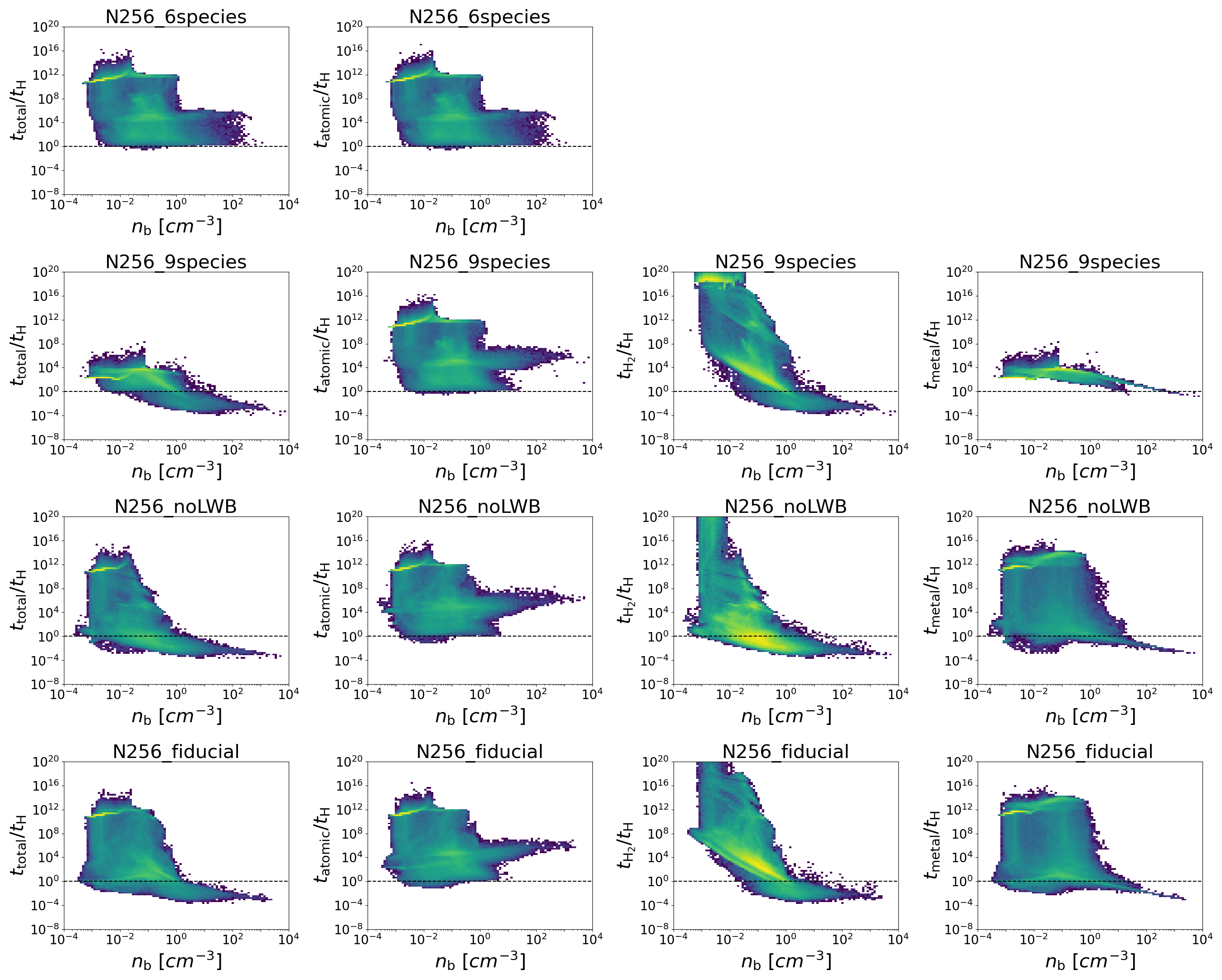}
    \caption{Comparison of diagrams of cooling time vs. baryon number density for the most massive halo in each of the $256^3$ simulations, except for \texttt{N256\_adiabatic}, which does not have gas cooling enabled. The $256^3$ simulations have the same initial conditions, so this is effectively a comparison of the same halo under different physical conditions. From left to right, total cooling time, atomic cooling time, H$_2$ cooling time, and metal cooling time are shown. Cooling times are normalized by the Hubble time at the final redshift.}
    \label{fig:tcool_vs_n_mmh_comparison}
\end{figure*}

\begin{figure*}
    \centering
    \includegraphics[width=0.96\textwidth]{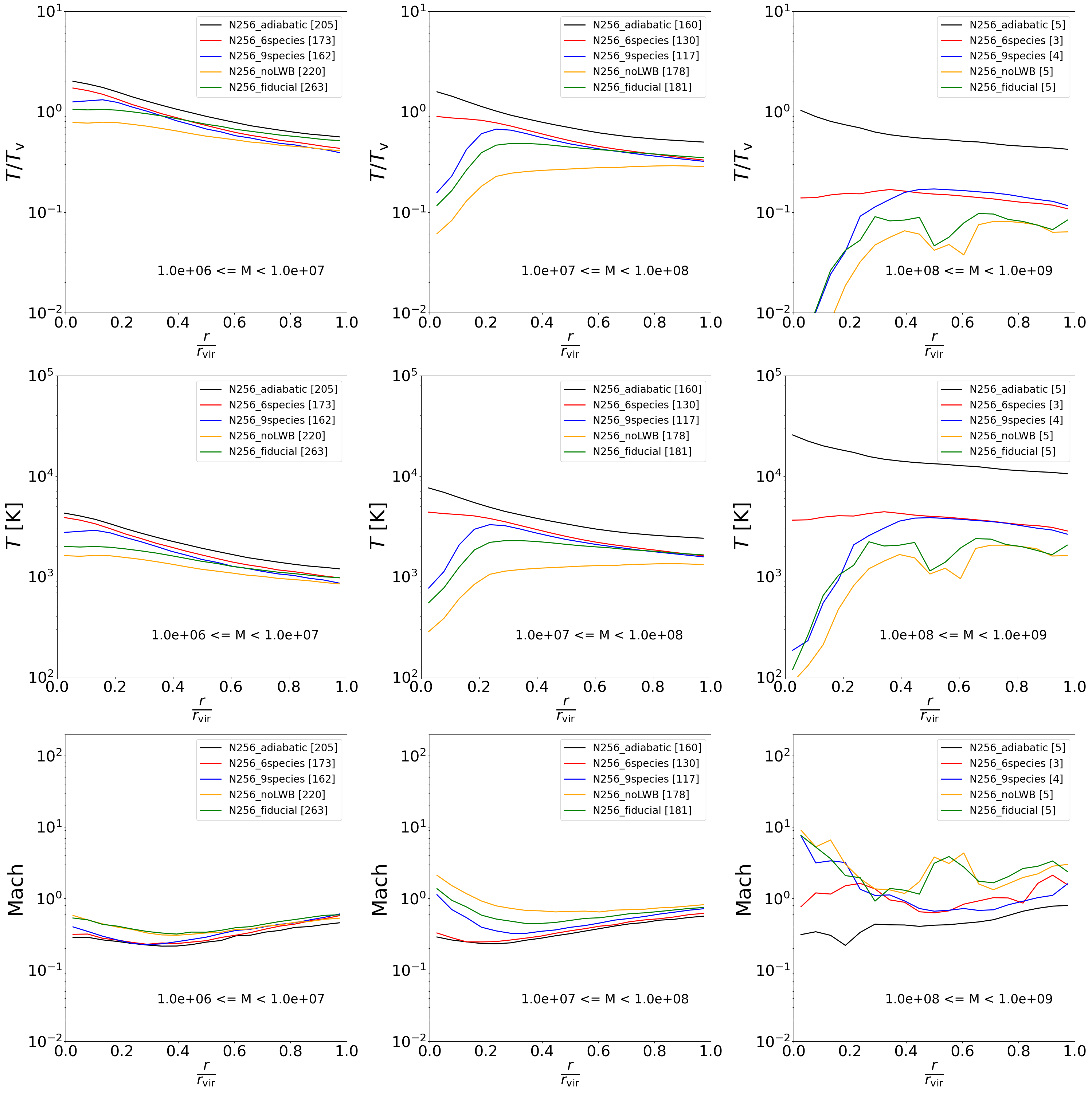}
 \caption{Average radial profiles for all halos in separate mass bins in the series of $256^3$ simulations. The top row shows gas temperature, normalized by the virial temperature. The middle row shows the un-normalized gas temperature. The bottom row shows mach number. The effect of using a uniform metallicity floor can be seen in the sharp drop in temperatures below $r/R_\mathrm{vir}=0.2$ in the blue temperature curves, which correspond to the \texttt{N256\_9species} simulation. Note that the halos samples for this figure include both cooling and non-cooling halos.}
    \label{fig:comparison_average_gas_properties}
\end{figure*}

To investigate how gas cooling is effected by varying feedback and chemistry prescriptions, we show phase diagrams of cooling time, normalized by the Hubble time ($t_\mathrm{cool}/t_\mathrm{H}$), versus baryon number density for the most massive halo in each of the $256^3$ simulations with gas cooling in Figure \ref{fig:tcool_vs_n_mmh_comparison}; the virial mass for each is listed in Table \ref{table: params}. These simulations have identical initial conditions, so Figure \ref{fig:tcool_vs_n_mmh_comparison} shows the same halo under different conditions for chemistry and cooling. In each of the panels in Figure \ref{fig:tcool_vs_n_mmh_comparison}, gas that is able to cool has $t_\mathrm{cool}/t_\mathrm{H} \leq 1$. This is generally true for gas with $n_\mathrm{b} > 10^0\,\,\mathrm{cm^{-3}}$ in three of the four simulations plotted, with \texttt{N256\_6species} being the exception. It can be seen in the second column that $t_\mathrm{cool}/t_\mathrm{H}=1$ is the approximate lower limit for atomic cooling, meaning that atomic cooling is effectively inactive for this halo. In support of Figure \ref{fig:cooling_times_vs_radius_average}, we can see that H$_2$ and metal cooling dominate in the densest gas for \texttt{N256\_fiducial} and \texttt{N256\_noLWB}. Interestingly, the \texttt{N256\_9species} case, which employs a metallicity floor of $Z_\mathrm{floor}=10^{-5.5}\,\,Z_\odot$, seems to under-predict the effects of metal cooling in dense gas, while over-predicting metal cooling in low-density gas (fourth column of Figure \ref{fig:tcool_vs_n_mmh_comparison}). The effects of the LWB can be seen in the H$_2$ cooling times (third column), where introducing a LWB increases H$_2$ cooling times by $\sim5$ orders of magnitude in low-density gas.

Slices of baryon number density, temperature, electron fraction, and entropy through the most massive halo in each of the $256^3$ simulations are shown in Figure \ref{fig:MMH_density_temperature_comparison}. The conventional wisdom is that gas that spherically accretes onto the halo (i.e. hot mode accretion) will shock heat to the virial temperature near the virial radius, whereas gas that flows in through filaments (i.e. cold mode accretion) will heat further in towards the center \citep{Keres2005} by weak shocks and compressional heating. The gas at the shock boundary will experience a sharp increase in entropy. This can be observed in the entropy slices of Figure \ref{fig:MMH_density_temperature_comparison}, most clearly in the adiabatic case. The shock boundary in the adiabatic case occurs at $r\approx R_\mathrm{vir}$, which is inline with the conventional wisdom. For all other cases, the shock boundary is notably less pronounced and is located at roughly $r=0.3\,R_\mathrm{vir}-0.5\,R_\mathrm{vir}$, in agreement with the early findings of \cite{WiseAbel2007}. For the non-adiabatic cases, the shock boundaries coincide with increased electron fractions due to collisional ionization. As more types of cooling are introduced (moving downward in Figure \ref{fig:MMH_density_temperature_comparison}), accretion flows onto the central region increase, and fragmentation can be observed.

Figure \ref{fig:comparison_average_gas_properties} shows a comparison between average profiles for all halos (i.e. not just cooling halos) in the $256^3$ simulations for temperature and Mach number. Central temperatures are highest for the \texttt{N256\_adiabatic} and \texttt{N256\_6species} cases across all mass ranges, as expected. These two cases are characterized by subsonic central flows. The lowest central temperatures are for \texttt{N256\_fiducial} and \texttt{N256\_noLWB}, which have supersonic central flows due to the low sound speed in the cold gas. The \texttt{N256\_9species} curves deviate from the rest at $r/r_\mathrm{vir} < 0.2$ because the inclusion of a metallicity floor allows all halos access to metal-line cooling, even those that would otherwise be pristine.  

%% file: Discussion.tex
\section{Discussion}
\label{sec:discussion}

\subsection{Cold Mode Accretion At High Redshifts}

Accretion histories of galaxies at $z \lesssim 4$ are studied in detail in \cite{Keres2005} and \cite{Keres2009} (K05 and K09 hereafter), where it is found that there are two distinct modes of accretion: a ``hot mode" and a ``cold mode". Hot mode accretion is typically experienced in high mass halos ($M_\mathrm{vir}\sim10^{12}\,\,M_\odot$), and is the case where gas accretes spherically into the halo, and gets shock-heated to the virial temperature in the process. Halos of lower masses tend to experience cold mode accretion, which is the case where gas flows into the halo smoothly through filaments. Because the accretion is smooth in the cold mode case, the inflowing gas is not shock-heated to the virial temperature during the process of accretion. The majority of the gas that ultimately coalesces in Milky Way-like galaxies by $z=0$ enters through cold mode accretion.

Our simulations probe accretion and cooling in galaxies during an entirely different phase of the universe than the simulations in K05 and K09. For one, we simulate a universe before reionization, when the IGM has not yet been heated to temperatures above the atomic cooling limit of $10^4$ K by the buildup of a global UV background. Additionally, the halos we resolve have lower masses, with values ranging between $10^6\,\,M_\odot$ and $10^9\,\,M_\odot$, as opposed to the $10^{10}\,\,M_\odot$ to $10^{14}\,\,M_\odot$ halos in K05 and K09. Regardless, cold mode accretion best describes the case that we observe in our simulations. In fact, the lack of shock-heated gas can also partially explain why we do not observe atomic cooling halos in our simulations, even though we resolve hundreds of halos with virial temperatures above the atomic cooling limit.

\cite{Greif2008} tracks the accretion history of a $5\times10^7\,\,M_\odot$ halo at $z\sim10$ within a $\sim700$ kpccm cosmological box. They do not include models for stellar feedback, yet allow the gas to undergo atomic and H$_2$ cooling. This is similar to our model, though we include Pop III feedback and subsequent metal cooling. In their study, they examine hot versus cold accretion in the halo's history, and find that there is an initial phase of hot accretion, with cold accretion taking over soon after. The hot accretion does manage to shock-heat a small amount of gas to $T\approx10^4$ K at the virial radius, but the gas quickly cools to $T\approx10^2$ K as it flows into the center, at which point the flows become supersonic and turbulent. This is consistent with our observations in Figures \ref{fig:gas_properties_vs_radius_average_0}, \ref{fig:mach_scatter}, \ref{fig:halo_phases}, and \ref{fig:comparison_average_gas_properties}. Figure 7 in \cite{Greif2008} shows phase diagrams of gas temperature and H$_2$ fraction for the $5\times10^7\,\,M_\odot$ halo, which can be compared with the gas phase diagrams for our five most massive halos in Figure \ref{fig:halo_phases}. There are some differences in the H$_2$ fractions at low densities due to the effects of the LWB in our simulation, but in both cases, H$_2$ fractions approach $10^{-3}$ in dense gas.

%%%%%%%%%%%%%%%%%%%%%%%%%%%%%%%%%%%%%

\subsection{Population III Star Formation Histories}

\cite{Trenti_2009b} derives the following equation that describes the minimum virial mass for a pristine minihalo to be able to cool in less than a Hubble time via H$_2$ cooling:
\begin{equation}
    \label{eq:trenti_min_mass}
    M_{t_\mathrm{H}-cool}\approx 1.54\times10^5\,M_\odot\left(\frac{1+z}{31}\right)^{-2.074}.
\end{equation}
At $z=11.92$, Equation \ref{eq:trenti_min_mass} evaluates to $9.5\times10^5\,\,M_\odot$, which is approximately equal to our minimum resolved halo mass of $10^6\,\,M_\odot$. From Figure \ref{fig:hmf_cooling}, we can see that there are a few H$_2$ cooling halos around $10^6\,\,M_\odot$, which is in good agreement with the prediction of Equation \ref{eq:trenti_min_mass}. Most of the cooling halos that we observe with $M_\mathrm{vir} \lesssim 3\times10^6\,\,M_\odot$ actually cool via metals. These low mass metal cooling halos are primarily enriched externally (Fig. \ref{fig:hmf_enrichment}), though we do see some internal Pop III star formation in a few of these halos (Fig. \ref{fig:popIII_halo_stats}).

\cite{Visbal_2020} uses semianalytic models to compute Pop III star formation histories for a set of dark matter halos extracteed from dark matter-only simulations of comparable mass resolution to our own, including a variety of different feedback effects. Specifically, they consider the effects of the LWB, internal and external metal enrichment, and reionization feedback. We do not include radiative feedback explicitly, meaning that our model is most comparable to the ``LW+M($Z_\mathrm{crit}=10^{-6}\,\,Z_\odot$)" case in \cite{Visbal_2020}, where $Z_\mathrm{crit}$ is the chosen metallicity that distinguishes Pop III star formation from metal enriched Pop II star formation. They find Pop III star formation occuring in halos with virial masses as low as $2\times10^6\,\,M_\odot$, which is consistent with our results in Figure \ref{fig:popIII_halo_stats}, as well as with the minimum H$_2$ cooling mass computed using Equation \ref{eq:trenti_min_mass}. Our computed Pop III SFRD in Figure \ref{fig:popIII_SFR} is generally in good agreement with those in \cite{Visbal_2020}, though our rates are higher than the ``LW+M" rates by a factor of 2-3 by the final output. Due to the many differences between our models and implementations, it is hard to say exactly where the deviation in our derived SFRDs originates, but the most likely reason is that we take into account H$_2$ self-shielding, while \cite{Visbal_2020} does not. H$_2$ self-shielding works against the H$_2$ photodissociating LWB, which works to suppress Pop III star formation. The inclusion of H$_2$ self-shielding therefore increases Pop III SFRs. They cite inhomogeneous reionization and pollution of pristine halos by metal-enriched winds from nearby galaxies for the general flattening of their Pop III SFRD below z=15. We do not include these effects in our simulations, which is an additional explanation for why our SFRD overshoots theirs at lower redshifts.  

%%%%%%%%%%%%%%%%%%%%%%%%%%%%%%%%%%%%%%%

\subsection{The Role of Primordial Chemical Enrichment}

Primordial chemical enrichment is an essential ingredient for galaxy formation. After all, all main-sequence stars that have been observed have nonzero metallicity, so it stands to reason that the gas in which they formed was metal enriched. We have replicated previous results that suggest that the primary mechanism for halo enrichment in the early universe is external enrichment, which is the case where metals are sourced from supernovae occuring outside the virial radius \citep{Hicks_2021, smith2015}.

As a part of the \textit{Birth of a Galaxy} series, \cite{wise2014} (hereafter W14) simulates chemical enrichment by Pop III stars explicitly, and finds a population of low mass halos ($10^6\,\,M_\odot < M_\mathrm{vir} < 10^8\,\,M_\odot$) that are able to form metal enriched stars despite the fact that their virial temperatures are below the atomic cooling limit. These halos are directly comparable to the population of metal cooling halos that we identify in our \texttt{N512\_fiducial} simulation. W14 performs a similar cooling analysis to the one we describe in Section \ref{sec:accretion_cooling}, and finds a clean transition from H$_2$ cooling to metal cooling at $M_\mathrm{vir} \gtrsim 10^7\,\,M_\odot$, and another transition from metal cooling to atomic cooling at $T_\mathrm{vir} = 10^4$ K (see Figure 2 of W14). In contrast, we find that H$_2$ cooling is overall dominant for halos with $M_\mathrm{vir} > 10^7\,\,M_\odot$, and do not find any atomic cooling halos in the sample. 

The main reason for our differing results with W14 is likely that W14 includes a full physics suite with Pop II star formation and feedback, including radiative feedback. This suggests that the relative importance of the three cooling rates we consider is modulated by stellar feedback from metal enriched stars. Since we do not observe any atomic cooling halos in our sample, it is possible that stellar feedback from metal enriched stars is actually a requirement for atomic cooling halos to exist. This is contrary to the usual story, which is typically recited in the literature in a way that suggests the opposite---that metal enriched star formation is not a cause, but a consequence of efficient atomic cooling in halos with $T_\mathrm{vir} > 10^4$ K (e.g., \cite{Greif2015}).

%%%%%%%%%%%%%%%%%%%%%%%%%%%%%%%%%%%%%%%%

\subsection{H$_2$ Cooling and Self-Shielding}

\cite{ahvazi2023} computes occupation fractions for halos of a given mass to form a dwarf galaxy, and find that H$_2$ cooling is the essential ingredient for the occupation fractions to match up with predictions (see Figure 2 in \cite{ahvazi2023}). The analysis used in the study takes the gas metallicity to be uniform throughout the box. Our results corroborate with this key finding. We have found that H$_2$ cooling tends to dominate over metal cooling within R$_{1000}$ for the majority of halos in our sample.

The strength of H$_2$ cooling is partially reliant on H$_2$ self-shielding, which protects the innermost regions of a halo from H$_2$ photodissociation by the LWB in halos with H$_2$ column densities greater than $10^{14}\,\,\mathrm{cm}^{-2}$ \citep{Draine1996}. We have seen in Figure \ref{fig:gas_properties_vs_radius_average_1} that the vast majority of cooling halos with $M_\mathrm{vir} > 10^{7}\,\,M_\odot$ in the \texttt{N512\_fiducial} simulation have $N_\mathrm{H_2} > 10^{14}\,\,\mathrm{cm^{-2}}$. The effect of H$_2$ self-shielding can be seen more explicitly when comparing the H$_2$ cooling times for the most massive halos in the \texttt{N256\_fiducial} and \texttt{N256\_noLWB} simulations in Figure \ref{fig:tcool_vs_n_mmh_comparison}. The impact of H$_2$ self-shielding on Pop III star formation are studied in \cite{Skinner2020}, where it is found that self-shielding lowers the minimum halo mass necessary for Pop III star formation to values as low as $3\times10^5\,\,M_\odot$. As a result, primordial chemical enrichment can occur at earlier redshifts, which is the precursor to metal enriched star formation.

%%%%%%%%%%%%%%%%%%%%%%%%%%%%%%%%%%%%%%%%

\subsection{Implications for Galaxy Formation Models}

Cooling is a driver for star formation within a halo. The different cooling processes considered in this study operate efficiently at different gas temperatures. H atomic line cooling is effective at cooling gas down to $10^4$ K. Below this temperature, atomic cooling quickly becomes inefficient as collisional excitations become rare, and H$_2$ cooling is required to cool the gas further \citep{ABN00,BrommLarson2004}. H$_2$ cooling will cool the gas down to $\sim200$ K. In pristine gas, this minimum temperature of $\sim 200$ K will result in a large Jeans mass, and will thus produce high mass Pop III stars \citep{ABN02}. If metals are present, the gas will cool and condense further before reaching stellar densities. Semianalytic models for galaxy formation in the early universe are often applied to pure dark-matter-only simulations, and thus do not have the ability to directly measure the various cooling rates in the way that we have in this study. Instead, the virial temperature is typically used as a metric to label halos as atomic cooling or otherwise. Even in hydrodynamical simulations that do include explicit chemistry and cooling, it is common practice to identify halos as atomic cooling if $T_\mathrm{vir} > 10^4$ K. 

\cite{chen2014} derives a series of scaling relations for galaxies formed in the \textit{Renaissance Simulations}, specifically in the ``Rarepeak" region at $z=15$. Most notably, the relationship between SFR and dark matter virial mass is piecewise, with a different slope on either side of $10^8\,\,M_\odot$ (see Fig. 3 in \citealt{chen2014}). All measured relationships in the study that depend on the SFR also have this feature. It is erroneously stated in the study that the discontinuity at $M_\mathrm{vir} \simeq 10^8\,\,M_\odot$ corresponds to $T_\mathrm{vir}\simeq10^4\,\,\mathrm{K}$, when the virial mass corresponding to $T_\mathrm{vir}\simeq10^4$ K at $z=15$ is actually $M_\mathrm{vir}\simeq2\times10^7\,\,M_\odot$. We speculate that the observed star formation efficiency at $M_\mathrm{vir} \simeq 10^8\,\,M_\odot$ is the result of self-enrichment of the halo by Pop II supernova feedback. Regardless, it is clear that the $10^7\,\,M_\odot-10^9\,\,M_\odot$ mass range marks a transition from discrete star-forming events separated by quenching due to stellar feedback to sustained Pop II star formation. This is demonstrated in Figure 22 of \cite{Xu16c}, which shows the fraction of halos in the \textit{Renaissance Simulations} that are actively forming stars at the final redshift for each simulated region. The curve is discontinuous, showing increased active star formation in halos with $M_\mathrm{vir} \gtrsim 10^8\,\,M_\odot$. It is clear from our results that while atomic cooling likely plays a role in counteracting gas heating from feedback, atomic cooling alone is not strong enough to mark a transition to efficient star formation in halos. It can be observed in Figure \ref{fig:hmf_enrichment} that virtually all halos with $T_\mathrm{vir} > 10^4$ K are chemically enriched by $z\sim12$, and are thus all candidates for Pop II star formation. These halos are generally massive enough to be able to contain outflows generated by Pop II supernovae through gravity, which could lead to the sustained star formation that has been observed in the \textit{Renaissance Simulations}. For now, this is left as a point of speculation. We will explore this topic further in a followup study.

We have shown that the virial temperature is not an accurate indicator of gas temperature for halos in the early universe with virial masses between $10^6\,\,M_\odot$ and $10^9\,\,M_\odot$, and as such, semianalytic arguments that rely on the virial theorem for assessing the state of the gas in the regime we study should be made with caution. Furthermore, defining an atomic cooling halo as a halo with a virial temperature of $T_\mathrm{vir}>10^4$ K is potentially misleading, as it is not at all guaranteed that the gas within such a halo would cool primarily through atomic H and He transitions. Our results suggest that gas heating from stellar feedback must be taken into account for efficient atomic cooling to take place in any capacity.

%%%%%%%%%%%%%%%%%%%%%%%%%%%%%%%%%%%%%%%%

\subsection{Caveats}

There are a few shortcomings in our analysis that should be noted. As mentioned throughout the text, we do not include Pop II stellar feedback in our simulation. Pop II supernovae within a star-forming halo would dramatically impact the dynamics of the gas within $R_{1000}$. The relative importance of atomic, H$_2$, and metal cooling within star-forming halos would also be affected, though exactly how is not clear. Atomic cooling rates would definitely become non-negligible due to the sharp increase in temperature in the wake of a supernova, though the energy injected into the gas would also induce collisional ionization of atomic hydrogen and helium, which would work against the increase in atomic cooling. The increased electron fractions that result from the collisional ionization of hydrogen and helium combined with the increased metal fractions due to stellar winds and metal ejection during supernovae would also serve to increase metal cooling rates. H$_2$ cooling rates would similarly be affected.

We also do not include the effects of radiative feedback. Radiation from young massive stars will ionize atomic hydrogen and helium, as well as heat the gas. In particular, local radiation from Pop II stars in the Lyman-Werner bands will photodissociate H$_2$ in star-forming halos. As a result, we expect that H$_2$ cooling will become less important for the cooling of the gas once a halo has achieved Pop II star formation.

The statistics for primordial stellar populations predicted by \texttt{StarNet} (number of Pop III stars, stellar masses, etc.) are randomly pulled from distribution functions derived from the \textit{Phoenix Simulations}. While this approach is sound for the sake of global statistics, the properties of individual stellar populations do not directly depend on the state of the alleged star-forming gas. While topics related to the Pop III stellar IMF and primordial stellar populations remain highly uncertain in the literature, it should be noted that this is also point of uncertainty in our model. Regardless, good agreement for a variety of halo-averaged properties were shown in \cite{Wells_2022b} between simulations that directly resolve Pop III star formation and feedback, and simulations that use \texttt{StarNet} to predict its effects.

%% file: Conclusion.tex
\section{Conclusion}
\label{sec:conclusion}

We have studied the chemical history and dynamics of gas within a large sample of dark matter halos with virial masses on the range $10^6\,\,M_\odot < M_\mathrm{vir} < 10^9\,\,M_\odot$ in the absence of Pop II star formation and feedback at high redshifts. The halos are generated in a $(5.12\,\,\mathrm{cMpc})^3$ volume using a large \texttt{Enzo-E} AMR cosmology simulation that includes models for gravity, multispecies hydrodynamics, gas chemistry and cooling, an H$_2$ photodissociating LWB, and Pop III chemical enrichment. The simulation is run down to $z=11.92$. The Pop III chemical enrichment is handled using the machine learning surrogate model, \texttt{StarNet}, which is called inline with the simulation. \texttt{StarNet} deposits metals from one or more Pop III supernovae in a composite supernova remnant near its terminal stage of expansion. This novel approach allows us to simulate larger volumes for improved halo statistics by bypassing the early expansion phase of Pop III remnants.   

We find that $16\%$ of halos are chemically enriched by the final output, and thus have access to metal-line cooling. We then calculate average cooling rates for gas within $R_{1000}$ for each halo, determine which halos contain a significant amount of gas that is able to cool in less than a Hubble time, and classify these halos as either (1) atomic cooling halos, (2) H$_2$ cooling halos, or (3) metal-cooling halos based on which cooling process is dominant. $12\%$ of halos in our sample are able to cool by any means. The main conclusions from our analysis are:
\begin{enumerate}
    \item H$_2$ is the dominant coolant for halos in our resolved mass range of $10^6\,\,M_\odot < M_\mathrm{vir} < 10^9\,\,M_\odot$, even for halos that are metal enriched. Metal cooling also makes a significant contribution; however, only 18\% of the cooling halos are identified as ``metal cooling" at the final redshift. For the most massive halos, it is typical for H$_2$ and metal cooling to contribute roughly equal amounts to the total cooling within $R_{1000}$.
    \item True atomic cooling halos (cooling by H and He line excitation) are exceedingly rare for halos in our resolved mass range, even for halos with $T_\mathrm{vir} > 10^4$ K. In fact, we find that there are no atomic cooling halos in our sample at all.  What little atomic cooling that is present is insufficient to induce the gravitational collapse of the first Pop II stars in a halo. Rather, it is likely that feedback from Pop II stars is actually required to heat the gas to temperatures sufficient for radiative transitions in atomic hydrogen and helium to take place. We will explore this topic in a followup study.
    \item For the majority of halos with $M_\mathrm{vir} < 10^9\,\,M_\odot$, the gas contained within the virial radius is not thermally virialized. Rather, consistent with earlier findings, \citep{WiseAbel2007,Greif2008}, we find that virial equilibrium is achieved by the additional kinetic energy of bulk flows and turbulence within the virial radius. Thanks to our large sample of halos, we find that the relative contribution of bulk kinetic energy to thermal energy of the gas systematically increases with virial mass (see Fig. \ref{fig:virial_check}), independent of the cooling properties of the halo.  In addition, the mean temperature within the virial radius for halos with $M_\mathrm{vir} > 10^7\,\,M_\odot$ is typically much less than the virial temperature due to runaway cooling near the center. In the absence of gas cooling, mean temperatures are still less than the virial temperature by a factor of $\sim 2$.
    \item Gas accretes onto halos primarily by cold mode accretion along filaments. Gaseous filaments at $z\sim12$ have temperatures between $10^2$ K and $5\times10^3$ K due to strong adiabatic shocks bounding the filaments. 
%    \item For the majority of halos, gas flows into the virial radius cold, with temperatures less than $10^3$ K. This gas is also typically not cooling efficiently by the time it crosses the virial radius.
    \item H$_2$ self-shielding from the LWB is active for halos across our entire resolved mass range. The effectiveness of H$_2$ self-shielding generally increases as halo mass increases. For halos with $M_\mathrm{vir} > 10^8\,\,M_\odot$, gas generally flows in with larger H$_2$ fractions than for lower mass halos. As a result, the average radial profile of H$_2$ fraction for $M_\mathrm{vir} > 10^8\,\,M_\odot$ lacks the sharp increase at $r/R_\mathrm{vir}\approx0.2$ that is typical for the lower mass halos, yet still reach values at the centers that are an order of magnitude higher than the lower mass halos.
    \item Assuming an homogeneous metallicity floor of $Z_\mathrm{floor}=10^{-5.5}\,\,Z_\odot$ in place of a model for Pop III star formation and feedback leads to over-estimated metal cooling in low density gas and under-estimated metal cooling in high density gas. Additionally, it gives all halos access to some amount of metal cooling, while, according to our fiducial simulation, only 16\% of halos are chemically enriched by $z\sim12$.
    \item Through a post-processing approach, we estimate Pop II stellar masses and SFRs for our sample of halos in our fiducial simulation and obtain values that are in good agreement with those seen in the \textit{Renaissance Simulations}. Whether a halo is an H$_2$ cooling or a metal cooling halo appears to have no effect on stellar mass or SFR.
\end{enumerate}

We have found that the commonly assumed progression of halos from H$_2$ cooling to metal cooling to atomic cooling needs to be analyzed further at high redshifts, and is likely not reflective of the truth. In reality, things are much more complicated. Caution should be taken when relying on the virial theorem to assess the thermal state of the gas within halos at $z\gtrsim10$, as it is not at all guaranteed that the gas within a given halo of $M_\mathrm{vir} < 10^9\,\,M_\odot$ is thermalized. Rather, we have found that kinetic motion of the gas (i.e. turbulence) contributes increasingly to the virial energy budget as halo mass increases. In a followup study, we will re-do our analysis in the presence of Pop II star formation and feedback to see what results change and what stays the same. We will place a focus on the properties of the galaxies that form, and will discuss any connections that can be made with recently observed JWST galaxies at $z>10$.

%% file: Validating.tex
\label{sec:comparisons}
Two sets of three cosmological simulations were run in order to compare the results and performance of \texttt{Enzo-E} with \texttt{Enzo} using simulations with identical parameters. First is an AMR simulation with a box size of 12 Mpc/h, a root grid of $64^3$ cells, and a maximum refinement level of 4. Refinement is triggered in a block when a cell satisfies $M>\delta_\mathrm{thresh}(\Delta x_\mathrm{root})^3$, where $\delta_\mathrm{thresh}=8$ is the chosen overdensity threshold, including both the dark matter and baryon components. Hydrodynamic fields are evolved using a piecewise parabolic method \citep{Bryan1995, Colella1984}. Additional density fields corresponding to six chemical species (HI, HII, HeI, HeII, HeIII, and $e^-$) are advected, and chemistry and cooling is performed using \texttt{Grackle} \citep{Grackle}. A global UV background is implemented using photoionization and cooling rates from \cite{Puchwein_2019}.

Gravity is handled completely differently in \texttt{Enzo} and \texttt{Enzo-E}, owing to differences in the AMR mesh structures and timestepping. \texttt{Enzo} uses structured AMR \citep{BC89,Enzo2014}, in which an adaptive hierarchy of overset grid patches of varying size and shape are evolved over a global root grid mesh. Patches at different levels of refinement are evolved with different timesteps according to a W-cycle schedule. Field data at different levels of refinement are kept consistent through conservative interpolation and restriction operators, as well as through flux correction. \texttt{Enzo-E}, on the other hand, uses a composite multiresolution adaptive mesh generated by array-of-octree refinement \citep{bordner_2018}. There are no overset meshes of lower resolution as in \texttt{Enzo}. Additionally every mesh patch regardless of level is cubic and of equal size (in terms of mesh points). Currently, the solution is evolved using a single global timestep, although work is underway on block-adaptive timestepping. 

Given this background, we can now briefly describe how the Poisson equation (PE) is solved on the adaptive mesh to get the gravitational potential. In \texttt{Enzo}, first the PE is solved on the uniform, periodic root grid using FFTs. Then, all patches at the next level of refinement are evolved with a smaller timestep, and so on recursively until the finest level is reached with a W-cycle schedule. The gravitational potential on a refined patch is computed as an isolated PE problem using a geometric multigrid solve, interpolating its boundary conditions from its parent grid or root grid as the case may be. In \texttt{Enzo-E}, since global timesteps are used to evolve the solution at every level of refinement simultaneously, the PE must be solved everywhere on the multiresolution mesh. This is done in several steps. First, the density field is restricted to the root grid from the leaf nodes of the array-of-octree mesh. Second, the global PE is solved on the root grid using V-cycle geometric multigrid for the triply periodic domain. Third, the PE is solved on the multilevel mesh for each refined octree of the array using the BiCGStab algorithm, interpolating the global potential obtained in step 1 onto the boundary faces of the octree mesh. Fourth, a Jacobi smoothing step is applied to the global, multilevel potential. In both \texttt{Enzo} and \texttt{Enzo-E}, the density field is advanced a half timestep to properly time center the gravitational acceleration relative to the hydro solver. In both \texttt{Enzo} and \texttt{Enzo-E}, collisionless N-body dynamics is solved using the Particle-Mesh method with cloud-in-cell (CiC) spatial interpolation and leapfrog time integration. For both codes, ghost zones are included around grid boundaries to reduce artifacts in the computed gravitational potential at the interfaces between grids. 

\begin{figure}
    \centering
    \includegraphics[width=0.9\textwidth]{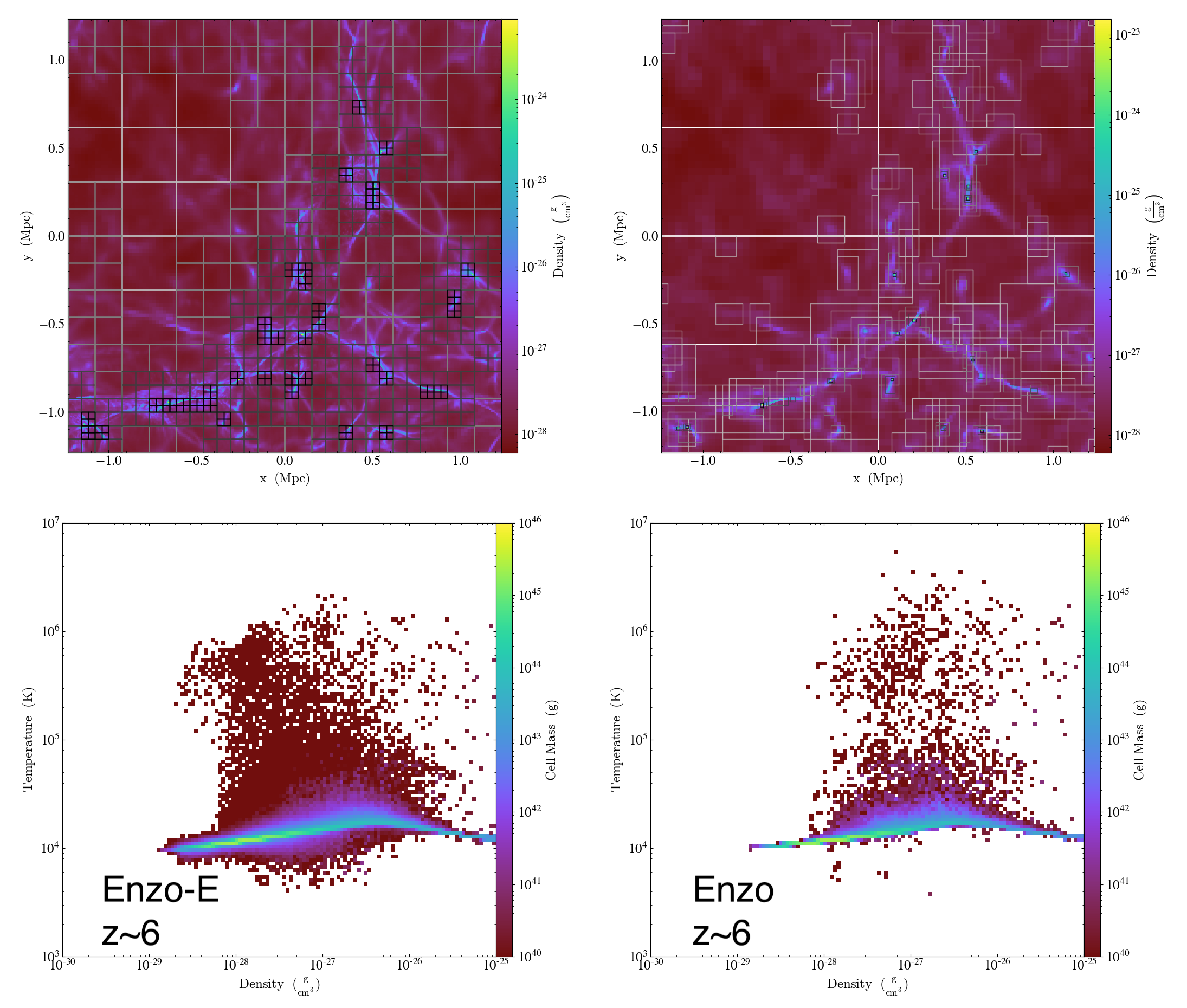}
    \caption{Comparison between \texttt{Enzo-E} (\textit{left}) and \texttt{Enzo} (\textit{right}) for a small cosmological simulation that includes a global UV background, showing density projections (\textit{top}) and phase diagrams (\textit{bottom}). The density projections have the grid overlaid to demonstrate the difference in AMR structure between the two codes. The \texttt{Enzo-E} simulation shows much more refinement overall, leading to better sampling of the density distribution, as shown by the phase diagrams.}
    \label{fig:Enzo_EnzoE_AMR}
\end{figure}

Figure \ref{fig:Enzo_EnzoE_AMR} shows density projections and phase diagrams at $z\sim6$ for each run. The density projections have the AMR mesh overlaid, demonstrating the difference between the two approaches. Note each square or rectangle corresponds to the boundary of a mesh block, not a single cell. \texttt{Enzo}'s structured AMR creates high-resolution patches in regions of interest that vary in size and dimension. This can be seen in the top right panel of Figure \ref{fig:Enzo_EnzoE_AMR}. The \texttt{Enzo-E} mesh is notably much more regular in its structure (top left of Fig. \ref{fig:Enzo_EnzoE_AMR}). \texttt{Enzo-E}'s AMR approach also results in more refined volume overall. By $z\sim6$, the \texttt{Enzo-E} simulation contains about 20 times more computational cells than the \texttt{Enzo} simulation. As a result, the phase diagram in the \texttt{Enzo-E} simulation is better sampled than the \texttt{Enzo} simulation, particularly in low-to-average-density regions. This comparison presents an important distinction: \texttt{Enzo-E} generates more data and can be more memory-intensive to use than \texttt{Enzo}, but it provides better resolution in regions at moderately high densities. Although \texttt{Enzo-E} generates more grid data, its parallel scaling in terms of memory is better than \texttt{Enzo}. This is because \texttt{Enzo}'s patchy grid structure requires dynamic reading and writing of AMR metadata for each grid patch (block center, dimensions, etc.). For sufficiently large problems, then, \texttt{Enzo-E} is actually less memory-intensive to use than \texttt{Enzo}. This, coupled with \texttt{Enzo-E}'s nearly ideal scaling makes \texttt{Enzo-E} ideal for large-box simulations that wish to resolve the physics of gas flowing into and out of dark matter halos (e.g. gaseous filaments, the CGM, etc.).

\begin{figure}
    \centering
    \begin{subfigure}{0.45\textwidth}
        \includegraphics[width=\textwidth]{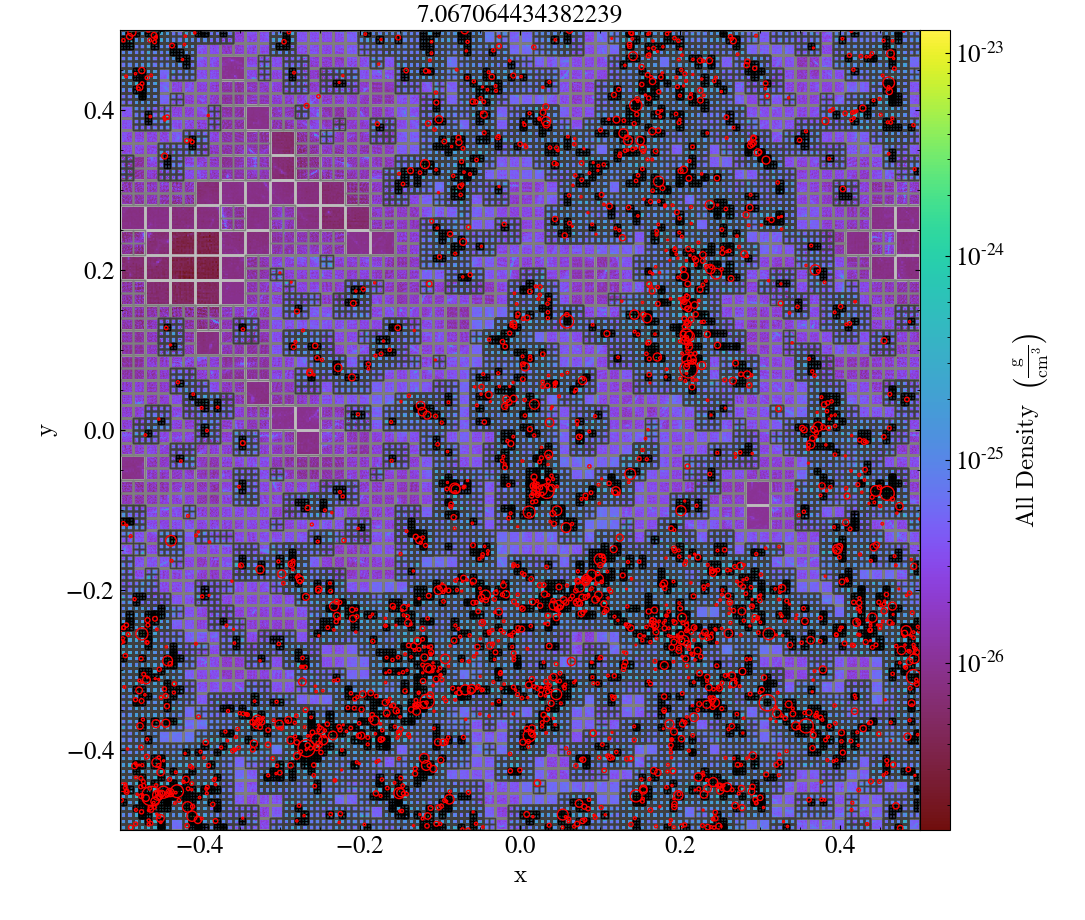}
    \end{subfigure}
    \begin{subfigure}{0.45\textwidth}
        \includegraphics[width=\textwidth]{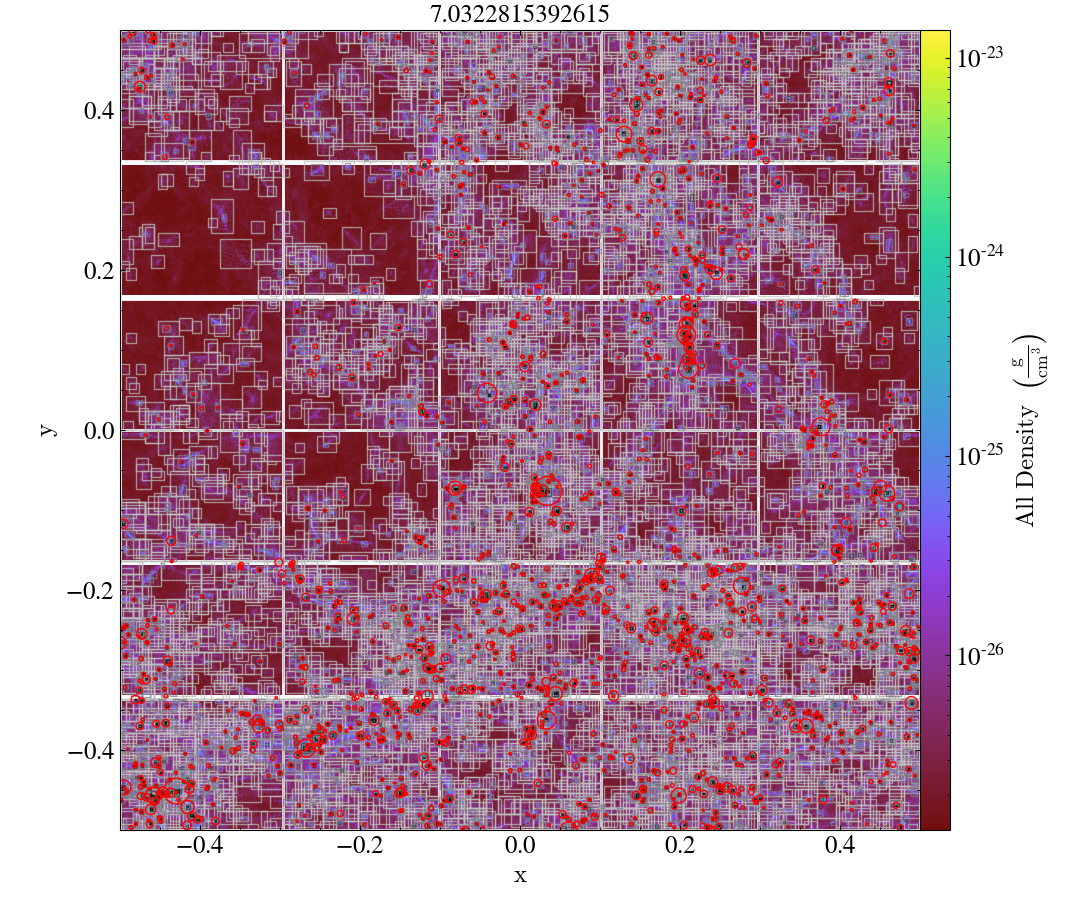}
    \end{subfigure}
    \caption{Dark matter density projections for identical $256^3$ dark-matter-only simulations run with \texttt{Enzo-E} (\textit{left}) and \texttt{Enzo} (\textit{right}). For each projection, the computational blocks are overlaid to demonstrate the difference in AMR structure. Halos are annotated as red circles, with radius equal to the virial radius.}
    \label{fig:Enzo_EnzoE_dm_only_projections}
\end{figure}

\begin{figure}
    \centering
    \begin{subfigure}{0.40\textwidth}
        \includegraphics[width=\textwidth]{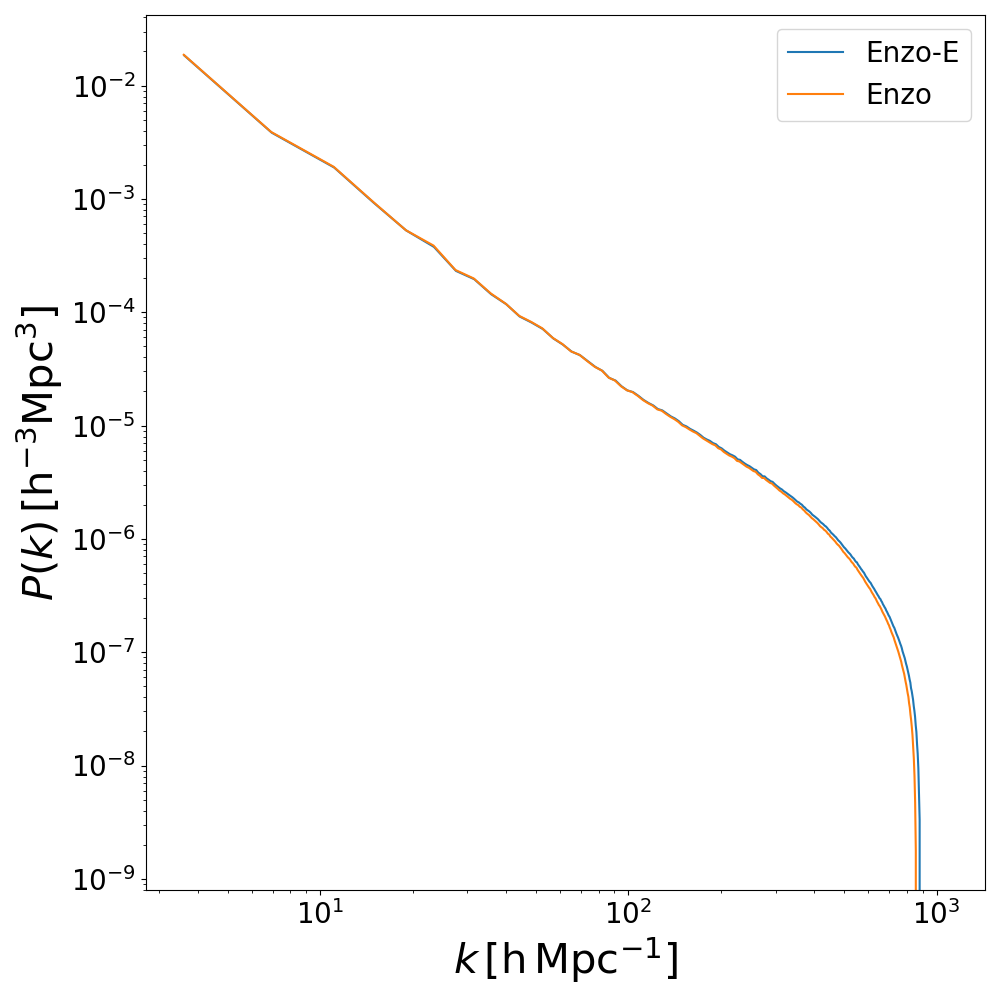}
    \end{subfigure}
    \begin{subfigure}{0.40\textwidth}
        \includegraphics[width=\textwidth]{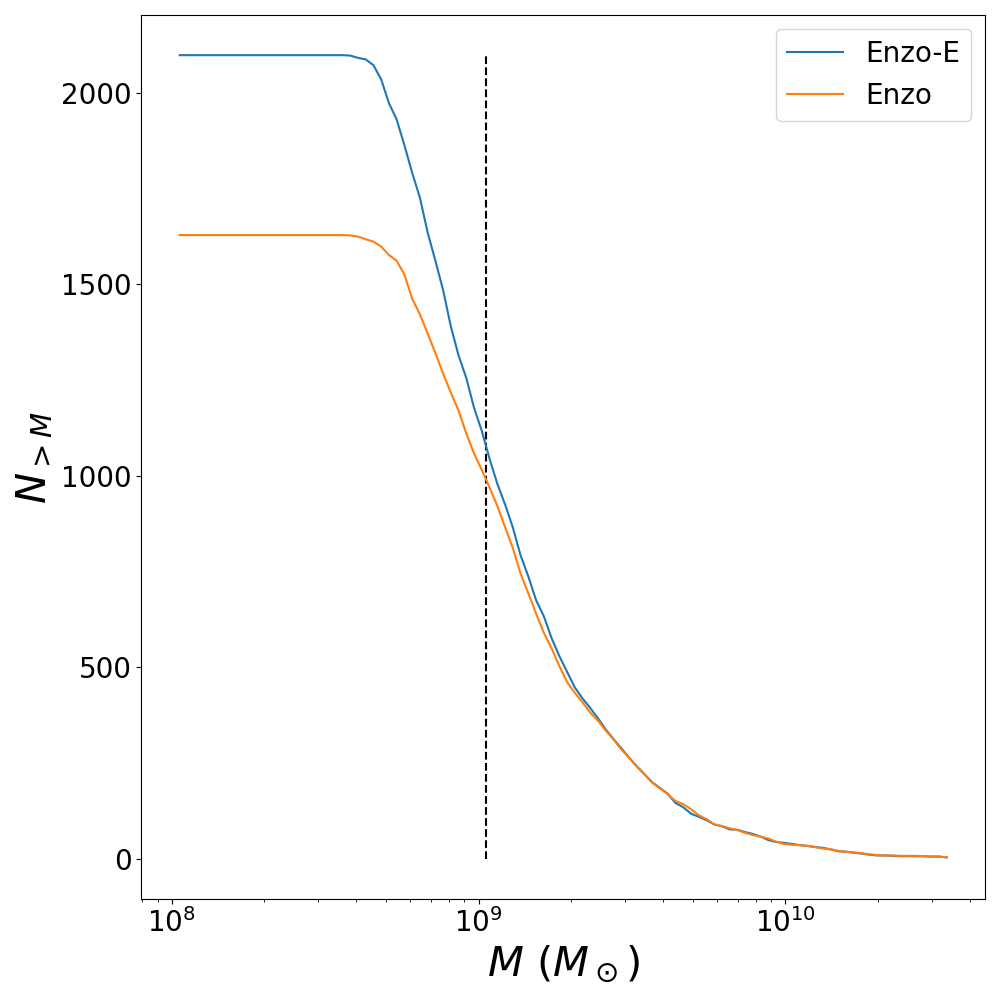}
    \end{subfigure}
    \caption{\textit{Left:} dark matter density power spectrum comparing \texttt{Enzo-E} and \texttt{Enzo} in the $256^3$ dark-matter-only simulations at $z\sim7$. \textit{Right:} Halo mass function for the same simulations at $z\sim7$. The ordinate axis shows the number of halos with virial mass greater than the value denoted by the abscissa. The vertical dashed line denotes a mass that is equal to $100\,M_\mathrm{p}$, where $M_\mathrm{p}$ is the dark matter particle mass. The difference in AMR approaches between the two codes shows some differences at small scales.}
    \label{fig:Enzo_EnzoE_dm_only_structure}
\end{figure}

To assess the impact, if any, the differences in AMR approaches and linear Poisson solvers have on structure formation, a dark-matter-only simulation was run using each of the codes. This simulation has a $256^3$ root grid, 4 levels of AMR, and a box size of 25 Mpc/h. Dark matter halos are identified using HOP \citep{HOP}. Mesh projections at $z\sim7$ are shown in Figure \ref{fig:Enzo_EnzoE_dm_only_projections}, with halos shown as red circles. The left panel of Figure \ref{fig:Enzo_EnzoE_dm_only_structure} shows a dark matter power spectrum for the \texttt{Enzo-E} and \texttt{Enzo} versions of the simulation. The two curves show very good agreement, with some slight differences at large $k$ (small scales). The differences at small scale can be seen more clearly in the right panel of Figure \ref{fig:Enzo_EnzoE_dm_only_structure}, which shows halo mass functions. The halo mass functions show that \texttt{Enzo-E} tends to generate more low-mass halos. This makes sense because the improved resolution in low-density regions due to \texttt{Enzo-E/Cello}'s AMR allows for a more accurate solution to the Poisson equation.

\begin{figure}
    \centering
    \includegraphics[width=0.9\textwidth]{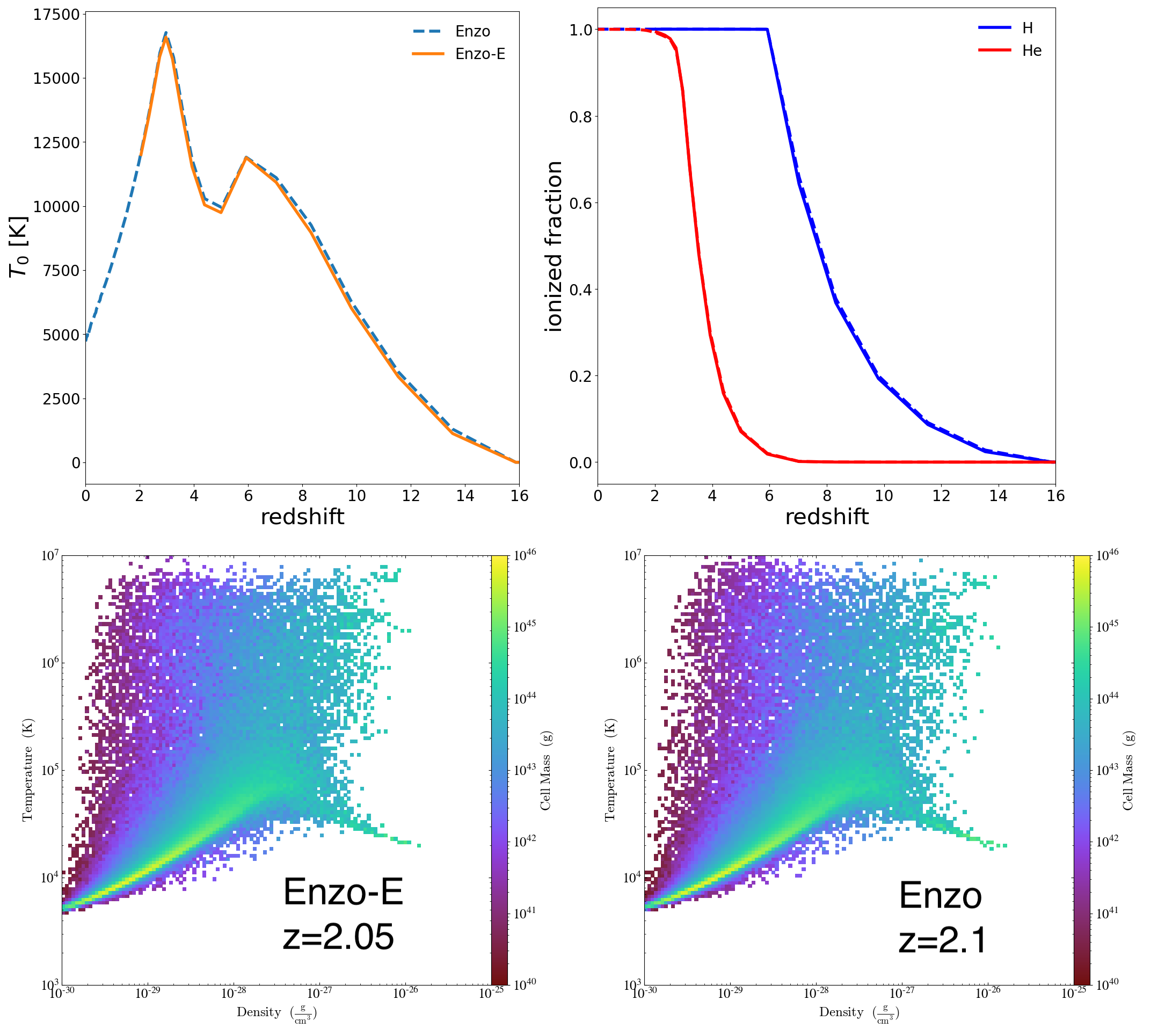}
    \caption{Unigrid comparison between \texttt{Enzo-E} and \texttt{Enzo} for a small cosmological simulation. \textit{Top left:} average temperature at mean density vs. redshift. $T_0$ is calculated as the median temperature of the subset of cells with density within 1 percent of the mean. \textit{Top right:} ionized fraction within the box for hydrogen and helium vs. redshift. Solid lines correspond to the \texttt{Enzo-E} simulation while the dashed line corresponds to the \texttt{Enzo} simulation. \textit{Bottom left:} phase diagram at $z\sim2$ for the \texttt{Enzo-E} simulation. \textit{Bottom right:} phase diagram at $z\sim2$ for the \texttt{Enzo} simulation.}
    \label{fig:Enzo_EnzoE_unigrid}
\end{figure}

A unigrid simulation was also run to check that \texttt{Enzo-E} and \texttt{Enzo} agree under circumstances where the computational mesh is identical between the two codes. This simulation uses all the same physics methods and models used in the previous AMR simulation, but has a $128^3$ root grid and a box size of 25 Mpc/h. Figure \ref{fig:Enzo_EnzoE_unigrid} shows the results of this unigrid simulation. The top left panel shows the median temperature of the box versus redshift. The temperature grows with time due to photoheating from the global UV background, with two features at $z\sim7$ and $z\sim3$ that coincide with the complete reionization of hydrogen ($z\sim7$) and helium ($z\sim3$) in the IGM. The decrease in temperature after reionization is primarily a result of cooling due to cosmological expansion. The top right panel shows this by plotting ionized fraction of hydrogen and helium versus redshift. The top left and right panels both show very good agreement between the \texttt{Enzo-E} and \texttt{Enzo} runs. The bottom left and right panels show phase diagrams of the \texttt{Enzo-E} (left) and \texttt{Enzo} (right) runs at $z\sim2$, which are nearly identical.

